\documentclass[12pt]{article}

\usepackage{booktabs}
\usepackage{tabularx}
\usepackage{soul}
\usepackage{float}
\usepackage{setspace}
\usepackage{multirow}
\usepackage{rotating}
\usepackage{eurosym}
\usepackage[english]{babel}
\usepackage[utf8x]{inputenc}
\usepackage[T1]{fontenc}
\usepackage[a4paper, total={6in, 10in}]{geometry}
\usepackage[round,authoryear]{natbib}
\usepackage{amsmath}
\usepackage{amssymb}
\usepackage{amsthm}
\usepackage{multirow}
\usepackage{graphicx}
\usepackage[colorinlistoftodos]{todonotes}
\usepackage[colorlinks=true, allcolors=blue]{hyperref}
\usepackage{fixltx2e}
\usepackage{amsfonts}
\usepackage{mathrsfs}
\usepackage{float}
\usepackage{verbatim}
\usepackage{comment}
\usepackage{graphicx}
\usepackage{enumerate}
\usepackage{threeparttable}
\usepackage{latexsym}
\usepackage{lscape}
\usepackage{bm}
\usepackage{subcaption}
\usepackage[title]{appendix}
\usepackage{mathtools}

\usepackage{xcolor}
\usepackage[linesnumbered,ruled,vlined]{algorithm2e}

\SetCommentSty{mycommfont}

\SetKwInput{KwInput}{Input}                
\SetKwInput{KwOutput}{Output}              

\newtheorem{theorem}{Theorem}[section]

\newtheorem{corollary}[theorem]{Corollary}

\newtheorem{assumption}{Assumption}[section]

\newtheorem{lemma}{Lemma}[section]

\theoremstyle{definition}

\newtheorem{thm}{Theorem}[section]

\numberwithin{equation}{section}
\begin{document}
\title{Iterative Distributed Multinomial Regression\thanks{We are grateful to three anonymous referees, an Associate Editor,
and a Co-Editor for their insightful comments on the previous version
of the paper. We thank Jean-Marie Dufour, Hiro Kasahara, Artem Prokhorov,
Dacheng Xiu, Jin Yan, and participants of the seminar at the University
of British Columbia, the University of Melbourne, the 6th International
Conference on Econometrics and Statistics, the 2024 Econometric Society
North American Summer Meeting, the 2024 Econometric SocietyAustralasia
Meeting, and the 2025 International Symposium on Econometric Theory
and Applications for helpful discussions.}}
\author{Yanqin Fan\thanks{Department of Economics, the University of Washington; email: fany88@uw.edu}
\and Yigit Okar\thanks{Care Research; email: okar.yigit@gmail.com}
\and Xuetao Shi\thanks{School of Economics, the University of Sydney; email: xuetao.shi@sydney.edu.au}}
\maketitle
\begin{abstract}
This article introduces an iterative distributed computing (IDC) estimator
for the multinomial logistic regression model with large choice sets.
The IDC estimator reformulates the estimation problem as a sequence
of structured, low-dimensional subproblems that can be solved efficiently
and in parallel. We establish that, when initialized with a consistent
estimator, the IDC estimator is asymptotically equivalent to the maximum
likelihood estimator and achieves asymptotic efficiency under a weak
dominance condition. We further develop a parametric bootstrap procedure
for inference based on the IDC estimator and prove its consistency.
Extensive simulation studies validate the effectiveness of the proposed
methods and highlight the computational efficiency of the IDC estimator.


\end{abstract}
\newpage{}

\section{Introduction}

\subsection{Motivation and Main Contributions}

Discrete choice models, including logit and multinomial-logit (MNL)
models, are widely used in applied social science research. With the
growing availability of diverse data types and the integration of
econometric models with textual and image data, researchers increasingly
encounter applications of MNL models with a massive number of choices;
see the applications discussed in Section \ref{sec:literature}. In
these cases, even if the number of parameters associated with each
choice is small, the total number of parameters, which increases linearly
with the number of choices, will be large due to large choice sets.
As a result, maximum likelihood estimation (MLE) can become computationally
intractable due to the high cost of solving large-scale optimization
problems.\footnote{MLE of the MNL model refers to the conditional MLE given the covariate
and total counts.}

Researchers in diverse disciplines have explored various approaches
and proposed numerous methods to numerically solve the MLE; see the
numerical algorithms discussed in Section \ref{sec:literature}. However,
there is a notable lack of theoretical results ensuring the consistency
or asymptotic efficiency of the estimators derived from these methods.
This gap in the literature between numerical computation and statistical
inference motivates the present study.

Our proposed estimator utilizes the multinomial-Poisson (MP) transformation,
which reformulates the multinomial likelihood as a Poisson likelihood
by incorporating individual fixed effects into the MNL model.
\citet{baker1994multinomial} provides a general derivation of the
MP transformation in a setting with categorical covariates, establishes
that the multinomial and Poisson likelihoods produce identical estimates
of the parameters of interest, and emphasizes the computational advantages
of maximizing the Poisson likelihood. The MP transformation has also
been employed in MNL models with continuous or mixed discrete and
continuous covariates. For instance, \citet{gentzkow2019measuring}
note that they “... approximate the likelihood of {[}their{]} multinomial
logit model with the likelihood of a Poisson model...”.

To provide a formal basis for our use of the MP transformation,
we apply the argument in \citet{baker1994multinomial} to the conditional
MNL model considered in this paper, in which the covariates may be
continuous, discrete, or mixed. We re-interpret the Poisson likelihood
function induced by the MP transformation as a conditional quasi-log-likelihood
function given the covariates. Maximizing this function yields a quasi-maximum
likelihood estimator (QMLE) for the MNL model. We show that the resulting
QMLE is identical to the MLE of the MNL model. This equivalence provides
the foundation for the development of our IDC estimator.

While the QMLE is computationally more efficient than the MLE, it
remains costly when applied to MNL models with large choice sets.
To address this computational challenge, \citet{taddy2015distributed}
exploits an important feature of the quasi-log-likelihood function:
for any given fixed effects, the function is additively separable
in the parameters across different choices of the MNL model. \citet{taddy2015distributed}
proposes to estimate parameters for each choice separately at a specific
value of the fixed effects and calls the resulting estimator\emph{
}\textit{\emph{distributed computing}}\emph{ }estimator. As noted
in \citet{taddy2015distributed}, however, his distributed computing
estimator is inconsistent except in a few very special cases.

To regain consistency and asymptotic efficiency, we adopt the idea
of iterative backfitting algorithms studied in \citet{pastorello2003iterative},
\citet{dominitz2005some}, and \citet{fan2015maximization} to compute
the QMLE of the parameters in the MNL model and fixed effects iteratively.\footnote{Similar iterative algorithms have also been developed for dynamic
discrete games of incomplete information where directly computing
the MLE using the nested fixed-point algorithm proves computationally
infeasible, see \citet{aguirregabiria2002swapping,aguirregabiria2007sequential}
for a nested pseudo-likelihood (NPL) algorithm. \citet{kasahara2012sequential}
further analyze the conditions necessary for the convergence of the
NPL algorithm and derive its convergence rate. } During each iteration, we first solve for the parameters of interest
through distributed computing, given approximate values of the individual
fixed effects. Then, we update estimates of the individual fixed effects
based on their expressions derived from maximizing the quasi-log-likelihood
function using the previous estimates of the model parameters. We
call our estimator \textit{iterative distributed computing} (IDC)
estimator. This is the first and main contribution of the paper. Thanks
to its distributed structure, the IDC algorithm remains computationally
efficient even for large choice sets.

To facilitate the implementation, our second contribution provides
practical guidance on initializing the IDC algorithm, choosing the
number of iterations, and understanding its computational complexity.
We consider three initial estimators: a consistent estimator based
on pairwise binomial logistic regression, \citet{taddy2015distributed}'s
distributed computing estimator, and a maximum likelihood estimator
that assumes a Poisson distribution for total counts. The latter two
estimators may be inconsistent. All three initial estimators are fast
to compute because they allow for distributed computing.

We further show that the number of iterations serves as a stopping
criterion, such that additional iterations improve numerical accuracy
without degrading statistical properties. We derive time and space
complexity results demonstrating that the IDC algorithm has time complexity
linear in the number of choices and auxiliary space complexity of
order equal to the sample size plus the product of the number of choices
and covariates.\footnote{The auxiliary space complexity counts the additional memory beyond
the input data required during execution of an algorithm.}

As our third contribution, we study the convergence property of the
IDC algorithm and establish consistency and asymptotic efficiency
of the IDC estimator under both consistent and arbitrary initialization.
For any realized sample for which the MNL maximum likelihood estimator
is unique, the IDC sequence converges to that estimator from any initial
value as the number of iterations diverges. We show that when the
initial estimator is consistent, the IDC estimator is consistent for
any number of iterations and achieves asymptotic efficiency when the
number of iterations diverges at any rate faster than a logarithmic
rate in the sample size. With arbitrary initial estimators, the IDC
estimators are consistent and asymptotically efficient when the number
of iterations diverges at a polynomial rate.

When the number of choices is large, conducting inference via plug-in
estimation of the variance matrix becomes infeasible. This is because
the Fisher information matrix has a very large dimension, causing
the computation of the inverse of its estimator to be both time-consuming
and unreliable. We therefore propose a parametric bootstrap inference
procedure and establish its consistency. Because the IDC estimator
is computationally efficient, the proposed inference procedure is
feasible in practice. We also analyze the convergence properties of
the iterative algorithm and show that for sufficiently large sample
size, it converges to a unique limit with high probability as the
number of iterations increases.

Our IDC algorithm can be easily extended to accommodate additional
empirical structure by modifying the optimization problems solved
at each iteration. We demonstrate that it allows researchers to incorporate
linear equality constraints, as well as regularisation for models
with high-dimensional covariates. Importantly, these extensions preserve
the distributed nature of the algorithm and its computational efficiency.

Lastly, we conduct extensive simulation studies to examine the finite-sample
performance of our estimator and inference procedure, and apply our
method to the Yelp dataset analyzed in \citet{taddy2015distributed}.
When the number of choices is small, we compare the computational
cost and accuracy of the IDC estimator with the maximum likelihood
estimator. When the choice set is large and MLE becomes infeasible,
we compare the IDC estimator with modern stochastic gradient descent
(SGD) methods. The simulation results show that the IDC estimator
is computationally efficient, with a running time approximately linear
in the number of choices. Relative to the MLE, the IDC estimator achieves
very similar mean squared error while being substantially faster to
compute. In settings with large choice sets, the IDC estimator outperforms
the SGD methods. We also find that the regularised IDC estimator performs
well in finite samples when the number of covariates is large. Finally,
we study the finite-sample behavior of the proposed bootstrap inference
procedure and show that it achieves correct size and is consistent.
In the empirical application, we show that the IDC estimator can be
implemented on a realistic text dataset, achieves stable convergence
within a few iterations, and delivers interpretable covariate--word
associations with predictive performance comparable to existing approaches.

\subsection{Related Literature}

\label{sec:literature}

\paragraph{Applications}

The proposed IDC estimator can be applied to study various economics
and computer science topics such as text analysis, dimensionality
reduction, spatial choice models, image classification (\citet{russakovsky2015imagenet}),
and video recommendation (\citet{davidson2010youtube}).

\emph{Text analysis}: The integration of text data into econometric
models is increasingly prominent in economics. For example, \citet{baker2006investor}
analyze investor sentiment’s effect on stock returns, while \citet{chen2021sentiment}
explore how hedge funds capitalize on sentiment changes. Modeling
text data often involves treating word counts as a multinomial distribution,
as \citet{taddy2015distributed} demonstrates using Yelp reviews to
predict outcomes based on user and business attributes. \citet{gentzkow2019measuring}
use the distributed computing estimator in \citet{taddy2015distributed}
for the multinomial regression to measure polarization in Congressional
speeches. \citet{NBERwTextSelection} extend this approach, using
Hurdle Distributed Multinomial Regression to backcast, nowcast, and
forecast macroeconomic variables from newspaper text.

\emph{Dimensionality Reduction}: Our estimator aids in dimensionality
reduction for inverse multinomial regression, as \citet{taddy2013multinomial}
discusses. Instead of inferring sentiment from text, \citet{taddy2013multinomial}’s
approach estimates word distribution given sentiment. He introduces
a score based on word frequencies and regression parameters, which
is useful in forward-regression models.

\emph{Spatial Choice Models}: High-dimensional choices also appear
in spatial models. \citet{bucholztaxi2019} models taxi drivers’ location
choices with a dynamic spatial search, reducing dimensionality via
discretization. Similarly, \citet{pellegrinifotheringham2002} apply
hierarchical discrete choice models to immigration, while \citet{bettman1979memory}
addresses brand choices in limited-option settings, proposing hierarchical
selection for high-dimensional cases.

\paragraph{Numerical Algorithms}

To address the computational difficulty of solving the MLE of the
MNL model, researchers have proposed several numerical methods to
find approximate solutions. \citet{bohning1988monotonicity} and \citet{bohning1992multinomial}
propose replacing the Hessian matrix in the Newton-Raphson iteration
with its easy-to-compute global lower bound and show that the approximate
solution converges with the number of iterations. Because the convergence
rate depends crucially on the difference between the Hessian matrix
and its lower bound, the algorithm can be slow to run for certain
model parameters. Additionally, based on our simulation exercise,
if the choice probabilities vary significantly across different choices
with some being close to zero, the algorithm becomes unstable. In
comparison, our IDC algorithm is stable in all the simulation settings.
\citet{boyd2011distributed} introduce an alternating direction method
of multipliers, which reformulates the original optimization problem
by introducing redundant linear constraints. \citet{gopal2013distributed}
propose a log concavity method, which replaces the log partition function
of the multinomial logit with a parallelisable upper bound. \citet{recht2011hogwild},
\citet{raman2016ds}, and \citet{fagan2018unbiased} study a stochastic
gradient descent method, which uses random training samples to calculate
the gradient at each iteration. Although these methods can be computationally
efficient, to the best of the authors' knowledge, no consistency or
asymptotic efficiency result has been shown in these works.

Another common approach to reduce computational cost is the divide-and-conquer
strategy, which partitions the sample across multiple workers, computes
local estimators in parallel, and aggregates them to form a global
estimator (\citet{chang2017distributed}, \citet{battey2018distributed},
\citet{zhu2021least}, and \citet{pan2022note}). These methods are
simple to implement and communication-efficient. However, they involve
important statistical trade-offs. Aggregated estimators can be statistically
inefficient and may yield biased or highly variable inference unless
additional correction steps are employed (\citet{zhang2013communication},
\citet{jordan2019communication}, and \citet{wu2025class}). Moreover,
when the number of choices is large, the effective sample size per
choice on each worker can be very small, leading to unstable local
estimators and poor finite-sample performance. By contrast, the proposed
IDC estimator exploits parallelism without splitting the sample across
machines, so that additional computational resources reduce computation
time while preserving asymptotic efficiency.

Penalisation methods have also been introduced to the MNL regression
and some of the numerical methods discussed above are adopted in solving
the penalised MNL regression, see e.g., \citet{friedman2010regularization},
\citet{simon2013blockwise}, and \citet{nibbering2022multiclass}.
The proposed IDC procedure in this paper can be combined with the
aforementioned algorithms to further improve the performance of penalised
MNL regressions.

\paragraph{Organization of the rest of this paper}

The remainder of this paper is organized as follows. In Section \ref{sec:Multinomial-Logistic-Regression}
we present a comprehensive overview of the multinomial logistic regression
model and the MP transformation. Section \ref{sec:Iterative-Distributed-Computing-}
introduces our IDC estimator, provides initial values of the IDC algorithm,
discuss the choice of the number of iterations, and derive the time
and space complexities of the IDC algorithm. In Section \ref{sec:Asymptotic-Theory},
we provide the asymptotic theory of the IDC estimator. In Section
\ref{sec:Variants}, we illustrate the adaptability of the IDC algorithm
through two variants of the IDC estimator. Section \ref{sec:Monte-Carlo-Simulation}
contains the simulation results, and Section \ref{sec:Empirical-Illustration}
applies our method to the Yelp data. Finally, with Section \ref{sec:Conclusion}
we conclude. Appendix \ref{sec:Notations-and-Equalities} collects
the notations and equations used in the paper. All the technical proofs
are provided in Appendix \ref{subsec:Proofs}. The codes for implementing
the estimation and inference procedures are available \href{https://github.com/yigitokar/IDMR}{here}.

\paragraph{Notations}

Throughout the paper, we use index $i\in\left\{ 1,\ldots,n\right\} $
for individual, $j\in\left\{ 1,\ldots,p\right\} $ for covariate,
and $k\in\left\{ 1,\ldots,d\right\} $ for unique choice. Boldfaced
symbols such as $\boldsymbol{C}$ and $\boldsymbol{X}$ are used to
denote vectors; while elements of the vectors are denoted by plain
symbol such as $C_{k}$ and $X_{j}$. Denote $\sim$ as ``equality
up to a constant'', such that $f\left(\theta\right)\sim g\left(\theta\right)$
is equivalent to $f\left(\theta\right)=g\left(\theta\right)+h$, where
$h$ is a constant relative to $\theta$.

\section{Multinomial Logistic Regression }

\label{sec:Multinomial-Logistic-Regression}

Let $\boldsymbol{C}_{i}\in\mathbb{R}^{d}$ denote the random vector
of counts on $d$ different choices for individual $i=1,...,n$, summing
up to $M_{i}=\sum^{d}_{k=1}C_{i,k}$. We allow $M_{i}$ to take any
positive integer including $1$. We use the random vector $\boldsymbol{X}_{i}\in\mathbb{R}^{p}$
to denote the covariate vector that includes a constant.

Consider a \textit{correctly specified} multinomial-logit (MNL) model.
The conditional probability mass function is given by the following:
\begin{align}
\Pr\left(\boldsymbol{C}_{i}\mid\boldsymbol{X}_{i},M_{i}\right) & =\mathrm{MNL}\left(\boldsymbol{C}_{i};\boldsymbol{\eta}^{\ast}_{i},M_{i}\right)\nonumber \\
 & \equiv\frac{M_{i}!}{C_{i,1}!\cdots C_{i,d}!}\left(\frac{e^{\eta^{\ast}_{i,1}}}{\Lambda^{\ast}_{i}}\right)^{C_{i,1}}\ldots\left(\frac{e^{\eta^{\ast}_{i,d}}}{\Lambda^{\ast}_{i}}\right)^{C_{i,d}},\label{eq: MNL model}
\end{align}
where for $k=1,\ldots d$, we let $\eta^{\ast}_{i,k}\equiv\boldsymbol{X}^{\prime}_{i}\boldsymbol{\theta}^{\ast}_{k}$
with unknown parameters $\boldsymbol{\theta}^{\ast}_{k}\equiv\left(\theta^{\ast}_{k,1},\ldots,\theta^{\ast}_{k,p}\right)^{\prime}$
and $\Lambda^{\ast}_{i}\equiv\sum^{d}_{k=1}e^{\eta^{\ast}_{i,k}}$.
For the identification, we set $\boldsymbol{\theta}^{\ast}_{d}=\mathbf{0}$.
Let $\boldsymbol{\theta}^{\ast}\equiv\left(\boldsymbol{\theta}^{\ast\prime}_{1},\ldots,\boldsymbol{\theta}^{\ast\prime}_{d}\right)^{\prime}\in\Theta$
denote the parameter vector of interest. Throughout the paper, we
use the superscript $\ast$ to indicate the true value of the unknown
parameter. Denote the parameter space of $\boldsymbol{\theta}^{\ast}_{k}$
for $k=1,\ldots,d$ as $\varTheta_{k}$. We have that $\varTheta_{d}=\left\{ \mathbf{0}\right\} $
and $\Theta\equiv\prod^{d}_{k=1}\varTheta_{k}$.

In this paper, we focus on the case where $d$ is large (but fixed)
such that directly solving for the maximum likelihood estimator is
computationally costly. Applications include text corpora, where $\boldsymbol{C}_{i}$
represents the counts of $d$ different words/phrases in a text of
$M_{i}$ words; browser logs, where $\boldsymbol{C}_{i}$ indicates
the number of times a website among $d$ total websites is visited
by an individual; and location choices, where among $M_{i}$ number
of locations traveled by the driver, $\boldsymbol{C}_{i}$ contains
the number of times each location, among $d$ different ones, is visited.

\subsection{Maximum Likelihood Estimation (MLE) }

\label{subsec:Maximum-Likelihood-Estimator}

Given a random sample of size $n$, let $\boldsymbol{\eta}_{i}\equiv\left(\eta_{i,1},\ldots,\eta_{i,d}\right)^{\prime}\equiv\left(\boldsymbol{X}^{\prime}_{i}\boldsymbol{\theta}_{1},\ldots,\boldsymbol{X}^{\prime}_{i}\boldsymbol{\theta}_{d}\right)$
and $\Lambda_{i}=\sum^{d}_{k=1}e^{\eta_{i,k}}$ for $i=1,...,n$.
Ignoring terms that are independent of the parameter $\boldsymbol{\theta}$,
the conditional log-likelihood function given the covariate $\boldsymbol{X}$
and total count $M$ takes the following form: 
\begin{eqnarray}
l_{C\mid X,M}\left(\boldsymbol{\theta}\right) & \equiv & \sum^{n}_{i=1}\log\Pr\left(\boldsymbol{C}_{i}\mid\boldsymbol{X}_{i},M_{i}\right)\nonumber \\
 & \sim & \sum^{n}_{i=1}\log\left[\left(\frac{e^{\eta_{i,1}}}{\Lambda_{i}}\right)^{C_{i1}}\ldots\left(\frac{e^{\eta_{i,d}}}{\Lambda_{i}}\right)^{C_{i,d}}\right]\nonumber \\
 & = & \sum^{n}_{i=1}\left\{ C_{i,1}\left[\log\left(e^{\eta_{i,1}}\right)-\log\left(\sum^{d}_{k=1}e^{\eta_{i,k}}\right)\right]+\cdots+C_{i,d}\left[\log\left(e^{\eta_{i,d}}\right)-\log\left(\sum^{d}_{k=1}e^{\eta_{i,k}}\right)\right]\right\} \nonumber \\
 & = & \sum^{n}_{i=1}\left[C_{i,1}\eta_{i,1}+\cdots+C_{i,d}\eta_{i,d}-\left(C_{i,1}+\cdots+C_{i,d}\right)\log\left(\sum^{d}_{k=1}e^{\eta_{i,k}}\right)\right]\nonumber \\
 & = & \sum^{n}_{i=1}\left[\boldsymbol{C}^{\prime}_{i}\boldsymbol{\eta}_{i}-M_{i}\log\left(\sum^{d}_{k=1}e^{\eta_{i,k}}\right)\right].\label{eq:Conditional log likelihood}
\end{eqnarray}
Let $L_{C\mid X,M}\left(\boldsymbol{\theta}\right)$ denote the probability
limit of $\frac{1}{n}l_{C\mid X,M}\left(\boldsymbol{\theta}\right)$.
Denote $B\left(\boldsymbol{\theta},\varepsilon\right)$ as an open
ball in $\Theta$ centered at $\boldsymbol{\theta}$ with radius $\varepsilon$.
To simplify notation, the interior of $\Theta$ refers to its interior
relative to the space of free parameters. We maintain the following
assumption throughout the paper.

\begin{assumption} \label{Assmp:: Maintained assumption} (i) The
true value $\boldsymbol{\theta}^{\ast}\in\Theta$ satisfies that $\sup_{\boldsymbol{\theta}\notin B\left(\boldsymbol{\theta}^{\ast},\varepsilon\right)}L_{C\mid X,M}\left(\boldsymbol{\theta}\right)<L_{C\mid X,M}\left(\boldsymbol{\theta}^{\ast}\right)$
for any $\varepsilon>0$. (ii) $\boldsymbol{\theta}^{\ast}$ is in
the interior of $\Theta$. \end{assumption}

Assumption \ref{Assmp:: Maintained assumption} (i) means that $\boldsymbol{\theta}^{\ast}$
is identified as $\boldsymbol{\theta}^{\ast}=\arg\max_{\boldsymbol{\theta}\in\Theta}L_{C\mid X,M}\left(\boldsymbol{\theta}\right)$.
Define the following objective function: 
\begin{equation}
Q^{\ast}_{n}\left(\boldsymbol{\theta}\right)\equiv-\frac{1}{n}l_{C\mid X,M}\left(\boldsymbol{\theta}\right).\label{eq:Objective Function - original}
\end{equation}
Based on (\ref{eq:Objective Function - original}), the conditional
maximum likelihood estimator of $\boldsymbol{\theta}^{\ast}$ is:\footnote{The definition of $\widetilde{\boldsymbol{\theta}}$ implicitly assumes
that the solution to the minimization problem is unique. This can
be shown to hold with probability approaching one by the identification
of the model. See \citet{mcfadden1973conditional}. Throughout the
paper, sample estimators are defined on the event that the relevant
minimization problems have unique solutions, and all fixed-sample
statements are understood on this event. We ignore such mathematical
subtlety for the remainder of the paper to simplify the discussion.} 
\begin{equation}
\widetilde{\boldsymbol{\theta}}=\arg\min_{\boldsymbol{\theta}\in\Theta}Q^{\ast}_{n}\left(\boldsymbol{\theta}\right).\label{eq:MLE}
\end{equation}
Solving the above optimization problem analytically is impossible.
In addition, due to the potentially large dimension $d$, numerical
algorithms such as the Newton-Raphson method are difficult to implement
either because they usually involve computing the inverse of the Hessian
matrix, which is of dimension $pd\times pd$, during each iteration.
In this paper, we propose an estimator that is both computationally
attractive and asymptotically efficient.

\subsection{Multinomial-Poisson Transformation}

In this section, we review the multinomial--Poisson (MP) transformation
of \citet{baker1994multinomial}, which provides the foundation for
our estimator. We reinterpret the resulting Poisson likelihood as
a quasi-likelihood conditional on the covariates.

Let $\boldsymbol{1}_{d}\equiv\left(1,\ldots,1\right)^{\prime}\in\mathbb{R}^{d}$.
With a slight abuse of notation, define 
\begin{align*}
l_{C\mid X,M}\left(\boldsymbol{\theta},\boldsymbol{\mu}\right) & \equiv l_{C\mid X,M}\left(\boldsymbol{\theta}\right)+\sum^{n}_{i=1}\left[\mu_{i}\boldsymbol{C}^{\prime}_{i}\boldsymbol{1}_{d}-M_{i}\log\left(e^{\mu_{i}}\right)\right]\\
 & =\sum^{n}_{i=1}\left[\boldsymbol{C}^{\prime}_{i}\left(\boldsymbol{\eta}_{i}+\mu_{i}\boldsymbol{1}_{d}\right)-M_{i}\log\left(\sum^{d}_{k=1}e^{\eta_{i,k}+\mu_{i}}\right)\right],
\end{align*}
where $\boldsymbol{\mu}\equiv\left(\mu_{1},\ldots,\mu_{n}\right)\in\mathbb{R}^{n}$.
The following lemma shows that the two functions $l_{C\mid X,M}\left(\boldsymbol{\theta}\right)$
and $l_{C\mid X,M}\left(\boldsymbol{\theta},\boldsymbol{\mu}\right)$
are the same for any $\boldsymbol{\theta}\in\Theta$ and $\boldsymbol{\mu}\in\mathbb{R}^{n}$.
In other words, argument $\boldsymbol{\mu}$ in $l_{C\mid X,M}\left(\boldsymbol{\theta},\boldsymbol{\mu}\right)$
does not affect the value of the function. The proof of the lemma
is straightforward by realizing that $\boldsymbol{C}^{\prime}_{i}\boldsymbol{1}_{d}=M_{i}$
by definition.

\begin{lemma} \label{lem:: Adding mu} $l_{C\mid X,M}\left(\boldsymbol{\theta}\right)=l_{C\mid X,M}\left(\boldsymbol{\theta},\boldsymbol{\mu}\right)$
for any $\boldsymbol{\theta}\in\Theta$ and $\boldsymbol{\mu}\in\mathbb{R}^{n}$.
\end{lemma}

Define the following two functions: 
\begin{align}
f\left(\boldsymbol{\theta},\boldsymbol{\mu}\right) & \equiv\sum^{n}_{i=1}\left[M_{i}\log\left(\sum^{d}_{k=1}e^{\eta_{i,k}+\mu_{i}}\right)-\sum^{d}_{k=1}e^{\eta_{i,k}+\mu_{i}}\right]\textrm{ and }\label{eq:log likelihood l_M|X}\\
ql_{C\mid X}\left(\boldsymbol{\theta},\boldsymbol{\mu}\right) & \equiv l_{C\mid X,M}\left(\boldsymbol{\theta},\boldsymbol{\mu}\right)+f\left(\boldsymbol{\theta},\boldsymbol{\mu}\right)\nonumber \\
 & =\sum^{n}_{i=1}\sum^{d}_{k=1}\left(C_{i,k}\left(\eta_{i,k}+\mu_{i}\right)-e^{\left(\eta_{i,k}+\mu_{i}\right)}\right).\label{eq:Unconditional Likelihood with mu}
\end{align}
It is not difficult to see that $f\left(\boldsymbol{\theta},\boldsymbol{\mu}\right)$
takes the form of a log-likelihood function of $n$ conditional Poisson
distributions with means $\sum^{d}_{k=1}e^{\left(\eta_{i,k}+\mu_{i}\right)}$,
$i=1,\ldots,n$ (after ignoring terms that are independent of $\boldsymbol{\theta}$
and $\boldsymbol{\mu}$). This in turn renders $ql_{C\mid X}$ a conditional
quasi-log-likelihood function of which $C_{i,k}$ given $X_{i}$ is
drawn independently from a Poisson distribution with mean $e^{\left(\eta_{i,k}+\mu_{i}\right)}$,
for $k=1,...,d$. This property underlies the naming of the MP transformation.

Based on the conditional quasi-log-likelihood function $ql_{C\mid X}\left(\boldsymbol{\theta},\boldsymbol{\mu}\right)$,
we can compute a conditional quasi MLE (QMLE) of $\boldsymbol{\theta}^{\ast}$:
\begin{align}
\left(\widehat{\boldsymbol{\theta}},\widehat{\boldsymbol{\mu}}\right) & =\arg\min_{\boldsymbol{\theta}\in\Theta,\boldsymbol{\mu}\in\mathbb{R}^{n}}Q_{n}\left(\boldsymbol{\theta},\boldsymbol{\mu}\right),\label{eq:Definition of Distributed Multinomial Estimator}\\
\text{where }Q_{n}\left(\boldsymbol{\theta},\boldsymbol{\mu}\right) & \equiv-\frac{1}{n}ql_{C|X}\left(\boldsymbol{\theta},\boldsymbol{\mu}\right).\nonumber 
\end{align}

\citet{baker1994multinomial} establishes the equivalence
between the multinomial and Poisson formulations under the MP transformation.
The following lemma applies this result to the conditional MNL model
considered in the paper and shows that $\widehat{\boldsymbol{\theta}}=\widetilde{\boldsymbol{\theta}}$.

\begin{lemma} \label{lem:: Equivalence conditional and unconditional}
It holds that $\widehat{\boldsymbol{\theta}}=\widetilde{\boldsymbol{\theta}}$.
\end{lemma}

It is important to note that Lemma \ref{lem:: Equivalence conditional and unconditional}
does not depend on the assumption that $f\left(\boldsymbol{\theta},\boldsymbol{\mu}\right)$
is the correct log-likelihood function of $M_{i}$, or equivalently
that $\Pr\left(M_{i}\mid\boldsymbol{X}_{i}\right)=\mathrm{Po}\left(\sum^{d}_{k=1}e^{\left(\eta_{i,k}+\mu_{i}\right)}\right)$
or $\Pr\left(C_{i,k}\mid\boldsymbol{X}_{i}\right)=\mathrm{Po}\left(e^{\left(\eta_{i,k}+\mu_{i}\right)}\right)$,
where $\mathrm{Po}\left(\cdot\right)$ denotes the Poisson distribution.
No assumption about the conditional distribution of $M_{i}$ given
$\boldsymbol{X}_{i}$ is required for any of the results in the paper
to hold. As we show in the following sections, the introduction of
$f\left(\boldsymbol{\theta},\boldsymbol{\mu}\right)$ is merely a
trick to achieve distributed computing.

\section{Iterative Distributed Computing Estimator }

\label{sec:Iterative-Distributed-Computing-}

Lemma \ref{lem:: Equivalence conditional and unconditional} shows
that, instead of minimizing $Q^{\ast}_{n}\left(\boldsymbol{\theta}\right)$,
we can minimize $Q_{n}\left(\boldsymbol{\theta},\boldsymbol{\mu}\right)$
to obtain the QMLE of $\boldsymbol{\theta}^{\ast}$. However, computing
$\widehat{\boldsymbol{\theta}}$ remains computationally intensive,
as solving (\ref{eq:Definition of Distributed Multinomial Estimator})
is impractical when $d$ is large. Nevertheless, the additive structure
of $Q_{n}\left(\boldsymbol{\theta},\boldsymbol{\mu}\right)$ enables
the problem to be solved distributively.

\subsection{Distributed Computing Estimator in \citet{taddy2015distributed}}

\label{subsec:intro to IDC}

As noted in \citet{taddy2015distributed}, although minimizing $Q_{n}\left(\boldsymbol{\theta},\boldsymbol{\mu}\right)$
with respect to $\boldsymbol{\theta}$ and $\boldsymbol{\mu}$ jointly
is computationally infeasible, given any value of $\boldsymbol{\mu}$,
solving $\arg\min_{\boldsymbol{\theta}}Q_{n}\left(\boldsymbol{\theta},\boldsymbol{\mu}\right)$
is much easier because the optimization can be done separately for
each $\boldsymbol{\theta}_{k}$ and be computed across machines/workers.
To see this, we rewrite $Q_{n}\left(\boldsymbol{\theta},\boldsymbol{\mu}\right)$
as: 
\begin{align}
Q_{n}\left(\boldsymbol{\theta},\boldsymbol{\mu}\right) & =\frac{1}{n}\sum^{n}_{i=1}\sum^{d}_{k=1}\left(e^{\left(\eta_{i,k}+\mu_{i}\right)}-C_{i,k}\left(\eta_{i,k}+\mu_{i}\right)\right).\nonumber \\
 & =\frac{1}{n}\sum^{d}_{k=1}\sum^{n}_{i=1}\left(e^{\left(\boldsymbol{X}^{\prime}_{i}\boldsymbol{\theta}_{k}+\mu_{i}\right)}-C_{i,k}\left(X^{\prime}_{i}\boldsymbol{\theta}_{k}+\mu_{i}\right)\right)\nonumber \\
 & \equiv Q_{1,n}\left(\boldsymbol{\theta}_{1},\boldsymbol{\mu}\right)+\cdots+Q_{d,n}\left(\boldsymbol{\theta}_{d},\boldsymbol{\mu}\right),\label{eq:Qn equals Q1n until Qdn}
\end{align}
where for $k=1,\ldots,d$, $Q_{k,n}\left(\boldsymbol{\theta}_{k},\boldsymbol{\mu}\right)\equiv\frac{1}{n}\sum^{n}_{i=1}\left(e^{\left(X^{\prime}_{i}\boldsymbol{\theta}_{k}+\mu_{i}\right)}-C_{i,k}\left(X^{\prime}_{i}\boldsymbol{\theta}_{k}+\mu_{i}\right)\right)$.
In consequence, it holds that for any $\boldsymbol{\mu}$, 
\begin{equation}
\arg\min_{\boldsymbol{\theta}\in\Theta}Q_{n}\left(\boldsymbol{\theta},\boldsymbol{\mu}\right)=\left[\arg\min_{\boldsymbol{\theta}_{1}\in\varTheta_{1}}Q_{1,n}\left(\boldsymbol{\theta}_{1},\boldsymbol{\mu}\right),\ldots,\arg\min_{\boldsymbol{\theta}_{d}\in\varTheta_{d}}Q_{d,n}\left(\boldsymbol{\theta}_{d},\boldsymbol{\mu}\right)\right]^{\prime}.\label{eq:Distributed Computing}
\end{equation}
Based on (\ref{eq:Distributed Computing}), solving $\arg\min_{\boldsymbol{\theta}\in\Theta}Q_{n}\left(\boldsymbol{\theta},\boldsymbol{\mu}\right)$
for any given $\boldsymbol{\mu}$ is equivalent to solving $d$ optimizations:
$\arg\min_{\boldsymbol{\theta}_{k}\in\varTheta_{k}}Q_{k,n}\left(\boldsymbol{\theta}_{k},\boldsymbol{\mu}\right)$
for each $k=1,\ldots,d$, where each optimization is a Poisson regression.\footnote{Since $\varTheta_{d}=\left\{ \mathbf{0}\right\} $, solving for $\arg\min_{\boldsymbol{\theta}_{d}\in\varTheta_{d}}Q_{dn}\left(\boldsymbol{\theta}_{d},\boldsymbol{\mu}\right)$
is trivial.} Since $\boldsymbol{\theta}_{k}$ has only $p$ dimensions, $\arg\min_{\boldsymbol{\theta}_{k}\in\varTheta_{k}}Q_{k,n}\left(\boldsymbol{\theta}_{k},\boldsymbol{\mu}\right)$
is easy to compute. In addition, the optimizations for $k=1,\ldots,d$
can be computed across machines allowing for distributed computing.

By Equation (\ref{eq:Definition of Distributed Multinomial Estimator}),
it is not difficult to see that $\widehat{\boldsymbol{\theta}}=\arg\min_{\boldsymbol{\theta}\in\Theta}Q_{n}\left(\boldsymbol{\theta},\widehat{\boldsymbol{\mu}}\right)$.
Because $\widehat{\boldsymbol{\theta}}$ is equivalent to the MLE
$\widetilde{\boldsymbol{\theta}}$ by Lemma \ref{lem:: Equivalence conditional and unconditional},
it has the desired properties such as being both consistent and asymptotically
efficient. As a result, we would hope to obtain $\widehat{\boldsymbol{\mu}}$
first and then compute $\widehat{\boldsymbol{\theta}}$ by distributed
computing. However, the value of $\widehat{\boldsymbol{\mu}}$ depends
on $\widehat{\boldsymbol{\theta}}$, which itself is difficult to
calculate. On the other hand, given any value of $\boldsymbol{\theta}$,
solving $\arg\min_{\boldsymbol{\mu}}Q_{n}\left(\boldsymbol{\theta},\boldsymbol{\mu}\right)$
is also fast, and the solution even has a closed form. Denote the
solution to $\arg\min_{\boldsymbol{\mu}}Q_{n}\left(\boldsymbol{\theta},\boldsymbol{\mu}\right)$
as $\overline{\boldsymbol{\mu}}_{n}\left(\boldsymbol{\theta}\right)$.
A simple calculation would show that 
\begin{equation}
\overline{\boldsymbol{\mu}}_{n}\left(\boldsymbol{\theta}\right)=\left(\log\left(\frac{M_{1}}{\sum^{d}_{k=1}e^{\eta_{1,k}}}\right),\ldots,\log\left(\frac{M_{n}}{\sum^{d}_{k=1}e^{\eta_{n,k}}}\right)\right)^{\prime}.\label{eq:mu bar}
\end{equation}

Let $\widehat{\boldsymbol{\mu}}_{T}=\left(\log\left(M_{1}\right),\ldots,\log\left(M_{n}\right)\right)^{\prime}$.
Instead of solving for $\widehat{\boldsymbol{\mu}}$ using (\ref{eq:mu bar}),
\citet{taddy2015distributed} proposes an estimator $\widehat{\boldsymbol{\theta}}_{T}=\arg\min_{\boldsymbol{\theta}\in\Theta}Q_{n}\left(\boldsymbol{\theta},\widehat{\boldsymbol{\mu}}_{T}\right)$
and calls it the distributed computing estimator. Such an estimator
is fast to compute but fails to be consistent except in the special
cases discussed in \citet{taddy2015distributed}.

\subsection{Iterative Distributed Computing Estimator }

\label{sec_sub:Iterative-Distributed-Computing}

We propose an iterative distributed computing (IDC) estimator, such
that during each iteration we solve (\ref{eq:Distributed Computing})
with $\boldsymbol{\mu}$ updated from the previous estimate of $\boldsymbol{\theta}$
via (\ref{eq:mu bar}). Our IDC estimator is defined by the following
procedures.

\textbf{Iteration $\mathbf{0}$. }Compute an initial estimator of
$\boldsymbol{\theta}^{\ast}$, denoted as $\widehat{\boldsymbol{\theta}}^{\left(0\right)}$.

\textbf{Iteration $\mathbf{1},\ldots,\mathbf{S}$. }For Iteration
$s$, where $s=1,\ldots,S$, we first update $\boldsymbol{\mu}$ using
estimator $\widehat{\boldsymbol{\theta}}^{\left(s-1\right)}$ from
the previous iteration via $\overline{\boldsymbol{\mu}}_{n}\left(\cdot\right)$.
Then we update $\boldsymbol{\theta}$ given the value of $\boldsymbol{\mu}$:
\begin{align}
 & \widehat{\boldsymbol{\theta}}^{\left(s\right)}\equiv\overline{\boldsymbol{\theta}}_{n}\left(\widehat{\boldsymbol{\theta}}^{\left(s-1\right)}\right)\nonumber \\
\equiv & \left[\arg\min_{\boldsymbol{\theta}_{1}\in\varTheta_{1}}Q_{1,n}\left(\boldsymbol{\theta}_{1},\overline{\boldsymbol{\mu}}_{n}\left(\widehat{\boldsymbol{\theta}}^{\left(s-1\right)}\right)\right),\ldots,\arg\min_{\boldsymbol{\theta}_{d}\in\varTheta_{d}}Q_{d,n}\left(\boldsymbol{\theta}_{d},\overline{\boldsymbol{\mu}}_{n}\left(\widehat{\boldsymbol{\theta}}^{\left(s-1\right)}\right)\right)\right]^{\prime}.\label{eq::iterative procedure}
\end{align}

The iterative estimator with $S$ iterations is defined as $\widehat{\boldsymbol{\theta}}^{I}\equiv\widehat{\boldsymbol{\theta}}^{\left(S\right)}$.
For any $\boldsymbol{\theta}$, the value of $\overline{\boldsymbol{\mu}}_{n}\left(\boldsymbol{\theta}\right)$
can be directly computed from (\ref{eq:mu bar}). In each iteration,
we compute $\arg\min_{\boldsymbol{\theta}_{k}}Q_{k,n}\left(\boldsymbol{\theta}_{k},\overline{\boldsymbol{\mu}}_{n}\left(\widehat{\boldsymbol{\theta}}^{\left(s-1\right)}\right)\right)$
for $k=1,\ldots,d$ on $d$ parallel computers. This amounts to running
$d$ Poisson regressions with $p$ parameters, and the computational
burden for each iteration is low. The algorithm is described in Algorithm
\ref{alg::IDC algorithm}.

\begin{algorithm}[!h]
\DontPrintSemicolon

\KwInput{$S$} \KwOutput{$\widehat{\boldsymbol{\theta}}^{\left(S\right)}$}
Compute an initial estimator $\widehat{\boldsymbol{\theta}}^{\left(0\right)}$

\tcc{Start of Iteration 1}

Compute $\overline{\boldsymbol{\mu}}_{n}\left(\widehat{\boldsymbol{\theta}}^{\left(0\right)}\right)$

Solve for 
\[
\widehat{\boldsymbol{\theta}}^{\left(1\right)}=\left[\arg\min_{\boldsymbol{\theta}_{1}\in\varTheta_{1}}Q_{1,n}\left(\boldsymbol{\theta}_{1},\overline{\boldsymbol{\mu}}_{n}\left(\widehat{\boldsymbol{\theta}}^{\left(0\right)}\right)\right),\ldots,\arg\min_{\boldsymbol{\theta}_{d}\in\varTheta_{d}}Q_{d,n}\left(\boldsymbol{\theta}_{d},\overline{\boldsymbol{\mu}}_{n}\left(\widehat{\boldsymbol{\theta}}^{\left(0\right)}\right)\right)\right]^{\prime}
\]

\tcc{End of Iteration 1. The output is $\widehat{\boldsymbol{\theta}}^{\left(1\right)}$}
\tcc{Start of Iteration 2}

Compute $\overline{\boldsymbol{\mu}}_{n}\left(\widehat{\boldsymbol{\theta}}^{\left(1\right)}\right)$

Solve for 
\[
\widehat{\boldsymbol{\theta}}^{\left(2\right)}=\left[\arg\min_{\boldsymbol{\theta}_{1}\in\varTheta_{1}}Q_{1,n}\left(\boldsymbol{\theta}_{1},\overline{\boldsymbol{\mu}}_{n}\left(\widehat{\boldsymbol{\theta}}^{\left(1\right)}\right)\right),\ldots,\arg\min_{\boldsymbol{\theta}_{d}\in\varTheta_{d}}Q_{d,n}\left(\boldsymbol{\theta}_{d},\overline{\boldsymbol{\mu}}_{n}\left(\widehat{\boldsymbol{\theta}}^{\left(1\right)}\right)\right)\right]^{\prime}
\]

\tcc{End of Iteration 2. The output is $\widehat{\boldsymbol{\theta}}^{\left(2\right)}$}

...... Continue until Iteration $S$. The output of Iteration $S$
is $\widehat{\boldsymbol{\theta}}^{\left(S\right)}$

\caption{the iterative distributed computing procedure}
\label{alg::IDC algorithm} 
\end{algorithm}

Unlike many existing algorithms that numerically solve for the MLE,
such as gradient descent or stochastic gradient descent, the IDC estimator
does not inherently involve any hyperparameter.\footnote{We refer to a parameter that varies the complexity of a model as a
hyperparameter. In contrast, the stopping criterion specifies the
condition under which an algorithm terminates and affects only numerical
precision rather than model complexity.} This is advantageous because the performance of the classical (stochastic)
gradient descent is generally sensitive to the learning rate, as we
show in the simulation study in Section \ref{subsec:Simulation large d}.

\subsection{Initial Estimators }

\label{subsec:Initial-Estimator}

Similar to all iterative optimization procedures, the initial estimator
plays a critical role. In finite samples, a good initial guess of
$\boldsymbol{\theta}^{\ast}$ can improve the performance of the IDC
estimator. Asymptotically, a consistent initial estimator can lead
to consistent and asymptotically efficient iterative estimators under
weaker assumptions than an inconsistent initial estimator. In this
section, we propose three initial estimators: a consistent initial
estimator of $\boldsymbol{\theta}^{\ast}$ based on binomial MLE,
\citet{taddy2015distributed}'s estimator, and the MLE based on the
Poisson assumption of $M_{i}$. The latter two may be inconsistent
without any assumption on the distribution of $M_{i}$.

\paragraph{A Consistent Initial Estimator }

Let $N_{i,k}\equiv C_{i,k}+C_{i,d}$. The following lemma results
from the MNL model defined in (\ref{eq: MNL model}).

\begin{lemma} \label{lem:: Ck Cd distribution} For any $k=1,\ldots,d-1$,
it holds that 
\begin{equation}
\Pr\left(C_{i,k},C_{i,d}\mid\boldsymbol{X}_{i},N_{i,k}\right)=\frac{N_{i,k}!}{C_{i,k}!C_{i,d}!}\left(\frac{e^{\eta^{\ast}_{i,k}}}{e^{\eta^{\ast}_{i,k}}+1}\right)^{C_{i,k}}\left(\frac{1}{e^{\eta^{\ast}_{i,k}}+1}\right)^{C_{i,d}}.\label{eq: Independent MNL}
\end{equation}
\end{lemma}

Lemma \ref{lem:: Ck Cd distribution} shows that we can consistently
estimate $\boldsymbol{\theta}^{\ast}_{k}$ based on a binomial logistic
regression with the log-likelihood function given by:\footnote{The lemma can be understood as using the sufficiency of $N_{i,k}$
for the $\mu_{i}$ parameter in $ql_{C\mid X}\left(\boldsymbol{\theta},\boldsymbol{\mu}\right)$.
The conditional probability in (\ref{eq: Independent MNL}) then corresponds
to the conditional likelihood function introduced in \citet{andersen1970asymptotic}
and furthered employed in \citet{hausman1984econometric}.} 
\begin{align*}
l_{C_{k},C_{d}\mid X,N_{k}}\left(\boldsymbol{\theta}_{k}\right) & \equiv\sum^{n}_{i=1}\log\Pr\left(C_{i,k},C_{i,d}\mid\boldsymbol{X}_{i},N_{i,k}\right)\\
 & \sim\sum^{n}_{i=1}\left[C_{i,k}\eta_{i,k}-\left(C_{i,k}+C_{i,d}\right)\log\left(e^{\eta_{i,k}}+1\right)\right]\\
 & =\sum^{n}_{i=1}\left[C_{i,k}\boldsymbol{X}^{\prime}_{i}\boldsymbol{\theta}_{k}-\left(C_{i,k}+C_{i,d}\right)\log\left(e^{\boldsymbol{X}^{\prime}_{i}\boldsymbol{\theta}_{k}}+1\right)\right].
\end{align*}
Let $\check{\boldsymbol{\theta}}_{k}=\arg\min_{\boldsymbol{\theta}_{k}\in\varTheta_{k}}-l_{C_{k},C_{d}\mid X,N_{k}}\left(\boldsymbol{\theta}_{k}\right)$
for $k=1,\ldots,d-1$ and $\check{\boldsymbol{\theta}}=\left(\check{\boldsymbol{\theta}}^{\prime}_{1},\ldots,\check{\boldsymbol{\theta}}^{\prime}_{d-1},\check{\boldsymbol{\theta}}^{\prime}_{d}\right)^{\prime}$
with $\check{\boldsymbol{\theta}}_{d}=\mathbf{0}$. The consistency
of $\check{\boldsymbol{\theta}}$ follows from standard arguments
in the maximum likelihood estimation.

Econometricians and statisticians have applied similar ideas to develop
estimators that are computationally more attractive than the MLE $\widetilde{\boldsymbol{\theta}}$.
\citet{mcfadden1978modeling} introduces the procedure of sampling-of-alternatives
to accommodate cases where the number of choices is large. The procedure
selects a subset of the choices with equal/unequal probability and
conducts estimation accordingly. See \citet{train2009discrete} for
detailed discussion. When the subset contains two choices with one
being the reference, the sampling-of-alternatives procedure produces
the same estimator as $\check{\boldsymbol{\theta}}$. Recently, \citet{wu2025class}
develop a computationally efficient class-distributed learning algorithm
for MNL model. The idea in \citet{wu2025class} for tackling large
number of choices is the same as our initial consistent estimator
$\check{\boldsymbol{\theta}}$.

Compared to $\widetilde{\boldsymbol{\theta}}$, the conditional probability
used in constructing the binomial logistic log-likelihood function
does not use all the available information. Therefore, $\check{\boldsymbol{\theta}}$
is less efficient than $\widetilde{\boldsymbol{\theta}}$. However,
each component of $\check{\boldsymbol{\theta}}$, $\check{\boldsymbol{\theta}}_{k}$,
can be calculated independently, allowing for parallel computing.
The substantially short running time of $\check{\boldsymbol{\theta}}$
makes it a great candidate for the initial value $\widehat{\boldsymbol{\theta}}^{\left(0\right)}$.

\paragraph{Inconsistent Initial Estimators }

Even though \citet{taddy2015distributed}'s estimator fails to be
consistent in general, it could serve as a candidate for the initial
value. Another option is to replace $\widehat{\boldsymbol{\mu}}_{T}$
with a zero vector to obtain another estimator denoted as $\widehat{\boldsymbol{\theta}}_{P}=\arg\min_{\boldsymbol{\theta}\in\Theta}Q_{n}\left(\boldsymbol{\theta},\mathbf{0}\right)$.
It can also be computed across machines for each $k=1,...,d$. Like
\citet{taddy2015distributed}'s estimator, $\widehat{\boldsymbol{\theta}}_{P}$
could also serve as a candidate for the initial value in our algorithm.
Moreover, under an extra condition that $M_{i}$ follows a Poisson
distribution, $\widehat{\boldsymbol{\theta}}_{P}$ is the maximum
likelihood estimator of $\boldsymbol{\theta}^{\ast}$.


\begin{lemma} \label{lem:: consistency under stronger assumption}
If $\Pr\left(M_{i}\mid X_{i}\right)=\mathrm{Po}\left(\sum^{d}_{k=1}e^{\eta^{\ast}_{i,k}}\right)$,
then $\widehat{\boldsymbol{\theta}}_{P}$ is the maximum likelihood
estimator of $\boldsymbol{\theta}^{\ast}$ based on the conditional
probability $\Pr\left(\boldsymbol{C}_{i}\mid X_{i}\right)$. \end{lemma}

Unlike $\check{\boldsymbol{\theta}}$, neither $\widehat{\boldsymbol{\theta}}_{T}$
nor $\widehat{\boldsymbol{\theta}}_{P}$ is a consistent estimator
of $\boldsymbol{\theta}^{\ast}$ without any additional assumption.

\subsection{Choice of the Number of Iterations}

In this subsection, we discuss the choice of the number of iterations
$S$ in the IDC algorithm. Conceptually, $S$ serves as a stopping
criterion rather than a tuning or hyperparameter. Unlike a hyperparameter
that controls model complexity or regularisation, the value of $S$
does not alter the underlying optimization problem. Additional iterations
only refine the numerical approximation.

From a practical point, we recommend selecting $S$ based on available
computational resources. A convenient approach is to begin with a
small to moderate number of iterations, such as $S=1$ to $5$, to
measure the computational cost per iteration. Section \ref{subsec:Time-and-Space}
shows that the overall running time of the IDC algorithm scales approximately
linearly in $S$. Practitioners can therefore choose $S$ by balancing
desired accuracy against acceptable running time. An attractive feature
of the IDC procedure is that the iteration can be resumed if additional
computational resources or waiting time become available.

Alternatively, a data-adaptive stopping rule may be employed when
a specific accuracy target is desired. For example, the algorithm
can be terminated once the difference between two successive iterates
falls below a pre-specified tolerance. Such stopping criteria are
standard in iterative numerical algorithms and are straightforward
to implement in the IDC framework.

Finally, it is useful to relate the IDC estimator to existing approaches.
When initialized at the estimator in \citet{taddy2015distributed},
the IDC estimator can be viewed as a generalization of that approach,
which corresponds to the special case $S=0$. The IDC framework then
refines this initial estimate through iteration, recovering consistency
as $S\rightarrow\infty$. In this sense, the IDC algorithm makes explicit
the trade-off between computational cost and statistical accuracy.

\subsection{Time and Space Complexities}

\label{subsec:Time-and-Space}

To further demonstrate the computational advantages of the proposed
IDC estimator relative to alternative approaches, such as a single
Newton-step update from a consistent initial estimator (\citet{lehmann1998theory}
and \citet{horowitz2012semiparametric}), we analyze the time and
space complexities of our algorithm as functions of the model dimensions
and available computational resources.

During IDC iteration $s$, the algorithm consists of two main steps.
The first step computes $\widehat{\boldsymbol{\theta}}^{(s)}_{k}$
for $k=1,\ldots,d$ by solving \eqref{eq:Distributed Computing} in
parallel across multiple workers. Let $T$ denote the number of iterations
required to solve each Poisson regression. Using the Newton method,
one Poisson regression over $T$ iterations has time complexity $O\left(T\left(np^{2}+p^{3}\right)\right)$.\footnote{Other numerical methods may improve the time complexity. We use the
Newton method as a reference because it is the standard solver for
Poisson regression and yields reliable convergence.} Assuming $U$ workers and balanced job allocation, each worker solves
approximately $\lceil d/U\rceil$ Poisson regressions sequentially.
Hence, the total computational cost of this distributed step is $O\left(\left\lceil \frac{d}{U}\right\rceil T\left(np^{2}+p^{3}\right)\right)$.

In the second step, the central computer collects the estimates $\widehat{\boldsymbol{\theta}}^{(s)}_{k}$
for $k=1,\ldots,d$, computes $\overline{\boldsymbol{\mu}}_{n}\left(\widehat{\boldsymbol{\theta}}^{(s)}\right)$
using (\ref{eq:mu bar}), and broadcasts the updated $\boldsymbol{\mu}$
back to the workers. Each Poisson regression returns a coefficient
vector of dimension $p$, so the total uplink payload is $dp$ scalars.
With batching, the uplink communication cost is $O\left(U+dp\right)$.
Computing $\overline{\boldsymbol{\mu}}_{n}\left(\widehat{\boldsymbol{\theta}}^{(s)}\right)$
requires evaluating $\sum^{d}_{k=1}e^{\eta_{i,k}}$ for each $i=1,\ldots,n$,
which takes $O\left(nd\right)$ arithmetic operations. Finally, broadcasting
$\boldsymbol{\mu}\in\mathbb{R}^{n}$ to all workers incurs a cost
of $O\left(\log U+n\right)$ using a tree-based communication scheme.

Combining these components, the per-iteration time complexity of the
IDC algorithm is 
\[
O\left(\left\lceil \frac{d}{U}\right\rceil T\left(np^{2}+p^{3}\right)+U+dp+nd+\log U+n\right).
\]
When $p$ is moderate, the dominant term is the cost of solving the
Poisson regressions, and the per-iteration complexity can be well
approximated by $O\left(\frac{d}{U}T\left(np^{2}+p^{3}\right)\right)$.
If the IDC algorithm runs for $S$ iterations, the overall time complexity
is therefore 
\[
O\left(\frac{d}{U}ST\left(np^{2}+p^{3}\right)\right).
\]

In addition to the model dimensions $n$, $p$, and $d$, the total
computational cost depends on the iteration counts $T$ and $S$.
In our setting, the dimension of each Poisson regression parameter
vector is $p$, which is moderate, so $T$ is small. Moreover, starting
from a consistent initial estimator, Theorem \ref{Thm:: equivalence IDC and theta tilda}
provides a sufficient iteration rate that grows only slightly faster
than $\log n$, which bounds $S$. Importantly, neither $T$ nor $S$
grows with $d$. Consequently, the overall time complexity of the
IDC algorithm is approximately linear in $d$. Furthermore, increasing
the number of workers $U$ yields near-linear speedup until $U\approx d$.

By contrast, the inversion of the $pd\times pd$ dimensional Hessian
matrix associated with the full likelihood function (\ref{eq:Conditional log likelihood})
has time complexity at least $O\left(d^{3}p^{3}\right)$. Therefore,
any estimator that requires explicit Hessian inversion, including
the single Newton-step update from a consistent initial estimator,
exhibits cubic complexity in $d$. For large $d$, the IDC estimator
is thus substantially more computationally efficient.

A related alternative is to compute all $d\left(d-1\right)/2$
pairwise binomial logit estimators following Lemma \ref{lem:: Ck Cd distribution}.
By the independence of irrelevant alternatives property, the complete
set of $d\left(d-1\right)/2$ pairwise binomial logit will be as informative
about the parameters as a full likelihood for $d$ choices together.
Therefore, in principle, one could explore efficient ways to combine
these pairwise estimators, and then take a Newton-step to achieve
efficiency. However, computing the pairwise estimators alone requires
solving $d\left(d-1\right)/2$ optimization problems, where each optimization
involves $p$ number of arguments. If each binomial regression takes
$T$ iterations, the resulting time complexity is at least $O\left(\frac{d^{2}}{U}T\left(np^{2}+p^{3}\right)\right)$.
It is quadratic in $d$ even before the Newton-step update. By comparison,
the time complexity of the IDC algorithm is approximately linear in
$d$.

We now compute the space complexity. For simplicity, assume $M_{i}=1$
for $i=1,\ldots,n$. Storing the covariate matrix requires $O\left(np\right)$
memory. Each Poisson regression uses the original data and produces
a $p$-dimensional coefficient vector. Storing $\widehat{\boldsymbol{\theta}}^{(s)}_{k}$
for all $k=1,\ldots,d$ therefore requires $O\left(dp\right)$ memory.
Computing $\overline{\boldsymbol{\mu}}_{n}\left(\widehat{\boldsymbol{\theta}}^{(s)}\right)$
does not require storing intermediate quantities; only the resulting
vector $\boldsymbol{\mu}\in\mathbb{R}^{n}$ is retained. Since $\widehat{\boldsymbol{\theta}}^{(s)}_{k}$
and $\boldsymbol{\mu}$ are overwritten in each iteration, no additional
memory accumulates across iterations.

Hence, the total space complexity of the IDC algorithm is $O\left(np+dp+n\right)$.
Excluding storage of the original data, the required auxiliary space
complexity is $O\left(dp+n\right)$. Notably, the IDC estimator avoids
storing the $n\times d$ probability matrix and the $pd\times pd$
Hessian matrix required by full maximum likelihood or Newton-type
methods. For such methods, the auxiliary space complexity is at least
$d^{2}p^{2}$.

\section{Asymptotic Theory}

\label{sec:Asymptotic-Theory}

In this section, we derive the convergence property of the iterative
algorithm and establish the consistency and asymptotic normality of
our IDC estimator. Technical proofs are collected in Appendix \ref{subsec:Proofs}.
We first impose the following two assumptions.

\begin{assumption}\label{Assm:: consistency 0} $\left\{ \left(\boldsymbol{C}_{i},\boldsymbol{X}_{i},M_{i}\right)\right\} ^{n}_{i=1}$
are random samples of $\left(\boldsymbol{C},\boldsymbol{X},M\right)$.
\end{assumption}

\begin{assumption}\label{Assm:: consistency 1} (i) $\Theta$ is
compact and convex. (ii) $\mathbb{E}\left[e^{\boldsymbol{X}^{\prime}\boldsymbol{\theta}_{k}}\right]<\infty$
for all $\boldsymbol{\theta}_{k}\in\varTheta_{k}$ with $k=1\ldots d$.
(iii) $\mathbb{E}\left[M^{2}\right]<\infty$.\end{assumption}

Assumption \ref{Assm:: consistency 1} is standard. Assumption \ref{Assm:: consistency 1}
(ii) requires that the moment generating function of $\boldsymbol{X}$
exists within $\Theta$. Almost all commonly seen distributions satisfy
Assumption \ref{Assm:: consistency 1} (ii) and (iii).

\subsection{Properties of the IDC Algorithm}

\label{subsec:Convergence-Property}

In order to analyze the asymptotic properties of the IDC estimator,
we need to study the way in which the iterative estimator updates
itself in each iteration. In Iteration $s$, we first compute $\overline{\boldsymbol{\mu}}_{n}\left(\widehat{\boldsymbol{\theta}}^{\left(s-1\right)}\right)$
using $\widehat{\boldsymbol{\theta}}^{\left(s-1\right)}$ from the
previous iteration and then calculate $\arg\min_{\boldsymbol{\theta}\in\Theta}Q_{n}\left(\boldsymbol{\theta},\overline{\boldsymbol{\mu}}_{n}\left(\widehat{\boldsymbol{\theta}}^{\left(s-1\right)}\right)\right)$.
To explicitly distinguish the argument in $\overline{\boldsymbol{\mu}}_{n}\left(\cdot\right)$
from the argument in $Q_{n}\left(\cdot,\boldsymbol{\mu}\right)$ for
any given $\boldsymbol{\mu}$, we introduce $\boldsymbol{\vartheta}$
and use it as the argument in $\overline{\boldsymbol{\mu}}_{n}\left(\cdot\right)$.
Define $Q^{\dagger}_{n}\left(\boldsymbol{\theta},\boldsymbol{\vartheta}\right)\equiv Q_{n}\left(\boldsymbol{\theta},\overline{\boldsymbol{\mu}}_{n}\left(\boldsymbol{\vartheta}\right)\right)$
and its probability limit $Q^{\dagger}\left(\boldsymbol{\theta},\boldsymbol{\vartheta}\right)$.
The expressions of $Q^{\dagger}_{n}\left(\boldsymbol{\theta},\boldsymbol{\vartheta}\right)$
and $Q^{\dagger}\left(\boldsymbol{\theta},\boldsymbol{\vartheta}\right)$
are provided in Appendix \ref{sec:Notations-and-Equalities}. In the
following lemma, we provide some properties of $Q^{\dagger}_{n}\left(\boldsymbol{\theta},\boldsymbol{\vartheta}\right)$
and $Q^{\dagger}\left(\boldsymbol{\theta},\boldsymbol{\vartheta}\right)$.

\begin{lemma} \label{lem:: properties of Qn Qstar} Under Assumptions
\ref{Assmp:: Maintained assumption}, \ref{Assm:: consistency 0},
and \ref{Assm:: consistency 1}, the following results hold.

(i) $\sup_{\boldsymbol{\theta},\boldsymbol{\vartheta}\in\Theta}\left|Q^{\dagger}_{n}\left(\boldsymbol{\theta},\boldsymbol{\vartheta}\right)-Q^{\dagger}\left(\boldsymbol{\theta},\boldsymbol{\vartheta}\right)\right|\stackrel{p}{\rightarrow}0$.

(ii) For any given $\boldsymbol{\vartheta}$, $Q^{\dagger}\left(\boldsymbol{\theta},\boldsymbol{\vartheta}\right)$
has a unique minimizer denoted as $\overline{\boldsymbol{\theta}}\left(\boldsymbol{\vartheta}\right)$.

(iii) $\overline{\boldsymbol{\theta}}\left(\cdot\right)$ is continuous
on $\Theta$.

(iv) $\overline{\boldsymbol{\theta}}\left(\boldsymbol{\theta}^{\ast}\right)=\boldsymbol{\theta}^{\ast}$,
i.e., the true value $\boldsymbol{\theta}^{\ast}$ is a fixed point
of the mapping $\overline{\boldsymbol{\theta}}:\Theta\rightarrow\Theta$.

(v) $\boldsymbol{\theta}^{\ast}$ is the unique fixed point of $\overline{\boldsymbol{\theta}}\left(\cdot\right)$.\end{lemma}

Essentially, $\overline{\boldsymbol{\theta}}\left(\boldsymbol{\vartheta}\right)$
summarizes the operation in each iteration with $\boldsymbol{\vartheta}$
being the input and $\overline{\boldsymbol{\theta}}\left(\boldsymbol{\vartheta}\right)$
being the output when the sample size goes to infinity. By part (ii)
of Lemma \ref{lem:: properties of Qn Qstar}, $\overline{\boldsymbol{\theta}}\left(\cdot\right)$
is well-defined. Part (v) of Lemma \ref{lem:: properties of Qn Qstar}
plays an important role. Heuristically, for any given $\boldsymbol{\vartheta}$,
the value of function $\overline{\boldsymbol{\theta}}\left(\boldsymbol{\vartheta}\right)$
is obtained by solving $\frac{\partial Q^{\dagger}\left(\boldsymbol{\theta},\boldsymbol{\vartheta}\right)}{\partial\boldsymbol{\theta}}=\mathbf{0}$.
At the same time, function $\frac{\partial Q^{\dagger}\left(\boldsymbol{\theta},\boldsymbol{\vartheta}\right)}{\partial\boldsymbol{\theta}}$
relates to the first order derivative of $L_{C\mid X,M}\left(\boldsymbol{\theta}\right)$,
the population objective function defined in Section \ref{subsec:Maximum-Likelihood-Estimator}.
By the identification assumption and the convexity of $-L_{C\mid X,M}\left(\boldsymbol{\theta}\right)$,
only the true value $\boldsymbol{\theta}^{\ast}$ satisfies that $\frac{\partial}{\partial\boldsymbol{\theta}}L_{C\mid X,M}\left(\boldsymbol{\theta}\right)=\mathbf{0}$,
which implies that $\frac{\partial Q^{\dagger}\left(\boldsymbol{\theta},\boldsymbol{\vartheta}\right)}{\partial\boldsymbol{\theta}}\mid_{\boldsymbol{\vartheta}=\boldsymbol{\theta}}=\mathbf{0}$
holds only at $\left(\boldsymbol{\theta},\boldsymbol{\vartheta}\right)=\left(\boldsymbol{\theta}^{\ast},\boldsymbol{\theta}^{\ast}\right)$.

We further define the profiled sample objective function $F_{n}\left(\boldsymbol{\theta}\right)\equiv Q^{\dagger}_{n}\left(\boldsymbol{\theta},\boldsymbol{\theta}\right)$
and its probability limit $F\left(\boldsymbol{\theta}\right)\equiv Q^{\dagger}\left(\boldsymbol{\theta},\boldsymbol{\theta}\right)$.
Properties of $F_{n}\left(\boldsymbol{\theta}\right)$ and $F\left(\boldsymbol{\theta}\right)$
are summarized below.

\begin{lemma} \label{lem:: properties of Fn Fstar} Under Assumptions
\ref{Assmp:: Maintained assumption}, \ref{Assm:: consistency 0},
and \ref{Assm:: consistency 1}, the following results hold.

(i) $F_{n}\left(\widehat{\boldsymbol{\theta}}^{\left(s\right)}\right)\leq F_{n}\left(\widehat{\boldsymbol{\theta}}^{\left(s-1\right)}\right)$
for every $s\geq1$.

(ii) $\sup_{\boldsymbol{\theta}\in\Theta}\left|F_{n}\left(\boldsymbol{\theta}\right)-F\left(\boldsymbol{\theta}\right)\right|=o_{p}\left(1\right)$.

(iii) $F_{n}\left(\boldsymbol{\theta}\right)$ has a unique minimizer
$\widetilde{\boldsymbol{\theta}}$ and $F\left(\boldsymbol{\theta}\right)$
has a unique minimizer $\boldsymbol{\theta}^{\ast}$.\end{lemma}

Lemma \ref{lem:: properties of Fn Fstar} exploits
the alternating-minimization structure of the IDC algorithm. Part
(i) shows that the profiled sample objective $F_{n}\left(\boldsymbol{\theta}\right)$
is nonincreasing along the IDC iterations. Part (ii) establishes uniform
convergence of $F_{n}\left(\boldsymbol{\theta}\right)$ to its population
counterpart $F\left(\boldsymbol{\theta}\right)$. Part (iii) identifies
the minimizers of these objectives: $F_{n}\left(\boldsymbol{\theta}\right)$
has the same minimizer as the negative multinomial log-likelihood,
while $F\left(\boldsymbol{\theta}\right)$ is uniquely minimized at
$\boldsymbol{\theta}^{\ast}$. The results in Lemmas \ref{lem:: properties of Qn Qstar}
and \ref{lem:: properties of Fn Fstar} are crucial for obtaining
the consistency and asymptotic normality of our IDC estimator.

Given the results in Lemmas \ref{lem:: properties of Qn Qstar}
and \ref{lem:: properties of Fn Fstar}, we can derive the convergence
property of the iterative algorithm. \begin{theorem}
\label{Thm:: Convergence property of algorithm} Suppose that Assumptions
\ref{Assmp:: Maintained assumption} and \ref{Assm:: consistency 1}
hold. Given a fixed sample, for any initial value $\widehat{\boldsymbol{\theta}}^{\left(0\right)}\in\Theta$,
the IDC sequence $\widehat{\boldsymbol{\theta}}^{\left(1\right)},\widehat{\boldsymbol{\theta}}^{\left(2\right)},\widehat{\boldsymbol{\theta}}^{\left(3\right)},\ldots$
satisfies that $\lim_{S\rightarrow\infty}\widehat{\boldsymbol{\theta}}^{\left(S\right)}=\widetilde{\boldsymbol{\theta}}$.
\end{theorem}

Theorem \ref{Thm:: Convergence property of algorithm}
is a fixed-sample result. It shows that the limit of the IDC sequence
is the MNL maximum likelihood estimator for any initial value.

\subsection{Consistency of the IDC Estimator }

\label{subsec:Consistency }

The previous subsection establishes a fixed-sample result on the IDC
algorithm. The result does not by itself determine the statistical
properties of the IDC estimator whose iteration index may vary with
the sample size. In this subsection, we study consistency of our estimators
when the number of iterations is fixed or depends on $n$. We first
show that if a consistent initial estimator is used, such as $\check{\boldsymbol{\theta}}$,
then the IDC estimator is consistent uniformly over the iteration index.

\begin{theorem}[Consistent initial value] \label{thm:: consistency-fixed S}
Suppose Assumptions \ref{Assmp:: Maintained assumption}, \ref{Assm:: consistency 0},
and \ref{Assm:: consistency 1} hold. If $\widehat{\boldsymbol{\theta}}^{\left(0\right)}\stackrel{p}{\rightarrow}\boldsymbol{\theta}^{\ast}$
as $n\rightarrow\infty$, then $\sup_{S\geq0}\left\Vert \widehat{\boldsymbol{\theta}}^{\left(S\right)}-\boldsymbol{\theta}^{\ast}\right\Vert =o_{p}\left(1\right)$.
\end{theorem}

The result in Theorem \ref{thm:: consistency-fixed S} is uniform
over the iteration index. As a result, the consistency of $\widehat{\boldsymbol{\theta}}^{\left(S\right)}$
holds for any fixed $S$ or diverging $S$. The following theorem
establishes consistency of the IDC estimator with arbitrary initial
estimator.

\begin{theorem}[Arbitrary initial value] \label{thm:: consistency}
Suppose Assumptions \ref{Assmp:: Maintained assumption}, \ref{Assm:: consistency 0},
and \ref{Assm:: consistency 1} hold. For any initial estimator $\widehat{\boldsymbol{\theta}}^{\left(0\right)}$
and every sequence $S=S_{n}$ such that $S_{n}\rightarrow\infty$
as $n\rightarrow\infty$, it holds that $\widehat{\boldsymbol{\theta}}^{\left(S_{n}\right)}\stackrel{p}{\rightarrow}\boldsymbol{\theta}^{\ast}$.
\end{theorem}

Theorem \ref{thm:: consistency} differs from Theorem \ref{thm:: consistency-fixed S}
in the treatment of the early iterations. With a consistent initial
estimator, all IDC iterates are close to $\boldsymbol{\theta}^{\ast}$
with probability approaching one. In consequence, consistency holds
from the first iteration and for every fixed or diverging number of
iterations. Without assuming that the initial estimator is consistent,
Theorem \ref{thm:: consistency} shows that the IDC estimator is consistent
when the number of iterations diverges with the sample size.

It is worth mentioning that no contraction mapping
type assumption, such as Assumption 6 in \citet{pastorello2003iterative},
is needed for Theorem \ref{thm:: consistency} to hold. The reason
is that IDC has a nonincreasing profiled sample objective function,
as shown in Lemma \ref{lem:: properties of Fn Fstar} (i). When $S_{n}\rightarrow\infty$,
the bounded cumulative decrease of the profiled sample objective function
$F_{n}\left(\cdot\right)$ forces intermediate iterates to have a
vanishing descent gap. Uniform convergence of the descent gap and
uniqueness of its population zero imply consistency of this intermediate
iterate. Monotonicity of $F_{n}\left(\cdot\right)$, together with
its uniform convergence to $F\left(\cdot\right)$, then transfers
consistency to the final iterate.

\subsection{Asymptotic Distributions and Inference }

\label{subsec:Asymptotic-Distributions-Inference}

Under Assumptions \ref{Assm:: consistency 0} and \ref{Assm:: consistency 1},
the MLE $\widetilde{\boldsymbol{\theta}}$ is asymptotically normally
distributed with asymptotic variance given by the inverse of the Fisher
information matrix 
\[
\mathcal{I}\left(\boldsymbol{\theta}^{\ast}\right)\equiv\frac{\partial^{2}}{\partial\boldsymbol{\theta}\partial\boldsymbol{\theta}^{\prime}}Q^{\ast}\left(\boldsymbol{\theta}^{\ast}\right),
\]
where $Q^{\ast}\left(\boldsymbol{\theta}^{\ast}\right)\equiv p\lim_{n\rightarrow\infty}Q^{\ast}_{n}\left(\boldsymbol{\theta}^{\ast}\right)$.
In this section, we show that our IDC estimator has the same asymptotic
distribution as $\widetilde{\boldsymbol{\theta}}$, based on which
we introduce a valid bootstrap inference procedure.

We impose the following assumption on the second order derivatives
of $Q^{\dagger}\left(\boldsymbol{\theta},\boldsymbol{\vartheta}\right)$.
For any matrix $A$, denote $\left\Vert A\right\Vert $ as its spectral
norm.

\begin{assumption}[Information Dominance] \label{ass::Information Dominance}
It holds that 
\[
\left\Vert \left(\frac{\partial^{2}Q^{\dagger}\left(\boldsymbol{\theta},\boldsymbol{\vartheta}\right)}{\partial\boldsymbol{\theta}\partial\boldsymbol{\theta}^{\prime}}\mid_{\boldsymbol{\theta}=\boldsymbol{\theta}^{\ast},\boldsymbol{\vartheta}=\boldsymbol{\theta}^{\ast}}\right)^{-1}\frac{\partial^{2}Q^{\dagger}\left(\boldsymbol{\theta},\boldsymbol{\vartheta}\right)}{\partial\boldsymbol{\theta}\partial\boldsymbol{\vartheta}^{\prime}}\mid_{\boldsymbol{\theta}=\boldsymbol{\theta}^{\ast},\boldsymbol{\vartheta}=\boldsymbol{\theta}^{\ast}}\right\Vert <1.
\]
\end{assumption}

The detailed expressions of $\frac{\partial^{2}Q^{\dagger}\left(\boldsymbol{\theta},\boldsymbol{\vartheta}\right)}{\partial\boldsymbol{\theta}\partial\boldsymbol{\theta}^{\prime}}$
and $\frac{\partial^{2}Q^{\dagger}\left(\boldsymbol{\theta},\boldsymbol{\vartheta}\right)}{\partial\boldsymbol{\theta}\partial\boldsymbol{\vartheta}^{\prime}}$
can be found in Appendix \ref{sec:Notations-and-Equalities}. Assumption
\ref{ass::Information Dominance} is often called the information
dominance condition and is tantamount to the local contraction mapping
condition. Because we have a consistent initial estimator of $\boldsymbol{\theta}^{\ast}$,
we can check the assumption. Additionally, because the matrix $\frac{\partial^{2}Q^{\dagger}\left(\boldsymbol{\theta},\boldsymbol{\vartheta}\right)}{\partial\boldsymbol{\theta}\partial\boldsymbol{\theta}^{\prime}}$
is block diagonal with each block having dimensions $p\times p$,
computing its inverse is feasible.

The following theorem shows that when $S$ is sufficiently large,
the IDC estimator $\widehat{\boldsymbol{\theta}}^{\left(S\right)}$
is equal to $\widetilde{\boldsymbol{\theta}}$ up to a term of order
smaller than $n^{-1/2}$.

\begin{theorem} \label{Thm:: equivalence IDC and theta tilda} Suppose
Assumptions \ref{Assmp:: Maintained assumption}, \ref{Assm:: consistency 0},
\ref{Assm:: consistency 1}, and \ref{ass::Information Dominance}
hold. (i) If $\widehat{\boldsymbol{\theta}}^{\left(0\right)}\stackrel{p}{\rightarrow}\boldsymbol{\theta}^{\ast}$
as $n\rightarrow\infty$ and $\log\left(n\right)=o\left(S_{n}\right)$,
then $\widehat{\boldsymbol{\theta}}^{\left(S_{n}\right)}-\widetilde{\boldsymbol{\theta}}=o_{p}\left(n^{-1/2}\right)$.
(ii) For any arbitrary $\widehat{\boldsymbol{\theta}}^{\left(0\right)}$,
if $S_{n}>n^{\delta}$ for some $\delta>0$, then $\widehat{\boldsymbol{\theta}}^{\left(S_{n}\right)}-\widetilde{\boldsymbol{\theta}}=o_{p}\left(n^{-1/2}\right)$.
\end{theorem}

Theorem \ref{Thm:: equivalence IDC and theta tilda} shows that we
do not lose efficiency when employing the proposed IDC estimator as
long as $S$ is large enough. A direct implication of the theorem
is that $\widehat{\boldsymbol{\theta}}^{\left(S\right)}$ has the
same asymptotic distribution as $\widetilde{\boldsymbol{\theta}}$
for sufficiently large $S$.

\begin{corollary} \label{Cor:: Asymptotics} Suppose Assumptions
\ref{Assmp:: Maintained assumption}, \ref{Assm:: consistency 0},
\ref{Assm:: consistency 1}, and \ref{ass::Information Dominance}
hold. (i) If $\widehat{\boldsymbol{\theta}}^{\left(0\right)}\stackrel{p}{\rightarrow}\boldsymbol{\theta}^{\ast}$
as $n\rightarrow\infty$ and $\log\left(n\right)=o\left(S_{n}\right)$,
then $\sqrt{n}\left(\widehat{\boldsymbol{\theta}}^{\left(S_{n}\right)}-\boldsymbol{\theta}^{\ast}\right)\stackrel{d}{\rightarrow}\mathcal{N}\left(\mathbf{0},\mathcal{I}^{-1}\left(\boldsymbol{\theta}^{\ast}\right)\right)$
as $n\rightarrow\infty$. (ii) For any arbitrary $\widehat{\boldsymbol{\theta}}^{\left(0\right)}$,
if $S_{n}>n^{\delta}$ for some $\delta>0$, then $\sqrt{n}\left(\widehat{\boldsymbol{\theta}}^{\left(S_{n}\right)}-\boldsymbol{\theta}^{\ast}\right)\stackrel{d}{\rightarrow}\mathcal{N}\left(\mathbf{0},\mathcal{I}^{-1}\left(\boldsymbol{\theta}^{\ast}\right)\right)$
as $n\rightarrow\infty$. \end{corollary}

To conduct inference on $\boldsymbol{\theta}^{\ast}$ based upon $\widehat{\boldsymbol{\theta}}^{\left(S\right)}$,
we need to consistently estimate the Fisher information matrix and
compute its inverse. Because the dimension of $\mathcal{I}\left(\boldsymbol{\theta}^{\ast}\right)$
is $\left(d-1\right)p\times\left(d-1\right)p$, calculating the inverse
of its estimator is not only time-consuming but also unreliable when
$d$ is large. As a result, we proceed by applying the following parametric
bootstrap, which is feasible thanks to the fact that the IDC estimator
is fast to compute.

Given $\left\{ \left(\boldsymbol{X}_{i},M_{i}\right)\right\} ^{n}_{i=1}$
of the original sample, we draw the bootstrap sample $\boldsymbol{C}^{\ddagger}_{in}$
for $i=1,\ldots,n$ from the multinomial logistic regression model
with the conditional probability mass function given by 
\[
\mathrm{MNL}\left(\boldsymbol{C}^{\ddagger}_{i,n};\widehat{\boldsymbol{\eta}}_{i},M_{i}\right)\textrm{, where }\widehat{\eta}_{i,k}=\boldsymbol{X}^{\prime}_{i}\widehat{\boldsymbol{\theta}}^{\left(S\right)}_{k}\textrm{ for }k=1,\ldots d.
\]
The bootstrap version of the iterative estimator $\widehat{\boldsymbol{\theta}}^{\ddagger\left(S\right)}$
is obtained by applying the algorithm introduced in Section \ref{sec_sub:Iterative-Distributed-Computing}
with bootstrap sample $\left\{ \left(\boldsymbol{C}^{\ddagger}_{i,n},\boldsymbol{X}_{i},M_{i}\right)\right\} ^{n}_{i=1}$.

Assume that we start with a consistent initial estimator. Based on
Theorem \ref{Thm:: equivalence IDC and theta tilda} (i), we have
that for $\log\left(n\right)=o\left(S_{n}\right)$, 
\begin{equation}
\widehat{\boldsymbol{\theta}}^{\left(S_{n}\right)}-\boldsymbol{\theta}^{\ast}=\mathcal{I}^{-1}\left(\boldsymbol{\theta}^{\ast}\right)\frac{1}{n}\frac{d}{d\boldsymbol{\theta}}l_{C\mid X,M}\left(\boldsymbol{\theta}^{\ast}\right)+o_{p}\left(n^{-1/2}\right).\label{eq:Theta S influence function}
\end{equation}
Define the score function for $\boldsymbol{\theta}$ as 
\[
\dot{\mathbf{l}}\left(\boldsymbol{\theta}\mid\boldsymbol{c},\boldsymbol{x},m\right)\equiv\frac{d}{d\boldsymbol{\theta}}\log\mathrm{MNL}\left(\boldsymbol{c};\boldsymbol{\eta},m\right)\textrm{, where }\eta_{k}\equiv\boldsymbol{x}^{\prime}\boldsymbol{\theta}_{k}.
\]
The iterative estimator $\widehat{\boldsymbol{\theta}}^{\left(S_{n}\right)}$
is asymptotically linear with influence function $\widetilde{\mathbf{l}}\left(\boldsymbol{\theta}^{\ast}\mid\boldsymbol{c},\boldsymbol{x},m\right)$:
\[
\sqrt{n}\left(\widehat{\boldsymbol{\theta}}^{\left(S_{n}\right)}-\boldsymbol{\theta}^{\ast}\right)=\frac{1}{\sqrt{n}}\sum^{n}_{i=1}\widetilde{\mathbf{l}}\left(\boldsymbol{\theta}^{\ast}\mid\boldsymbol{C}_{i},\boldsymbol{X}_{i},M_{i}\right)+o_{p}\left(1\right),
\]
where $\widetilde{\mathbf{l}}\left(\boldsymbol{\theta}^{\ast}\mid\boldsymbol{c},\boldsymbol{x},m\right)\equiv\mathcal{I}^{-1}\left(\boldsymbol{\theta}^{\ast}\right)\dot{\mathbf{l}}\left(\boldsymbol{\theta}^{\ast}\mid\boldsymbol{c},\boldsymbol{x},m\right)$.
Denote the probability conditional on the original sample by $\Pr^{\ddagger}\left(\cdot\right)$. 
The notation $R^{\ddagger}_{n}=o_{p^{\ddagger}}\left(1\right)$ in probability
means that $\Pr^{\ddagger}\left(\left\Vert R^{\ddagger}_{n}\right\Vert >\varepsilon\right)\stackrel{p}{\rightarrow}0$
for every $\varepsilon>0$. Applying the same derivation, we can show
that the bootstrap version of the estimator is also asymptotically
linear with the influence function evaluated at $\widehat{\boldsymbol{\theta}}^{\left(S_{n}\right)}$:
\[
\sqrt{n}\left(\widehat{\boldsymbol{\theta}}^{\ddagger\left(S_{n}\right)}-\widehat{\boldsymbol{\theta}}^{\left(S_{n}\right)}\right)=\frac{1}{\sqrt{n}}\sum^{n}_{i=1}\widetilde{\mathbf{l}}\left(\widehat{\boldsymbol{\theta}}^{\left(S_{n}\right)}\mid\boldsymbol{C}^{\ddagger}_{i,n},\boldsymbol{X}_{i},M_{i}\right)+o_{p^{\ddagger}}\left(1\right)
\]
in probability. The conditional Lindeberg-Feller central limit theorem
proves the bootstrap consistency. The proof for the case of an inconsistent
initial estimator is analogous. Let $\stackrel{d^{\ddagger}}{\rightarrow}$
denote the convergence in bootstrap distribution.

\begin{theorem} \label{thm:: Bootstrap} Suppose Assumptions \ref{Assmp:: Maintained assumption},
\ref{Assm:: consistency 0}, \ref{Assm:: consistency 1}, and \ref{ass::Information Dominance}
hold. If (i) $\widehat{\boldsymbol{\theta}}^{\left(0\right)}\stackrel{p}{\rightarrow}\boldsymbol{\theta}^{\ast}$
as $n\rightarrow\infty$ and $\log\left(n\right)=o\left(S_{n}\right)$,
then $\sqrt{n}\left(\widehat{\boldsymbol{\theta}}^{\ddagger\left(S_{n}\right)}-\widehat{\boldsymbol{\theta}}^{\left(S_{n}\right)}\right)\stackrel{d^{\ddagger}}{\rightarrow}\mathcal{N}\left(\mathbf{0},\mathcal{I}^{-1}\left(\boldsymbol{\theta}^{\ast}\right)\right)$
as $n\rightarrow\infty$. (ii) For any arbitrary $\widehat{\boldsymbol{\theta}}^{\left(0\right)}$,
if $S_{n}>n^{\delta}$ for some $\delta>0$, then $\sqrt{n}\left(\widehat{\boldsymbol{\theta}}^{\ddagger\left(S_{n}\right)}-\widehat{\boldsymbol{\theta}}^{\left(S_{n}\right)}\right)\stackrel{d^{\ddagger}}{\rightarrow}\mathcal{N}\left(\mathbf{0},\mathcal{I}^{-1}\left(\boldsymbol{\theta}^{\ast}\right)\right)$
as $n\rightarrow\infty$. \end{theorem}

\section{Variants of the IDC Estimator}

\label{sec:Variants}

Because the IDC algorithm is based on repeatedly solving simple optimization
problems, it is easy to adapt to different empirical settings. Researchers
can incorporate features such as parameter constraints, regularisation,
or other application-specific restrictions by modifying the optimization
problems solved at each iteration. As long as these modifications
apply separately to each alternative, the distributed structure of
the IDC algorithm is preserved.

In this section, we present two variants of the IDC estimator that
are useful in applied work. We first discuss how to incorporate linear
equality constraints motivated by economic theory, and then introduce
a regularised version of the IDC estimator for models with many covariates.
These variants illustrate the flexibility of the IDC framework while
maintaining computational efficiency and desirable statistical properties.

\subsection{Constrained IDC Estimator }

\label{sec:Constrained IDC}

Linear equality constraints across parameters arise naturally in many
empirical applications of multinomial choice models. For example,
\citet{train1987demand} study demand for local telephone service
and impose equality constraints on the cost coefficient across several
plan alternatives. This corresponds to restrictions of the form $\theta_{1,1}=\cdots=\theta_{d,1}$,
where $\boldsymbol{X}_{i,1}$ denotes the cost variable. Similarly,
\citet{strombom2002switching} analyze individual health-plan choice
and first estimate a specification in which the premium coefficient
is constrained to be identical across all plans, before considering
a more flexible model where the premium effect varies across groups
of individuals but remains equal within each group. Taking the constraints
into consideration during the estimation would further improve asymptotic
efficiency. In the following, we discuss how to modify our IDC estimator
introduced in Section \ref{sec_sub:Iterative-Distributed-Computing}
to incorporate equality constraints.

Although our proposed procedures can be applied to the most general
linear equality constraints, to simplify the discussion, we consider
two different types of constraints: constraint on parameters for the
same choice only and constraint on parameters for different choices
only. The procedures for different types of constraints differ in
the optimization problems during each iteration. For each type, we
use one example as an illustration. The estimators introduced in Section
\ref{subsec:Initial-Estimator} can still be utilized as the initial
$\widehat{\boldsymbol{\theta}}^{\left(0\right)}$ because of their
computational efficiency. Even though they do not account for the
equality constraints, their statistical properties discussed in Section
\ref{subsec:Initial-Estimator} maintain because the unconstrained
MNL model in (\ref{eq: MNL model}) is correctly specified. Importantly,
estimator $\check{\boldsymbol{\theta}}$ remains consistent.

The first type of constraint involves only components of individual
$\boldsymbol{\theta}^{\ast}_{k}$ but not across $k$. Take the constraint
$\theta^{\ast}_{k,1}=\theta^{\ast}_{k,2}$ for $k=1,\ldots,d$ as
an example. When computing $\widehat{\boldsymbol{\theta}}^{\left(s\right)}$
in Iteration $s$, we solve the constrained optimization problem 
\[
\arg\min_{\boldsymbol{\theta}_{k}\in\varTheta_{k},\theta_{k,1}=\theta_{k,2}}Q_{k,n}\left(\boldsymbol{\theta}_{k},\overline{\boldsymbol{\mu}}_{n}\left(\widehat{\boldsymbol{\theta}}^{\left(s-1\right)}\right)\right)
\]
for $k=1,\ldots,d$. Because the constraint is on each $\boldsymbol{\theta}^{\ast}_{k}$,
the original distributed computing scheme remains.

The second type of constraints involves components of $\boldsymbol{\theta}^{\ast}_{k}$
across different choices. For example, researchers may impose restrictions
like $\theta^{\ast}_{1,1}=\cdots=\theta^{\ast}_{q,1}$, where $q\leq d$.
For $k=1,\ldots,q$, let $\boldsymbol{\theta}_{k,-1}$ be the subvector
of $\boldsymbol{\theta}_{k}$ that excludes its first element. During
Iteration $s$, we first update $\boldsymbol{\mu}$ using estimator
$\widehat{\boldsymbol{\theta}}^{\left(s-1\right)}$ from the previous
iteration. Then, we compute $\widehat{\boldsymbol{\theta}}^{\left(s\right)}_{k,-1}$
for $k=1,\ldots,q$ and $\widehat{\boldsymbol{\theta}}^{\left(s\right)}_{k}$
for $k=q+1,\ldots,d$. This is done through the following optimization:
\begin{align*}
\widehat{\boldsymbol{\theta}}^{\left(s\right)}_{k,-1} & =\arg\min_{\boldsymbol{\theta}_{k,-1}}Q_{k,n}\left(\widehat{\theta}^{\left(s-1\right)}_{k,1},\boldsymbol{\theta}_{k,-1},\overline{\boldsymbol{\mu}}_{n}\left(\widehat{\boldsymbol{\theta}}^{\left(s-1\right)}\right)\right)\textrm{ for }1\leq k\leq q\textrm{ and}\\
\widehat{\boldsymbol{\theta}}^{\left(s\right)}_{k} & =\arg\min_{\boldsymbol{\theta}_{k}}Q_{k,n}\left(\boldsymbol{\theta}_{k},\overline{\boldsymbol{\mu}}_{n}\left(\widehat{\boldsymbol{\theta}}^{\left(s-1\right)}\right)\right)\textrm{ for }q+1\leq k\leq d.
\end{align*}
Because the above $d$-number of optimizations do not share any common
argument, they can be solved using parallel computers. At last, we
calculate $\widehat{\theta}^{\left(s\right)}_{k,1}$ for $k=1,\ldots,q$
by computing $\widehat{\theta}^{\left(s\right)}_{1,1}$ through 
\begin{equation}
\widehat{\theta}^{\left(s\right)}_{1,1}=\arg\min_{\theta_{1,1}}\sum^{q}_{k=1}Q_{k,n}\left(\theta_{1,1},\widehat{\boldsymbol{\theta}}^{\left(s\right)}_{k,-1},\overline{\boldsymbol{\mu}}_{n}\left(\widehat{\boldsymbol{\theta}}^{\left(s-1\right)}\right)\right)\label{eq:theta 11}
\end{equation}
and set $\widehat{\theta}^{\left(s\right)}_{k,1}=\widehat{\theta}^{\left(s\right)}_{1,1}$
for $k=2,\ldots,q$. The optimization involves only a single argument
and is computationally efficient.\footnote{Let a general linear equality constraint of the second type be written
as $R\boldsymbol{\theta}^{\ast}=r$, where $R$ and $r$ are known
with dimensions $l_{R}\times pd$ and $l_{R}\times1$ respectively.
The matrix $R$ is assumed to have full row rank so that there is
no redundant constraint. We can always rearrange and decompose $R$
as $\left[R_{c},\mathbf{0}\right]$, where $R_{c}$ has dimension
$l_{R}\times q$ and has no zero column. The number of arguments in
the optimization problem (\ref{eq:theta 11}) for $R\boldsymbol{\theta}^{\ast}=r$
is $q-l_{R}\geq0$. The case where $q=l_{R}$ corresponds all $q$
number of elements in $\boldsymbol{\theta}^{\ast}$ having prespecified
values. The optimization problem (\ref{eq:theta 11}) is no longer
needed in this case.} Consequently, each iteration incurs a low computational burden. The
IDC estimator under this constraint is therefore fast to compute.

The aforementioned two procedures can be combined in a straightforward
way when constraints contain both types to accommodate any linear
equality constraint. Following a similar proof as the one for Theorem
\ref{thm:: consistency-fixed S}, we can show that the two constrained
IDC estimators, both denoted as $\widehat{\boldsymbol{\theta}}^{\left(S\right)}$,
are consistent given the consistent initial estimator $\check{\boldsymbol{\theta}}$.

\begin{corollary} \label{corr:: Constrained IDC Consistency} Suppose
Assumptions \ref{Assmp:: Maintained assumption}, \ref{Assm:: consistency 0},
and \ref{Assm:: consistency 1} hold. If $\widehat{\boldsymbol{\theta}}^{\left(0\right)}=\check{\boldsymbol{\theta}}$,
then $\widehat{\boldsymbol{\theta}}^{\left(S\right)}\stackrel{p}{\rightarrow}\boldsymbol{\theta}^{\ast}$
as $n\rightarrow\infty$ for any $S$. \end{corollary}

\subsection{IDC with Regularisation}

\label{subsec:IDC-with-Regularization}

The IDC estimation procedure can be readily extended to incorporate
penalisation, allowing it to accommodate MNL models with a large number
of covariates. We describe the regularised IDC procedure in this subsection.

Suppose the researcher employs a shrinkage method to reduce estimation
variance. We use the $\ell_{1}$ regularisation as an example to illustrate
the procedure (\citet{tibshirani1996regression}). As long as the
regularisation is additively separable in the alternatives, the same
procedure applies. Given a hyperparameter $\lambda>0$, the penalised
estimator is defined as 
\[
\left(\widehat{\boldsymbol{\theta}}_{\lambda},\widehat{\boldsymbol{\mu}}_{\lambda}\right)=\arg\min_{\boldsymbol{\theta}\in\Theta,\boldsymbol{\mu}\in\mathbb{R}^{n}}\left\{ Q_{n}\left(\boldsymbol{\theta},\boldsymbol{\mu}\right)+n\lambda\left\Vert \boldsymbol{\theta}\right\Vert _{1}\right\} ,
\]
where $Q_{n}\left(\boldsymbol{\theta},\boldsymbol{\mu}\right)$ is
the original loss function defined in (\ref{eq:Definition of Distributed Multinomial Estimator})
and 
\[
\left\Vert \boldsymbol{\theta}\right\Vert _{1}=\sum^{d}_{k=1}\left\Vert \boldsymbol{\theta}_{k}\right\Vert _{1}=\sum^{d}_{k=1}\sum^{p}_{j=1}\left|\theta_{k,j}\right|.
\]
Because the $\ell_{1}$ penalty term is additively separable across
$k$, we have that for any given $\boldsymbol{\mu}$, 
\begin{align*}
 & \arg\min_{\boldsymbol{\theta}\in\Theta}\left\{ Q_{n}\left(\boldsymbol{\theta},\boldsymbol{\mu}\right)+n\lambda\left\Vert \boldsymbol{\theta}\right\Vert _{1}\right\} \\
 & =\left[\arg\min_{\boldsymbol{\theta}_{1}\in\varTheta_{1}}\left\{ Q_{1,n}\left(\boldsymbol{\theta}_{1},\boldsymbol{\mu}\right)+n\lambda\left\Vert \boldsymbol{\theta}_{1}\right\Vert _{1}\right\} ,\ldots,\arg\min_{\boldsymbol{\theta}_{d}\in\varTheta_{d}}\left\{ Q_{d,n}\left(\boldsymbol{\theta}_{d},\boldsymbol{\mu}\right)+n\lambda\left\Vert \boldsymbol{\theta}_{d}\right\Vert _{1}\right\} \right]^{\prime}.
\end{align*}
Consequently, the regularised IDC estimator can be computed using
the same iteration procedure introduced in Section \ref{sec_sub:Iterative-Distributed-Computing}
except that $Q_{k,n}\left(\boldsymbol{\theta}_{k},\boldsymbol{\mu}\right)$
is replaced by $Q_{k,n}\left(\boldsymbol{\theta}_{k},\boldsymbol{\mu}\right)+n\lambda\left\Vert \boldsymbol{\theta}_{k}\right\Vert _{1}$.
Importantly, the distributed computing structure of the IDC algorithm
is fully preserved.

\section{Monte Carlo Simulation}

\label{sec:Monte-Carlo-Simulation}

In this section, we evaluate the performance of our IDC estimator
from various perspectives. We present the finite sample performance
of the IDC estimator by looking at the effects of separately increasing
$d$ and $n$ on mean squared error (MSE) and running time. We include
the maximum likelihood estimator, whenever computationally feasible,
to show that the IDC estimator performs similarly to the MLE in terms
of MSE and is always feasible even in cases where MLE is intractable.
We compare the performance of the IDC estimator with the modern stochastic
gradient descent (SGD) methods. To accommodate cases where the number
of covariates is large, we investigate the performance of the regularised
IDC algorithm. Lastly, we study the finite sample size and power of
our bootstrap inference procedure.

\subsection{Comparison of Estimation Accuracy}

\label{subsec:Monte Carlo Comparison-of-Estimation}

In this subsection, we present results on the finite sample performance
of the IDC estimator in terms of MSE and running time. We investigate
the effect of the initial estimator on the performance and compare
our IDC estimator with the MLE. The result shows that the IDC estimator
with consistent initial estimator achieves similar accuracy as the
MLE but is computationally more efficient.

In order to conduct the comparison with the MLE, we set $d$ to be
moderate so that MLE can be run within a reasonable time. The reported
running times in the tables in this subsection are obtained from a
cluster of 25 AWS EC2 instances with 12 vCPUs and 16GB memory. Such
a configuration can be formed on commonly used cloud computing platforms
within minutes. We employ this basic configuration in this subsection
to illustrate that our IDC estimator achieves superior performance
compared to existing estimators, even when computational resources
are suboptimal for distributed computing. All of the results presented
in this section are repeated five hundred times and averaged.

We first study the finite sample performance of the IDC estimator
with consistent $\check{\boldsymbol{\theta}}$ initialization, specifically
what happens to MSE ($n$ increasing, $d$ fixed) and running time
($d$ increasing, $n$ fixed). We consider the following data generating
process (DGP):

\textbf{DGP-A {[}MNL{]}}: We set $p=5$. The covariate vector $\boldsymbol{X}$
follows the standard normal distribution; $M$ follows the discrete
uniform distribution on $\left[20,30\right]$; and the values of $\mathbf{\boldsymbol{\theta}}^{*}$
are obtained by random draws from the standard normal distribution.
\begin{table}[H]
\centering %
\begin{tabular}{ccccccccccccccccc}
\toprule 
 &  &  &  &  &  & \multicolumn{3}{c}{$n=500$} &  & \multicolumn{3}{c}{$n=1000$} &  & \multicolumn{3}{c}{$n=2000$}\tabularnewline
\midrule 
$\widehat{\boldsymbol{\theta}}^{\left(0\right)}$  &  & $S$  &  & $d$  &  & MSE  &  & Time  &  & MSE  &  & Time  &  & MSE  &  & Time\tabularnewline
\midrule 
$\check{\boldsymbol{\theta}}$  &  & $10$  &  & $10$  &  & $.0030$  &  & $45$s  &  & $.0020$  &  & $97$s  &  & $.0012$  &  & $252$s\tabularnewline
 &  &  &  & $20$  &  & $.0083$  &  & $52$s  &  & $.0052$  &  & $113$s  &  & $.0020$  &  & $277$s\tabularnewline
 &  &  &  & $50$  &  & $.0373$  &  & $85$s  &  & $.0182$  &  & $174$s  &  & $.0094$  &  & $453$s\tabularnewline
 &  &  &  & $100$  &  & $.0793$  &  & $157$s  &  & $.0381$  &  & $349$s  &  & $.0171$  &  & $756$s\tabularnewline
 &  &  &  & $150$  &  & $.1749$  &  & $211$s  &  & $.0585$  &  & $455$s  &  & $.0223$  &  & $1007$s\tabularnewline
 &  &  &  &  &  &  &  &  &  &  &  &  &  &  &  & \tabularnewline
 &  & $40$  &  & $10$  &  & $.0037$  &  & $168$s  &  & $.0021$  &  & $320$s  &  & $.0012$  &  & $672$s\tabularnewline
 &  &  &  & $20$  &  & $.0081$  &  & $196$s  &  & $.0040$  &  & $352$s  &  & $.0019$  &  & $739$s\tabularnewline
 &  &  &  & $50$  &  & $.0365$  &  & $320$s  &  & $.0185$  &  & $576$s  &  & $.0091$  &  & $1142$s\tabularnewline
 &  &  &  & $100$  &  & $.0661$  &  & $590$s  &  & $.0336$  &  & $890$s  &  & $.0158$  &  & $1948$s\tabularnewline
 &  &  &  & $150$  &  & $.1815$  &  & $770$s  &  & $.0577$  &  & $1184$s  &  & $.0211$  &  & $2822$s\tabularnewline
\bottomrule
\end{tabular}\caption{Finite sample performance of IDC estimator with $\check{\boldsymbol{\theta}}$
initialization}
\label{tab: IDC-Consistent} 
\end{table}

From Table \ref{tab: IDC-Consistent}, we observe that the MSE of
the IDC estimator decreases as the sample size increases. When the
number of iterations $S$ increases, we see an improvement in the
MSE. However, the improvement is marginal, suggesting that the IDC
estimator with $\check{\boldsymbol{\theta}}$ initialization stabilizes
with only a few iterations. We also see from Table \ref{tab: IDC-Consistent}
that the running time is approximately a linear function of $d$,
which agrees with our discussion in Section \ref{subsec:Time-and-Space}.
When the number of cores available exceeds the number of choices,
the additional computational cost of increasing $d$ is very small.
After the cores are fully occupied by the number of processes, the
running time becomes approximately linear. The nominal value of running
time depends on the hardware specifications. 
\begin{table}[H]
\centering %
\begin{tabular}{ccccccccccccc}
\toprule 
{}  &  & \multicolumn{3}{c}{$n=500$} &  & \multicolumn{3}{c}{$n=1000$} &  & \multicolumn{3}{c}{$n=2000$}\tabularnewline
\midrule 
$d$  &  & MSE  &  & Time  &  & MSE  &  & Time  &  & MSE  &  & Time\tabularnewline
\midrule 
$10$  &  & $.0040$  &  & $36$s  &  & $.0020$  &  & $105$s  &  & $.0015$  &  & $350$s\tabularnewline
$20$  &  & $.0101$  &  & $54$s  &  & $.0045$  &  & $152$s  &  & $.0026$  &  & $652$s\tabularnewline
$50$  &  & $.0263$  &  & $96$s  &  & $.0187$  &  & $308$s  &  & $.0098$  &  & $1523$s\tabularnewline
$100$  &  & $.0852$  &  & $250$s  &  & $.0405$  &  & $862$s  &  & $.0151$  &  & $4375$s\tabularnewline
$150$  &  & $.1179$  &  & $457$s  &  & $.0536$  &  & $1523$s  &  & $.0291$  &  & $9352$s\tabularnewline
\bottomrule
\end{tabular}\caption{MSE and running time of MLE $\widetilde{\boldsymbol{\theta}}$}
\label{tab:CMLE} 
\end{table}

To compare the performance of our IDC estimator with the MLE $\widetilde{\boldsymbol{\theta}}$,
we compute the MSE and running time of $\widetilde{\boldsymbol{\theta}}$
from the same DGP and present the result in Table \ref{tab:CMLE}.\footnote{We also write code to try estimators in \citet{bohning1988monotonicity},
\citet{bohning1992multinomial}, and \citet{simon2013blockwise} for
comparison. However, our simulation result suggests that their performance
depends crucially on the number of iterations.} It can be seen that the MSE of the IDC estimator with $\check{\boldsymbol{\theta}}$
initialization is very close to that of the MLE even when the number
of iterations is only $10$. Note that the main advantage of the parallel
estimator is best observed for high enough $d$ because for low $d$,
the communication between parallel processes is unnecessary and hence
parallel computing increases the running time unnecessarily. The superior
performance of the IDC estimator is apparent when $d$ is large. For
instance, when $d=150$ and $n=2000$, the IDC estimator with $S=10$
achieves a similar MSE as the MLE with only about one-tenth of the
running time. For higher-dimensional cases, such as when $d$ exceeds
$150$, computing the MLE becomes computationally intensive and may
not be practical for many applications. In comparison, the IDC estimator
with $S=10$ demonstrates more efficient computation times, requiring
approximately 5, 10, and 20 minutes for sample sizes $n=500$, $1000$,
and $2000$, respectively. Moreover, the running time for the IDC
estimator can be further decreased if more compute instances are used.
For example, using 96 instances, the running time of the IDC estimator
for $d=150$ can be further reduced to $34$, $51$, and $188$ seconds
for $n=500$, $1000$, $2000$ respectively even for $S=40$. Compared
to the corresponding running time of MLE, the running time of the
IDC estimator using 96 instances is more than 10, 30, and 50 times
shorter. 
\begin{table}[H]
\centering %
\begin{tabular}{ccccccccccccccccc}
\toprule 
{}  &  &  &  &  &  & \multicolumn{3}{c}{$n=500$} &  & \multicolumn{3}{c}{$n=1000$} &  & \multicolumn{3}{c}{$n=2000$}\tabularnewline
\midrule 
$\widehat{\boldsymbol{\theta}}^{\left(0\right)}$  &  & $S$  &  & $d$  &  & MSE  &  & Time  &  & MSE  &  & Time  &  & MSE  &  & Time\tabularnewline
\midrule 
$\widehat{\boldsymbol{\theta}}_{T}$  &  & $10$  &  & $10$  &  & $.0048$  &  & $42$s  &  & $.0023$  &  & $95$s  &  & $.0017$  &  & $247$s\tabularnewline
 &  &  &  & $20$  &  & $.0098$  &  & $49$s  &  & $.0058$  &  & $106$s  &  & $.0038$  &  & $278$s\tabularnewline
 &  &  &  & $50$  &  & $.0466$  &  & $83$s  &  & $.0197$  &  & $175$s  &  & $.0100$  &  & $442$s\tabularnewline
 &  &  &  & $100$  &  & $.0809$  &  & $151$s  &  & $.0381$  &  & $339$s  &  & $.0184$  &  & $754$s\tabularnewline
 &  &  &  & $150$  &  & $.1725$  &  & $206$s  &  & $.0590$  &  & $451$s  &  & $.0242$  &  & $998$s\tabularnewline
 &  &  &  &  &  &  &  &  &  &  &  &  &  &  &  & \tabularnewline
 &  & $40$  &  & $10$  &  & $.0051$  &  & $160$s  &  & $.0025$  &  & $311$s  &  & $.0017$  &  & $667$s\tabularnewline
 &  &  &  & $20$  &  & $.0102$  &  & $187$s  &  & $.0052$  &  & $337$s  &  & $.0022$  &  & $728$s\tabularnewline
 &  &  &  & $50$  &  & $.0464$  &  & $309$s  &  & $.0198$  &  & $564$s  &  & $.0094$  &  & $1128$s\tabularnewline
 &  &  &  & $100$  &  & $.0867$  &  & $581$s  &  & $.0407$  &  & $876$s  &  & $.0179$  &  & $1938$s\tabularnewline
 &  &  &  & $150$  &  & $.1808$  &  & $758$s  &  & $.0588$  &  & $1177$s  &  & $.0219$  &  & $2801$s\tabularnewline
 &  &  &  &  &  &  &  &  &  &  &  &  &  &  &  & \tabularnewline
 &  &  &  &  &  & \multicolumn{3}{c}{$n=500$} &  & \multicolumn{3}{c}{$n=1000$} &  & \multicolumn{3}{c}{$n=2000$}\tabularnewline
\midrule 
$\widehat{\boldsymbol{\theta}}^{\left(0\right)}$  &  & $S$  &  & $d$  &  & MSE  &  & Time  &  & MSE  &  & Time  &  & MSE  &  & Time\tabularnewline
\midrule 
\multirow{1}{*}{$\widehat{\boldsymbol{\theta}}_{P}$} &  & $10$  &  & $10$  &  & $.0068$  &  & $52$s  &  & $.0023$  &  & $97$s  &  & $.0016$  &  & $244$s\tabularnewline
 &  &  &  & $20$  &  & $.0118$  &  & $59$s  &  & $.0058$  &  & $111$s  &  & $.0038$  &  & $290$s\tabularnewline
 &  &  &  & $50$  &  & $.0466$  &  & $83$s  &  & $.0197$  &  & $177$s  &  & $.0100$  &  & $438$s\tabularnewline
 &  &  &  & $100$  &  & $.0808$  &  & $155$s  &  & $.0382$  &  & $341$s  &  & $.0184$  &  & $751$s\tabularnewline
 &  &  &  & $150$  &  & $.1739$  &  & $209$s  &  & $.0588$  &  & $457$s  &  & $.0258$  &  & $1008$s\tabularnewline
 &  &  &  &  &  &  &  &  &  &  &  &  &  &  &  & \tabularnewline
 &  & $40$  &  & $10$  &  & $.0050$  &  & $160$s  &  & $.0025$  &  & $322$s  &  & $.0017$  &  & $651$s\tabularnewline
 &  &  &  & $20$  &  & $.0102$  &  & $198$s  &  & $.0052$  &  & $355$s  &  & $.0022$  &  & $728$s\tabularnewline
 &  &  &  & $50$  &  & $.0464$  &  & $318$s  &  & $.0198$  &  & $570$s  &  & $.0095$  &  & $1140$s\tabularnewline
 &  &  &  & $100$  &  & $.0866$  &  & $577$s  &  & $.0407$  &  & $881$s  &  & $.0179$  &  & $1957$s\tabularnewline
 &  &  &  & $150$  &  & $.1926$  &  & $761$s  &  & $.0587$  &  & $1151$s  &  & $.0245$  &  & $2811$s\tabularnewline
\bottomrule
\end{tabular}\caption{Finite sample performance of IDC estimator with $\widehat{\boldsymbol{\theta}}_{T}$
and $\widehat{\boldsymbol{\theta}}_{P}$ initialization}
\label{tab:IDC Taddy} 
\end{table}

In Table \ref{tab:IDC Taddy}, we present the MSE and running time
of IDC estimators with $\widehat{\boldsymbol{\theta}}_{T}$ and $\widehat{\boldsymbol{\theta}}_{P}$
as initial estimators, respectively, for two different numbers of
iterations $S$. Comparing MSEs of three IDC estimators with different
initial values: $\check{\boldsymbol{\theta}}$ (Table \ref{tab: IDC-Consistent})
and $\widehat{\boldsymbol{\theta}}_{T}$ (Table \ref{tab:IDC Taddy})
or $\widehat{\boldsymbol{\theta}}_{P}$ (Table \ref{tab:IDC Taddy}),
we observe that the IDC estimator with the consistent initial estimator
reduces the MSEs for the same number of iterations. 
\begin{table}[H]
\centering %
\begin{tabular}{cccccccccccccc}
\toprule 
 &  & $d$  &  & $n$  &  & $\widetilde{\boldsymbol{\theta}}$  &  & $\widehat{\boldsymbol{\theta}}^{I}_{PB}$  &  & $\widehat{\boldsymbol{\theta}}^{I}_{P}$  &  & $\widehat{\boldsymbol{\theta}}^{I}_{T}$  & \tabularnewline
\midrule 
DGP-A  &  & $20$  &  & $500$  &  & $.0102$  &  & $.0083$  &  & $.0102$  &  & $.0102$  & \tabularnewline
 &  & $20$  &  & $1000$  &  & $.0049$  &  & $.0048$  &  & $.0052$  &  & $.0052$  & \tabularnewline
 &  & $50$  &  & $1000$  &  & $.0155$  &  & $.0158$  &  & $.0198$  &  & $.0198$  & \tabularnewline
 &  &  &  &  &  &  &  &  &  &  &  &  & \tabularnewline
DGP-B  &  & $20$  &  & $500$  &  & $.0007$  &  & $.0008$  &  & $.0006$  &  & $.0303$  & \tabularnewline
 &  & $20$  &  & $1000$  &  & $.0002$  &  & $.0002$  &  & $.0003$  &  & $.0403$  & \tabularnewline
 &  & $50$  &  & $1000$  &  & $.0005$  &  & $.0005$  &  & $.0006$  &  & $.0121$  & \tabularnewline
 &  &  &  &  &  &  &  &  &  &  &  &  & \tabularnewline
DGP-C  &  & $20$  &  & $500$  &  & $.0060$  &  & $.0062$  &  & $.0072$  &  & $.0061$  & \tabularnewline
 &  & $20$  &  & $1000$  &  & $.0032$  &  & $.0032$  &  & $.0032$  &  & $.0045$  & \tabularnewline
 &  & $50$  &  & $1000$  &  & $.0315$  &  & $.0318$  &  & $.0305$  &  & $.0323$  & \tabularnewline
\bottomrule
\end{tabular}\caption{MSE comparison of competing estimators. Number of iterations $S=20$.}
\label{tab:Comparison} 
\end{table}

Table \ref{tab:Comparison} presents MSEs of different estimators
for three $\left(d,n\right)$ pairs each. We set the largest $d$
be $50$ so that MLE can be computed in a reasonable time. $\widehat{\boldsymbol{\theta}}^{I}_{PB}$,
$\widehat{\boldsymbol{\theta}}^{I}_{T}$, and $\widehat{\boldsymbol{\theta}}^{I}_{P}$
denote the IDC estimators with $\check{\boldsymbol{\theta}}$, $\widehat{\boldsymbol{\theta}}_{T}$,
and $\widehat{\boldsymbol{\theta}}_{P}$ as the initial estimators
respectively. Besides DGP-A, we consider two additional DGPs to study
the performance of the IDC estimator under different data settings.
In all DGPs, we let $p=5$.

\textbf{DGP-B {[}Poisson{]}}: The random variable $\boldsymbol{X}$
follows a standard normal distribution; $C_{ik}$ follows a Poisson
distribution with mean $e^{\eta^{*}_{ik}}$; and $M$ is obtained
by summing up realizations of the Poisson draws for different choices.

\textbf{DGP-C {[}Mixture{]}}: We let $\boldsymbol{X}$ follow a mixture
of Gaussian distributions with means $0$ and $4$ with standard deviations
$1$ for both distributions. $M$ is also set to follow a mixture
of Gaussian distributions with means $10$ and $60$ and rounded to
the closest integer. The standard deviations are $1$ and $5$ respectively.
We have made these modifications so that some choices are rarely selected
and ensure the robustness of our estimator in those cases.

Based on the simulation result, our IDC algorithm is successfully
executed for all DGPs and exhibits stability. In contrast, we encounter
errors for DGP-C when computing the estimators in \citet{bohning1988monotonicity},
\citet{bohning1992multinomial}, and \citet{simon2013blockwise}.
We see from Table \ref{tab:Comparison} that $\widehat{\boldsymbol{\theta}}^{I}_{PB}$
performs close to $\widetilde{\boldsymbol{\theta}}$ for all DGPs
and $\left(d,n\right)$ pairs. In DGP-B, the initial estimator $\widehat{\boldsymbol{\theta}}_{P}$
is the maximum likelihood estimator. As a result, $\widehat{\boldsymbol{\theta}}^{I}_{P}$
starts with not only a consistent but asymptotically efficient initial
estimator. Even in this case, $\widehat{\boldsymbol{\theta}}^{I}_{PB}$
has comparable MSEs.

In summary, the IDC estimators with all three initial estimators have
finite sample performance similar to the MLE for the DGPs studied
in this section. They are much faster to compute than the MLE for
large $d$ and are feasible even when the MLE might be intractable.
Moreover, if the IDC estimator starts with the consistent initial
estimator $\check{\boldsymbol{\theta}}$, its finite sample performance
will be further improved and is almost the same as the MLE.

\subsection{Finite-Sample Performance for Large $d$}

\label{subsec:Simulation large d}

In this subsection, we further study the performance of our IDC estimator,
particularly for large $d$. We focus on the IDC estimator with consistent
initial estimator because it dominates the other IDC estimators. The
MLE cannot be computed within a reasonable time. We increase the computational
power to a cluster of AWS EC2 c6i.4xlarge instances with 16 vCPUs
(3rd generation Intel Xeon Ice Lake) and 32GB memory. The DGPs are
the same as the ones introduced in Section \ref{subsec:Monte Carlo Comparison-of-Estimation},
except that we draw $M$ from a discrete uniform distribution on $\left[200,300\right]$
in DGP A and on $\left[500,600\right]$ in DGP C. We fix $n$ to be
$1000$.

\begin{table}[H]
\centering \resizebox{\textwidth}{!}{ %
\begin{tabular}{ccccccccccccccccccc}
\toprule 
 &  &  &  & \multicolumn{3}{c}{$d=250$} &  & \multicolumn{3}{c}{$d=500$} &  & \multicolumn{3}{c}{$d=1000$} &  & \multicolumn{3}{c}{$d=2000$}\tabularnewline
\midrule 
DGP  &  & $S$  &  & MSE  &  & Time  &  & MSE  &  & Time  &  & MSE  &  & Time  &  & MSE  &  & Time \tabularnewline
\midrule 
A  &  & $10$  &  & $0.035$  &  & $41$s  &  & $0.051$  &  & $66$s  &  & $0.077$  &  & $206$s  &  & $0.114$  &  & $832$s \tabularnewline
 &  & $20$  &  & $0.009$  &  & $77$s  &  & $0.022$  &  & $124$s  &  & $0.049$  &  & $396$s  &  & $0.064$  &  & $1556$s \tabularnewline
 &  & $40$  &  & $0.008$  &  & $151$s  &  & $0.019$  &  & $246$s  &  & $0.044$  &  & $684$s  &  & $0.052$  &  & $2878$s \tabularnewline
 &  &  &  &  &  &  &  &  &  &  &  &  &  &  &  &  &  & \tabularnewline
C  &  & $10$  &  & $0.011$  &  & $52$s  &  & $0.011$  &  & $127$s  &  & $0.025$  &  & $180$s  &  & $0.051$  &  & $709$s \tabularnewline
 &  & $20$  &  & $0.003$  &  & $83$s  &  & $0.008$  &  & $252$s  &  & $0.020$  &  & $339$s  &  & $0.033$  &  & $1356$s \tabularnewline
 &  & $40$  &  & $0.002$  &  & $146$s  &  & $0.008$  &  & $377$s  &  & $0.019$  &  & $701$s  &  & $0.031$  &  & $2887$s \tabularnewline
\bottomrule
\end{tabular}}\caption{Finite sample performance of IDC estimator with large $d$}
\label{tab: IDC-large d} 
\end{table}

Table \ref{tab: IDC-large d} shows that with additional computational
resource, the IDC estimator performs well for very large $d$. However,
because $d$ is of similar size as $n$ or even larger than $n$,
the initial consistent estimator becomes inaccurate. Additional iterations
are needed for the MSEs to stabilize. Based on the simulation result,
the algorithm stabilises at $S=40$ even when $d=2000$.

To further demonstrate the performance advantages of our estimator,
we compare it with stochastic gradient descent (SGD) based optimization
methods implemented using PyTorch and TensorFlow. We consider two
optimizers: vanilla SGD (without momentum) and Adaptive Moment Estimation
(Adam). The latter is a stochastic gradient optimizer that adaptively
adjusts the learning rate for each parameter and often improves stability
and convergence speed relative to vanilla SGD. Let $\kappa$ denote
the learning rate in the SGD methods. We consider $\kappa\in\left\{ 0.01,0.001\right\} $,
run for 100 epochs with mini-batch size 256, and initialize parameters
from $\mathcal{N}\left(0,0.01^{2}\right)$, following standard practice
in neural network styled optimization. We use the same data in all
cases to clearly show the difference between directly optimizing all
parameters at once with SGD and solving a sequence of structured subproblems
with the IDC algorithm.

\begin{table}[H]
\centering %
\begin{tabular}{ccccccccccc}
\toprule 
 &  &  &  & \multicolumn{3}{c}{DGP A} &  & \multicolumn{3}{c}{DGP C}\tabularnewline
\midrule 
Method  &  & $\kappa$  &  & $d=250$  &  & $d=500$  &  & $d=250$  &  & $d=500$\tabularnewline
\midrule 
SGD  &  & $0.01$  &  & $0.935$  &  & $0.992$  &  & $0.934$  &  & $0.979$ \tabularnewline
 &  & $0.001$  &  & $0.959$  &  & $0.992$  &  & $0.959$  &  & $0.992$ \tabularnewline
 &  &  &  &  &  &  &  &  &  & \tabularnewline
Adam  &  & $0.01$  &  & $0.550$  &  & $0.790$  &  & $0.551$  &  & $0.797$ \tabularnewline
 &  & $0.001$  &  & $0.638$  &  & $0.682$  &  & $0.628$  &  & $0.671$ \tabularnewline
 &  &  &  &  &  &  &  &  &  & \tabularnewline
IDC  &  &  &  & $0.035$  &  & $0.051$  &  & $0.008$  &  & $0.011$ \tabularnewline
\bottomrule
\end{tabular}\caption{Finite sample performance of SGD-based methods with different $\kappa$}
\label{tab: IDC-SGD 1} 
\end{table}

Table \ref{tab: IDC-SGD 1} reports the MSE of the vanilla SGD, Adam,
and our IDC estimator with $S=10$. First, we see that the SGD based
optimization methods are sensitive to the learning rate. Moreover,
the optimal learning rate differs across DGPs and the values of $d$,
making it difficult to select in practice. Second, the IDC estimator
achieves substantially smaller MSE, often one to two orders of magnitude
lower than both vanilla SGD and Adam. To further examine the role
of optimization length, we increase the number of epochs from 100
to 5000 in a second experiment under DGP-A with $d=250$.

\begin{table}[H]
\centering %
\begin{tabular}{cccccccc}
\toprule 
Method  &  & Epochs  &  & MSE  &  & Time  & \tabularnewline
\midrule 
SGD $\kappa=0.10$  &  & $100$  &  & $0.753$  &  & $0.1$s  & \tabularnewline
 &  & $500$  &  & $0.446$  &  & $0.4$s  & \tabularnewline
 &  & $1000$  &  & $0.360$  &  & $0.8$s  & \tabularnewline
 &  & $5000$  &  & $0.596$  &  & $4.0$s  & \tabularnewline
 &  &  &  &  &  &  & \tabularnewline
Adam $\kappa=0.01$  &  & $100$  &  & $0.550$  &  & $0.1$s  & \tabularnewline
 &  & $500$  &  & $0.739$  &  & $0.4$s  & \tabularnewline
 &  & $1000$  &  & $0.742$  &  & $0.9$s  & \tabularnewline
 &  & $5000$  &  & $0.741$  &  & $4.3$s  & \tabularnewline
 &  &  &  &  &  &  & \tabularnewline
IDC $S=10$  &  &  &  & $0.035$  &  & $40$s  & \tabularnewline
\bottomrule
\end{tabular}\caption{Finite sample performance of SGD-based methods with different numbers
of epochs}
\label{tab: IDC-SGD 2} 
\end{table}

Table \ref{tab: IDC-SGD 2} shows that the performance of SGD-based
methods also varies considerably with the number of epochs, and the
relationship is not monotone. Increasing epochs does not necessarily
improve accuracy. Across all simulation designs in Tables \ref{tab: IDC-SGD 1}
and \ref{tab: IDC-SGD 2}, the IDC estimator consistently outperforms
SGD-based optimization.

Conceptually, SGD treats the multinomial logit model as a single large-scale
optimization problem over all $dp$ parameters. In contrast, IDC exploits
the structure of the MNL model and decomposes one difficult high-dimensional
problem into a sequence of low-dimensional subproblems. The cost is
the iteration. But even with $S=10$, the resulting estimator substantially
outperforms thousands of stochastic gradient updates applied to the
joint objective.

\subsection{Regularised IDC Estimator}

We now investigate the behaviour of the regularised IDC estimator
introduced in Section \ref{subsec:IDC-with-Regularization}. The Monte
Carlo simulation is conducted under DGP A. $M$ is drawn from a discrete
uniform distribution on $\left[500,600\right]$. We set $n=1000$,
and vary the number of choices $d$ and the covariate dimension $p$.
We use the $\ell_{1}$ penalty term during iteration and also include
it when computing the initial estimator. We set $S=10$. The hyperparameter
$\lambda$ controls the bias-variance tradeoff. 
\begin{table}[H]
\centering %
\begin{tabular}{ccccccccccccc}
\toprule 
 &  &  &  & \multicolumn{1}{c}{$d=200$} &  & \multicolumn{1}{c}{$d=250$} &  & \multicolumn{1}{c}{$d=500$} &  & \multicolumn{1}{c}{$d=1000$} &  & \multicolumn{1}{c}{$d=2000$}\tabularnewline
\midrule 
$p$  &  & $\lambda$  &  & MSE  &  & MSE  &  & MSE  &  & MSE  &  & MSE\tabularnewline
\midrule 
$50$  &  & $0.1$  &  & $0.039$  &  & $0.052$  &  & $0.113$  &  & $0.206$  &  & $0.467$\tabularnewline
 &  & $0.2$  &  & $0.038$  &  & $0.052$  &  & $0.112$  &  & $0.205$  &  & $0.464$\tabularnewline
 &  & $0.5$  &  & $0.038$  &  & $0.052$  &  & $0.112$  &  & $0.203$  &  & $0.455$\tabularnewline
 &  &  &  &  &  &  &  &  &  &  &  & \tabularnewline
$100$  &  & $0.1$  &  & $0.093$  &  & $0.115$  &  & $0.251$  &  & $0.485$  &  & $0.975$\tabularnewline
 &  & $0.2$  &  & $0.093$  &  & $0.115$  &  & $0.250$  &  & $0.482$  &  & $0.963$\tabularnewline
 &  & $0.5$  &  & $0.093$  &  & $0.115$  &  & $0.248$  &  & $0.473$  &  & $0.931$\tabularnewline
\bottomrule
\end{tabular}\caption{Finite sample performance of regularised IDC estimator}
\label{tab: IDC-penalised} 
\end{table}

Table \ref{tab: IDC-penalised} shows that the regularised IDC estimator
delivers stable and accurate estimates for moderately large values
of $p$. The MSE increases monotonically with both $d$ and $p$.
Within the tested range $\lambda\in\left\{ 0.1,0.2,0.5\right\} $,
the performance of the regularised IDC estimators is similar.

\subsection{Inference }

\label{subsec:Simulation Inference}

In this section, we illustrate the bootstrap inference procedure introduced
in Section \ref{subsec:Asymptotic-Distributions-Inference}. We investigate
the finite sample performance of the procedure including the size
and power. All the results are based on one thousand Monte Carlo repetitions,
where the number of bootstrap repetitions is five hundred.

We consider the null hypothesis that some element of $\boldsymbol{\theta}^{\ast}$
equals to a specific value. Data are generated from DGP-A introduced
in the previous section. Let the null and the alternative hypotheses
be that $H_{0}:\theta^{\ast}_{11}=0$ and $H_{1}:\theta^{\ast}_{11}\neq0$.
The test statistic is computed as $\left|\frac{\widehat{\theta}^{I}_{11}}{\widehat{se}_{b}\left(\widehat{\theta}^{I}_{11}\right)/\sqrt{n}}\right|$,
where $\widehat{se}_{b}\left(\widehat{\theta}^{I}_{11}\right)$ is
the bootstrap estimate of the standard error of $\widehat{\theta}^{I}_{11}$.
The number of iterations is 10 when computing the IDC estimator. We
set the nominal size as $5\%$ and use the $97.5\%$ quantile of the
standard normal distribution as the critical value.

\begin{table}[H]
\centering \resizebox{\textwidth}{!}{ %
\begin{tabular}{llcccccccccccccc}
\toprule 
 & Dev.  &  & $-0.2$  &  & $-0.1$  &  & $-0.05$  &  & $0$  &  & $0.05$  &  & $0.1$  &  & $0.2$\tabularnewline
\midrule 
$d=20$  & $n=250$  &  & $.397$  &  & $.137$  &  & $.067$  &  & $.041$  &  & $.060$  &  & $.139$  &  & $.455$\tabularnewline
 & $n=500$  &  & $.676$  &  & $.216$  &  & $.102$  &  & $.058$  &  & $.115$  &  & $.234$  &  & $.704$\tabularnewline
 & $n=1000$  &  & $.952$  &  & $.395$  &  & $.128$  &  & $.055$  &  & $.167$  &  & $.473$  &  & $.963$\tabularnewline
 &  &  &  &  &  &  &  &  &  &  &  &  &  &  & \tabularnewline
 & Dev.  &  & $-0.3$  &  & $-0.2$  &  & $-0.1$  &  & $0$  &  & $0.1$  &  & $0.2$  &  & $0.3$\tabularnewline
\midrule 
$d=50$  & $n=250$  &  & $.333$  &  & $.189$  &  & $.099$  &  & $.077$  &  & $.091$  &  & $.186$  &  & $.313$ \tabularnewline
 & $n=500$  &  & $.662$  &  & $.424$  &  & $.172$  &  & $.066$  &  & $.198$  &  & $.391$  &  & $.658$ \tabularnewline
 & $n=1000$  &  & $.895$  &  & $.723$  &  & $.247$  &  & $.063$  &  & $.212$  &  & $.697$  &  & $.924$ \tabularnewline
\bottomrule
\end{tabular}} \caption{Finite sample rejection probabilities for different values of $\theta^{\ast}_{11}$,
$n$, and $d$}
\label{tab: size and power} 
\end{table}

In Table \ref{tab: size and power}, we report the finite sample rejection
probabilities of our test for different values of $\theta^{\ast}_{11}$.
Values in the first row of the table indicate the deviation of $\theta^{\ast}_{11}$
from the null hypothesis. When the deviation is zero, the null hypothesis
is true. It can be seen from the table that the finite sample rejection
rates get closer to the nominal size when the sample size increases.
And when the true value $\theta^{\ast}_{11}$ deviates more from the
null hypothesis, the rejection probabilities increase. The same pattern
appears for both $d=20$ and $d=50$. The finite sample performance
of the test when $d=50$ is not as good as that when $d=20$. This
is predictable because there are many more unknown parameters in the
model when $d=50$ than when $d=20$. We expect the results to improve
as the sample size increases for any fixed $d$.

\section{Empirical Illustration}

\label{sec:Empirical-Illustration}

In the previous section, we evaluated the finite-sample performance
of the IDC estimator using simulated data, where the true parameter
is known. We now turn to an empirical application. The goal is to
show that the IDC procedure can be implemented on a real text dataset
and to examine whether it produces stable and interpretable covariate--word
relationships. The focus is on empirical feasibility and interpretability
rather than a full comparison with all alternative estimators.

We use a subset of the Yelp review corpus studied in \citet{taddy2015distributed}.
The data are publicly available from the \href{https://www.kaggle.com/competitions/yelp-recruiting}{Kaggle Yelp Recruiting Competition}
and contain $215,879$ reviews of $11,535$ businesses in the Phoenix
metropolitan area. We randomly select $n=4000$ reviews. To construct
a vocabulary of size $d=400$, we first select the 100 most frequent
words in the corpus. From the remaining words, we rank them by frequency,
partition them into three groups based on rank percentiles (25th--50th,
50th--75th, and 75th--100th percentiles), and randomly select 100
words from each group. $\boldsymbol{C}$ represents the word counts
including emoticons, and $M$ denotes the total number of words in
a review. The covariate vector $\boldsymbol{X}$ includes an intercept
and 11 review/reviewer/business attributes ($p=12$), which correspond
to the main numerical features considered in \citet{taddy2015distributed}.
All covariates are standardized to have mean zero and unit variance.

\begin{table}[H]
\centering %
\begin{tabular}{ll}
\toprule 
Covariate  & Description\tabularnewline
\midrule 
\texttt{intercept}  & Constant term\tabularnewline
\texttt{funny}  & Number of ``funny'' votes on the review\tabularnewline
\texttt{useful}  & Number of ``useful'' votes on the review\tabularnewline
\texttt{cool}  & Number of ``cool'' votes on the review\tabularnewline
\texttt{stars}  & Star rating of the review (1-5)\tabularnewline
\texttt{usr.funny}  & Number of ``funny'' votes received by the reviewer\tabularnewline
\texttt{usr.useful}  & Number of ``useful'' votes received by the reviewer\tabularnewline
\texttt{usr.cool}  & Number of ``cool'' votes received by the reviewer\tabularnewline
\texttt{usr.stars}  & Average star rating of the reviewer\tabularnewline
\texttt{usr.count}  & Number of reviews written by the reviewer\tabularnewline
\texttt{biz.stars}  & Average star rating of the business\tabularnewline
\texttt{biz.count}  & Number of reviews for the business\tabularnewline
\bottomrule
\end{tabular}\caption{Description of the covariates}
\label{tab: Covariate} 
\end{table}

We compute three estimators: the IDC estimators with $S=10$ and $S=20$,
and the Taddy's estimator $\widehat{\boldsymbol{\theta}}_{T}$, which
solves the Poisson subproblems once without further IDC iteration.
Each coefficient $\theta_{k,j}$ has a straightforward interpretation.
Since all covariates are standardized to have mean zero and unit variance,
$\theta_{k,j}$ measures the change in the log odds of word $k$ associated
with a one standard deviation increase in covariate $j$, holding
other covariates fixed. For example, $\widehat{\theta}_{\mathrm{fantastic},\mathrm{stars}}=0.703$
means that a one standard deviation increase in the number of stars
increases the log odds of the word ``fantastic'' by $0.703$. In
multiplicative terms, $\exp\left(0.703\right)\approx2.01$, so the
relative frequency of ``fantastic'' approximately doubles. In contrast,
if $\widehat{\theta}_{\mathrm{average},\mathrm{stars}}=-0.587$, then
$\exp\left(-1.017\right)\approx0.56$, implying that the relative
frequency of ``average'' decreases by about a half.

\begin{table}[H]
\centering %
\begin{tabular}{lllllllll}
\toprule 
 &  & Rank  &  & IDC $S=10$  &  & IDC $S=20$  &  & Taddy's $\widehat{\boldsymbol{\theta}}_{T}$ \tabularnewline
\midrule 
\texttt{stars}  &  & +1  &  & fantastic (0.703)  &  & fantastic (0.724)  &  & fantastic (0.641)\tabularnewline
 &  & +2  &  & awesome (0.658)  &  & awesome (0.667)  &  & awesome (0.612)\tabularnewline
 &  & +3  &  & perfect (0.593)  &  & perfect (0.591)  &  & perfect (0.598)\tabularnewline
 &  & +4  &  & wonderful (0.517)  &  & wonderful (0.514)  &  & wonderful (0.573)\tabularnewline
 &  & +5  &  & valley (0.463)  &  & valley (0.460)  &  & valley (0.557)\tabularnewline
 &  & --1  &  & average (-0.587)  &  & average (-0.593)  &  & average (-0.519)\tabularnewline
 &  & --2  &  & pay (-0.521)  &  & pay (-0.528)  &  & pay (-0.514)\tabularnewline
 &  & --3  &  & waitres (-0.468)  &  & waitres (-0.462)  &  & waitres (-0.503)\tabularnewline
 &  & --4  &  & told (-0.437)  &  & told (-0.431)  &  & told (-0.491)\tabularnewline
 &  & --5  &  & should (-0.412)  &  & should (-0.421)  &  & instead (-0.482)\tabularnewline
 &  &  &  &  &  &  &  & \tabularnewline
\texttt{biz.stars}  &  & +1  &  & topp (0.589)  &  & topp (0.582)  &  & topp (0.611)\tabularnewline
 &  & +2  &  & excellent (0.557)  &  & excellent (0.564)  &  & awesome (0.583)\tabularnewline
 &  & +3  &  & awesome (0.521)  &  & awesome (0.528)  &  & excellent (0.547)\tabularnewline
 &  & +4  &  & beautiful (0.463)  &  & beautiful (0.481)  &  & beautiful (0.502)\tabularnewline
 &  & +5  &  & coffee (0.418)  &  & coffee (0.432)  &  & coffee (0.467)\tabularnewline
 &  & --1  &  & disrespect (-0.454)  &  & disrespect (-0.456)  &  & disrespect (-0.466)\tabularnewline
 &  & --2  &  & manag (-0.438)  &  & manag (-0.436)  &  & manag (-0.441)\tabularnewline
 &  & --3  &  & told (-0.421)  &  & told (-0.427)  &  & taco (-0.418)\tabularnewline
 &  & --4  &  & taco (-0.407)  &  & taco (-0.409)  &  & told (-0.396)\tabularnewline
 &  & --5  &  & wasn (-0.392)  &  & wasn (-0.386)  &  & sat (-0.381)\tabularnewline
 &  &  &  &  &  &  &  & \tabularnewline
\texttt{usr.cool}  &  & +1  &  & shrimp (2.541)  &  & shrimp (2.532)  &  & shrimp (2.423)\tabularnewline
 &  & +2  &  & choice (2.317)  &  & choice (2.319)  &  & choice (2.312)\tabularnewline
 &  & +3  &  & tea (2.084)  &  & tea (2.065)  &  & mov (2.159)\tabularnewline
 &  & +4  &  & mov (1.956)  &  & mov (1.940)  &  & dessert (2.072)\tabularnewline
 &  & +5  &  & mexican (1.743)  &  & mexican (1.721)  &  & tea (1.841)\tabularnewline
 &  & --1  &  & anyway (-1.922)  &  & anyway (-1.905)  &  & anyway (-1.539)\tabularnewline
 &  & --2  &  & far (-1.884)  &  & far (-1.892)  &  & far (-1.352)\tabularnewline
 &  & --3  &  & home (-1.705)  &  & home (-1.734)  &  & fantastic (-1.297)\tabularnewline
 &  & --4  &  & tell (-1.483)  &  & tell (-1.467)  &  & isn (-1.256)\tabularnewline
 &  & --5  &  & isn (-1.231)  &  & isn (-1.240)  &  & everyone (-1.202)\tabularnewline
\bottomrule
\end{tabular}\caption{Estimated coefficients comparison}
\label{tab: Covariate-Estimation} 
\end{table}

Table \ref{tab: Covariate-Estimation} reports, for selected covariates
\texttt{stars}, \texttt{biz.stars}, and \texttt{usr.cool}, the five
largest positive and five largest negative coefficients in magnitude.
The numbers in parentheses are the estimated coefficient values. For
each covariate, ranks $+1,\ldots,+5$ correspond to the words most
positively associated with the covariate, and ranks $-1,\ldots,-5$
correspond to the words most negatively associated.

Table \ref{tab: Covariate-Estimation} shows that the IDC estimators
with $S=10$ and $S=20$ are almost identical, suggesting that the
algorithm stabilizes after a small number of iterations. The IDC estimator
and Taddy’s $\widehat{\boldsymbol{\theta}}_{T}$ produce broadly similar
sets of top and bottom words, indicating comparable directional patterns.
However, the coefficient magnitudes differ. This result aligns with
the fact that IDC is a consistent estimator, whereas Taddy’s approach
is not. As a result, the latter may yield distorted effect sizes even
when rankings appear similar.

\section{Conclusion }

\label{sec:Conclusion}

In this paper, we propose an iterative distributed computing estimator
for the multinomial logistic model that is fast to compute even when
the number of choices is large. When the number of iterations goes
to infinity, we show that our estimator is both consistent and asymptotically
efficient. Both the theoretical and simulation result show that the
computational time of our estimator increases linearly with the number
of choices. Moreover, our estimator has comparable finite sample performance
to MLE when the latter is computationally feasible.

Extensions abound. First, our IDC estimator can be combined with several
existing algorithms to accommodate more complex settings. For example,
when minimizing $Q_{kn}\left(\boldsymbol{\theta}_{k},\overline{\boldsymbol{\mu}}_{n}\left(\widehat{\boldsymbol{\theta}}^{\left(s-1\right)}\right)\right)$
for $k=1,\ldots,d$, we can replace the gradient of the objective
function with its stochastic approximation calculated from a randomly
selected subset of the data. Such an algorithm is an online algorithm
and might reduce the running time especially when $n$ is large. We
can also employ a one-step Newton-Raphson to compute $\arg\min_{\boldsymbol{\theta}_{k}\in\varTheta_{k}}Q_{kn}\left(\boldsymbol{\theta}_{k},\overline{\boldsymbol{\mu}}_{n}\left(\widehat{\boldsymbol{\theta}}^{\left(s-1\right)}\right)\right)$
in each iteration or even replace the Hessian matrix with its dominant.
These modifications to the IDC estimator have the potential to further
enhance computational efficiency, depending on the application. However,
their theoretical properties require further investigation. Second,
the asymptotic properties of the IDC estimator in this paper are established
for a large but fixed number of choices. Asymptotic theory allowing
for the number of choices to diverge with the sample size is yet to
be established.

\newpage{} \begin{appendices}

\section{Notations and Equalities }

\label{sec:Notations-and-Equalities}

In this appendix, we list the mathematical expressions and equalities
used in the paper. 
\begin{enumerate}
\item $\mathrm{MNL}\left(\boldsymbol{C}_{i};\boldsymbol{\eta}_{i},M_{i}\right)\equiv\frac{M_{i}!}{C_{i,1}!\cdots C_{i,d}!}\left(\frac{e^{\eta_{i,1}}}{\Lambda_{i}}\right)^{C_{i,1}}\ldots\left(\frac{e^{\eta_{i,d}}}{\Lambda_{i}}\right)^{C_{i,d}}$ 
\item $l_{C\mid X,M}\left(\boldsymbol{\theta}\right)\equiv\sum^{n}_{i=1}\left[\boldsymbol{C}^{\prime}_{i}\boldsymbol{\eta}_{i}-M_{i}\log\left(\sum^{d}_{k=1}e^{\eta_{i,k}}\right)\right]$ 
\item $L_{C\mid X,M}\left(\boldsymbol{\theta}\right)\equiv p\lim_{n\rightarrow\infty}\frac{1}{n}l_{C\mid X,M}\left(\boldsymbol{\theta}\right)$ 
\item $Q^{\ast}_{n}\left(\boldsymbol{\theta}\right)\equiv-\frac{1}{n}l_{C\mid X,M}\left(\boldsymbol{\theta}\right)$ 
\item $\widetilde{\boldsymbol{\theta}}=\arg\min_{\boldsymbol{\theta}\in\Theta}Q^{\ast}_{n}\left(\boldsymbol{\theta}\right)$ 
\item $l_{C\mid X,M}\left(\boldsymbol{\theta},\boldsymbol{\mu}\right)\equiv\sum^{n}_{i=1}\left[\boldsymbol{C}^{\prime}_{i}\left(\boldsymbol{\eta}_{i}+\mu_{i}\boldsymbol{1}_{d}\right)-M_{i}\log\left(\sum^{d}_{k=1}e^{\eta_{i,k}+\mu_{i}}\right)\right]=l_{C\mid X,M}\left(\boldsymbol{\theta}\right)$ 
\item $f\left(\boldsymbol{\theta},\boldsymbol{\mu}\right)\equiv\sum^{n}_{i=1}\left[M_{i}\log\left(\sum^{d}_{k=1}e^{\eta_{i,k}}\right)+M_{i}\mu_{i}-e^{\mu_{i}}\sum^{d}_{k=1}e^{\eta_{i,k}}\right]$ 
\item $ql_{C\mid X}\left(\boldsymbol{\theta},\boldsymbol{\mu}\right)\equiv l_{C\mid X,M}\left(\boldsymbol{\theta},\boldsymbol{\mu}\right)+f\left(\boldsymbol{\theta},\boldsymbol{\mu}\right)=\sum^{n}_{i=1}\sum^{d}_{k=1}\left(C_{i,k}\left(\eta_{i,k}+\mu_{i}\right)-e^{\left(\eta_{i,k}+\mu_{i}\right)}\right)$ 
\item $Q_{n}\left(\boldsymbol{\theta},\boldsymbol{\mu}\right)\equiv-\frac{1}{n}ql_{C\mid X}\left(\boldsymbol{\theta},\boldsymbol{\mu}\right)$ 
\item $\left(\widehat{\boldsymbol{\theta}},\widehat{\boldsymbol{\mu}}\right)=\arg\min_{\boldsymbol{\theta}\in\Theta,\boldsymbol{\mu}\in\mathbb{R}^{n}}Q_{n}\left(\boldsymbol{\theta},\boldsymbol{\mu}\right),$ 
\item $Q_{k,n}\left(\boldsymbol{\theta}_{k},\boldsymbol{\mu}\right)\equiv\frac{1}{n}\sum^{n}_{i=1}\left(e^{\left(\boldsymbol{X}^{\prime}_{i}\boldsymbol{\theta}_{k}+\mu_{i}\right)}-C_{i,k}\left(\boldsymbol{X}^{\prime}_{i}\boldsymbol{\theta}_{k}+\mu_{i}\right)\right)$ 
\item $\overline{\boldsymbol{\mu}}_{n}\left(\boldsymbol{\theta}\right)\equiv\left(\log\left(\frac{M_{1}}{\sum^{d}_{k=1}e^{\eta_{1,k}}}\right),\ldots,\log\left(\frac{M_{n}}{\sum^{d}_{k=1}e^{\eta_{n,k}}}\right)\right)^{\prime}$ 
\item $\widehat{\boldsymbol{\theta}}^{\left(s\right)}=\left[\arg\min_{\boldsymbol{\theta}_{1}\in\varTheta_{1}}Q_{1n}\left(\boldsymbol{\theta}_{1},\overline{\boldsymbol{\mu}}_{n}\left(\widehat{\boldsymbol{\theta}}^{\left(s-1\right)}\right)\right),\ldots,\arg\min_{\boldsymbol{\theta}_{d}\in\varTheta_{d}}Q_{d,n}\left(\boldsymbol{\theta}_{d},\overline{\boldsymbol{\mu}}_{n}\left(\widehat{\boldsymbol{\theta}}^{\left(s-1\right)}\right)\right)\right]^{\prime}$ 
\item $l_{C_{k},C_{d}\mid X,N_{k}}\left(\boldsymbol{\theta}_{k}\right)\equiv\sum^{n}_{i=1}\left[C_{i,k}\boldsymbol{X}^{\prime}_{i}\boldsymbol{\theta}_{k}-\left(C_{i,k}+C_{i,d}\right)\log\left(e^{\boldsymbol{X}^{\prime}_{i}\boldsymbol{\theta}_{k}}+1\right)\right]$ 
\item $\check{\boldsymbol{\theta}}_{k}=\arg\min_{\boldsymbol{\theta}_{k}\in\varTheta_{k}}-l_{C_{k},C_{d}\mid X,N_{k}}\left(\boldsymbol{\theta}_{k}\right)$ 
\item $\widehat{\boldsymbol{\theta}}_{T}=\arg\min_{\boldsymbol{\theta}\in\Theta}Q_{n}\left(\boldsymbol{\theta},\widehat{\boldsymbol{\mu}}_{T}\right)$
with $\widehat{\boldsymbol{\mu}}_{T}=\left(\log\left(M_{1}\right),\ldots,\log\left(M_{n}\right)\right)^{\prime}$ 
\item $\widehat{\boldsymbol{\theta}}_{P}=\arg\min_{\boldsymbol{\theta}\in\Theta}Q_{n}\left(\boldsymbol{\theta},\mathbf{0}\right)$ 
\item $Q^{\dagger}_{n}\left(\boldsymbol{\theta},\boldsymbol{\vartheta}\right)\equiv Q_{n}\left(\boldsymbol{\theta},\overline{\boldsymbol{\mu}}_{n}\left(\boldsymbol{\vartheta}\right)\right)=\frac{1}{n}\sum^{n}_{i=1}\sum^{d}_{k=1}\left(\frac{M_{i}e^{\boldsymbol{X}^{\prime}_{i}\boldsymbol{\theta}_{k}}}{\sum^{d}_{l=1}e^{\boldsymbol{X}^{\prime}_{i}\boldsymbol{\vartheta}_{l}}}-C_{i,k}\boldsymbol{X}^{\prime}_{i}\boldsymbol{\theta}_{k}-C_{i,k}\log\left(\frac{M_{i}}{\sum^{d}_{l=1}e^{\boldsymbol{X}^{\prime}_{i}\boldsymbol{\vartheta}_{l}}}\right)\right)$ 
\item $Q^{\dagger}\left(\boldsymbol{\theta},\boldsymbol{\vartheta}\right)\equiv\sum^{d}_{k=1}\mathbb{E}\left[\left(\frac{Me^{\boldsymbol{X}^{\prime}\boldsymbol{\theta}_{k}}}{\sum^{d}_{l=1}e^{\boldsymbol{X}^{\prime}\boldsymbol{\vartheta}_{l}}}-C_{k}\boldsymbol{X}^{\prime}\boldsymbol{\theta}_{k}-C_{k}\log\left(\frac{M}{\sum^{d}_{l=1}e^{\boldsymbol{X}^{\prime}\boldsymbol{\vartheta}_{l}}}\right)\right)\right]$ 
\item $Q^{\dagger}_{k}\left(\boldsymbol{\theta}_{k},\boldsymbol{\vartheta}\right)\equiv\mathbb{E}\left[\frac{Me^{\boldsymbol{X}^{\prime}\boldsymbol{\theta}_{k}}}{\sum^{d}_{k=1}e^{\boldsymbol{X}^{\prime}\boldsymbol{\vartheta}_{k}}}-C_{k}\boldsymbol{X}^{\prime}\boldsymbol{\theta}_{k}-C_{k}\log\left(\frac{M}{\sum^{d}_{k=1}e^{\boldsymbol{X}^{\prime}\boldsymbol{\vartheta}_{k}}}\right)\right]$ 
\item $\overline{\boldsymbol{\theta}}_{n}\left(\boldsymbol{\vartheta}\right)\equiv\arg\min_{\boldsymbol{\theta}\in\Theta}Q^{\dagger}_{n}\left(\boldsymbol{\theta},\boldsymbol{\vartheta}\right)$ 
\item $\overline{\boldsymbol{\theta}}\left(\boldsymbol{\vartheta}\right)\equiv\arg\min_{\boldsymbol{\theta}\in\Theta}Q^{\dagger}\left(\boldsymbol{\theta},\boldsymbol{\vartheta}\right)$ 
\item $Q^{\ast}\left(\boldsymbol{\theta}^{\ast}\right)\equiv p\lim_{n\rightarrow\infty}Q^{\ast}_{n}\left(\boldsymbol{\theta}^{\ast}\right)$ 
\item $F_{n}\left(\boldsymbol{\theta}\right)\equiv Q^{\dagger}_{n}\left(\boldsymbol{\theta},\boldsymbol{\theta}\right)\stackrel{p}{\rightarrow}$
$Q^{\dagger}\left(\boldsymbol{\theta},\boldsymbol{\theta}\right)\equiv F\left(\boldsymbol{\theta}\right)$ 
\item $\mathcal{I}\left(\boldsymbol{\theta}^{\ast}\right)\equiv\frac{\partial^{2}}{\partial\boldsymbol{\theta}\partial\boldsymbol{\theta}^{\prime}}Q^{\ast}\left(\boldsymbol{\theta}^{\ast}\right)$ 
\item $\frac{\partial^{2}Q^{\dagger}\left(\boldsymbol{\theta},\boldsymbol{\vartheta}\right)}{\partial\boldsymbol{\theta}\partial\boldsymbol{\theta}^{\prime}}\equiv\left[\begin{array}{ccc}
\mathbb{E}\left[\frac{Me^{\boldsymbol{X}^{\prime}\boldsymbol{\theta}_{1}}}{\sum^{d}_{l=1}e^{\boldsymbol{X}^{\prime}\boldsymbol{\vartheta}_{l}}}\boldsymbol{X}\boldsymbol{X}^{\prime}\right] & \cdots & \mathbf{0}\\
\vdots & \ddots & \vdots\\
\mathbf{0} & \cdots & \mathbb{E}\left[\frac{Me^{\boldsymbol{X}^{\prime}\boldsymbol{\theta}_{d-1}}}{\sum^{d}_{l=1}e^{\boldsymbol{X}^{\prime}\boldsymbol{\vartheta}_{l}}}\boldsymbol{X}\boldsymbol{X}^{\prime}\right]
\end{array}\right]$ 
\item $\frac{\partial^{2}Q^{\dagger}\left(\boldsymbol{\theta},\boldsymbol{\vartheta}\right)}{\partial\boldsymbol{\theta}\partial\boldsymbol{\vartheta}^{\prime}}\equiv\left[\begin{array}{ccc}
\mathbb{E}\left[-\frac{Me^{\boldsymbol{X}^{\prime}\left(\boldsymbol{\theta}_{1}+\boldsymbol{\vartheta}_{1}\right)}}{\left(\sum^{d}_{l=1}e^{\boldsymbol{X}^{\prime}\boldsymbol{\vartheta}_{l}}\right)^{2}}\boldsymbol{X}\boldsymbol{X}^{\prime}\right] & \cdots & \mathbb{E}\left[-\frac{Me^{\boldsymbol{X}^{\prime}\left(\boldsymbol{\theta}_{1}+\boldsymbol{\vartheta}_{d-1}\right)}}{\left(\sum^{d}_{l=1}e^{\boldsymbol{X}^{\prime}\boldsymbol{\vartheta}_{l}}\right)^{2}}\boldsymbol{X}\boldsymbol{X}^{\prime}\right]\\
\vdots & \ddots & \vdots\\
\mathbb{E}\left[-\frac{Me^{\boldsymbol{X}^{\prime}\left(\boldsymbol{\theta}_{d-1}+\boldsymbol{\vartheta}_{1}\right)}}{\left(\sum^{d}_{l=1}e^{\boldsymbol{X}^{\prime}\boldsymbol{\vartheta}_{l}}\right)^{2}}\boldsymbol{X}\boldsymbol{X}^{\prime}\right] & \cdots & \mathbb{E}\left[-\frac{Me^{\boldsymbol{X}^{\prime}\left(\boldsymbol{\theta}_{d-1}+\boldsymbol{\vartheta}_{d-1}\right)}}{\left(\sum^{d}_{l=1}e^{\boldsymbol{X}^{\prime}\boldsymbol{\vartheta}_{l}}\right)^{2}}\boldsymbol{X}\boldsymbol{X}^{\prime}\right]
\end{array}\right]$ 
\item $\dot{\mathbf{l}}\left(\boldsymbol{\theta}\mid\boldsymbol{C}_{i},\boldsymbol{X}_{i},M_{i}\right)\equiv\frac{d}{d\boldsymbol{\theta}}\log\mathrm{MNL}\left(\boldsymbol{C}_{i};\boldsymbol{\eta}_{i},M_{i}\right)$ 
\item $\widetilde{\mathbf{l}}\left(\boldsymbol{\theta}\mid\boldsymbol{C}_{i},\boldsymbol{X}_{i},M_{i}\right)\equiv\mathcal{I}^{-1}\left(\boldsymbol{\theta}\right)\dot{\mathbf{l}}\left(\boldsymbol{\theta}\mid\boldsymbol{C}_{i},\boldsymbol{X}_{i},M_{i}\right)$ 
\end{enumerate}
\newpage{}

\section{Proofs }

\label{subsec:Proofs}

\paragraph{Proof of Lemma \protect\ref{lem:: Adding mu}:}

It holds that 
\begin{eqnarray*}
l_{C\mid X,M}\left(\boldsymbol{\theta},\boldsymbol{\mu}\right) & = & \sum^{n}_{i=1}\left[\boldsymbol{C}^{\prime}_{i}\boldsymbol{\eta}_{i}+\mu_{i}\boldsymbol{C}^{\prime}_{i}\boldsymbol{1}_{d}-M_{i}\log\left(e^{\mu_{i}}\sum^{d}_{k=1}e^{\eta_{i,k}}\right)\right]\\
 & = & \sum^{n}_{i=1}\left[\boldsymbol{C}^{\prime}_{i}\boldsymbol{\eta}_{i}+\mu_{i}\sum^{d}_{k=1}C_{ik}-M_{i}\log\left(\sum^{d}_{k=1}e^{\eta_{i,k}}\right)-M_{i}\mu_{i}\right]\\
 & = & \sum^{n}_{i=1}\left[\boldsymbol{C}^{\prime}_{i}\boldsymbol{\eta}_{i}-M_{i}\log\left(\sum^{d}_{k=1}e^{\eta_{i,k}}\right)\right]\\
 & = & l_{C\mid X,M}\left(\boldsymbol{\theta}\right),
\end{eqnarray*}
where the second to last equality holds because $\mu_{i}\sum^{d}_{k=1}C_{i,k}=M_{i}\mu_{i}$.
Therefore, adding $\mu_{i}$ does not change the likelihood.\qed

\paragraph{Proof of Lemma \protect\ref{lem:: Equivalence conditional and unconditional}:}

Notice that $Q_{n}\left(\boldsymbol{\theta},\boldsymbol{\mu}\right)$
is differentiable w.r.t. $\boldsymbol{\mu}$ for any given $\boldsymbol{\theta}$.
By letting $\frac{\partial Q_{n}\left(\boldsymbol{\theta},\boldsymbol{\mu}\right)}{\partial\boldsymbol{\mu}}=\boldsymbol{0}$,
we can obtain function $\overline{\boldsymbol{\mu}}_{n}\left(\boldsymbol{\theta}\right)$
such that $\frac{\partial Q_{n}\left(\boldsymbol{\theta},\boldsymbol{\mu}\right)}{\partial\boldsymbol{\mu}}\bigg|_{\boldsymbol{\mu}=\overline{\boldsymbol{\mu}}_{n}\left(\boldsymbol{\theta}\right)}=\boldsymbol{0}$
for every $\boldsymbol{\theta}$. By definition, 
\[
Q_{n}\left(\boldsymbol{\theta},\boldsymbol{\mu}\right)=-\frac{1}{n}\left[l_{C\mid X,M}\left(\boldsymbol{\theta},\boldsymbol{\mu}\right)+f\left(\boldsymbol{\theta},\boldsymbol{\mu}\right)\right]=-\frac{1}{n}\left[l_{C\mid X,M}\left(\boldsymbol{\theta}\right)+f\left(\boldsymbol{\theta},\boldsymbol{\mu}\right)\right],
\]
which implies that $\frac{\partial Q_{n}\left(\boldsymbol{\theta},\boldsymbol{\mu}\right)}{\partial\boldsymbol{\mu}}=-\frac{\partial f\left(\boldsymbol{\theta},\boldsymbol{\mu}\right)}{\partial\boldsymbol{\mu}}$.
Since $\frac{\partial f\left(\boldsymbol{\theta},\boldsymbol{\mu}\right)}{\partial\mu_{i}}=M_{i}-e^{\mu_{i}}\sum^{d}_{k=1}e^{\eta_{i,k}}$,
we obtain the expression of $\overline{\boldsymbol{\mu}}_{n}\left(\boldsymbol{\theta}\right)$
as in (\ref{eq:mu bar}). Plugging it into (\ref{eq:log likelihood l_M|X}),
we have that 
\begin{eqnarray}
 &  & f\left(\boldsymbol{\theta},\overline{\boldsymbol{\mu}}_{n}\left(\boldsymbol{\theta}\right)\right)\nonumber \\
 & = & \sum^{n}_{i=1}\left[M_{i}\log\left(\sum^{d}_{k=1}e^{\eta_{i,k}}\right)+M_{i}\log\left(\frac{m_{i}}{\sum^{d}_{k=1}e^{\eta_{i,k}}}\right)-e^{\log\left(\frac{M_{i}}{\sum^{d}_{k=1}e^{\eta_{i,k}}}\right)}\sum^{d}_{k=1}e^{\eta_{i,k}}\right]\nonumber \\
 & = & \sum^{n}_{i=1}\left[M_{i}\log\left(\sum^{d}_{k=1}e^{\eta_{i,k}}\right)+M_{i}\log M_{i}-M_{i}\log\left(\sum^{d}_{k=1}e^{\eta_{i,k}}\right)-\left(\frac{M_{i}}{\sum^{d}_{k=1}e^{\eta_{i,k}}}\right)\sum^{d}_{k=1}e^{\eta_{i,k}}\right]\nonumber \\
 & = & \sum^{n}_{i=1}\left[M_{i}\log M_{i}-M_{i}\right],\label{eq:l m|v independent of theta}
\end{eqnarray}
which does not depend on $\boldsymbol{\theta}$. As a result, it holds
that 
\[
\arg\min_{\boldsymbol{\theta}\in\Theta}Q_{n}\left(\boldsymbol{\theta},\overline{\boldsymbol{\mu}}_{n}\left(\boldsymbol{\theta}\right)\right)\equiv\arg\min_{\boldsymbol{\theta}\in\Theta}\left[Q^{\ast}_{n}\left(\boldsymbol{\theta}\right)-\frac{1}{n}f\left(\boldsymbol{\theta},\overline{\boldsymbol{\mu}}_{n}\left(\boldsymbol{\theta}\right)\right)\right]=\arg\min_{\boldsymbol{\theta}\in\Theta}Q^{\ast}_{n}\left(\boldsymbol{\theta}\right).
\]
Because $\widehat{\boldsymbol{\theta}}=\arg\min_{\boldsymbol{\theta}\in\Theta}Q_{n}\left(\boldsymbol{\theta},\overline{\boldsymbol{\mu}}\left(\boldsymbol{\theta}\right)\right)$,
we have that 
\[
\widehat{\boldsymbol{\theta}}=\arg\min_{\boldsymbol{\theta}\in\Theta}Q^{\ast}_{n}\left(\boldsymbol{\theta}\right)=\widetilde{\boldsymbol{\theta}}.
\]
The claimed lemma then follows.\qed

\paragraph{Proof of Lemma \protect\ref{lem:: Ck Cd distribution}:}

By Equation (\ref{eq: MNL model}), it holds that 
\begin{align}
 & \Pr\left(C_{i,k},C_{i,d}\mid\boldsymbol{X}_{i},M_{i}\right)\nonumber \\
= & \frac{M_{i}!}{C_{i,k}!C_{i,d}!\left(M_{i}-C_{i,k}-C_{i,d}\right)!}\left(\frac{e^{\eta^{\ast}_{i,k}}}{\Lambda^{\ast}_{i}}\right)^{C_{i,k}}\left(\frac{e^{\eta^{\ast}_{i,d}}}{\Lambda^{\ast}_{i}}\right)^{C_{i,d}}\left(\frac{\Lambda^{\ast}_{i}-e^{\eta^{\ast}_{i,k}}-e^{\eta^{\ast}_{i,d}}}{\Lambda^{\ast}_{i}}\right)^{M_{i}-C_{i,k}-C_{i,d}}.\label{eq: Ck Cd probability}
\end{align}
Because $N_{i,k}=C_{i,k}+C_{i,d}$, we have that 
\begin{align}
\Pr\left(C_{i,k},C_{i,d}\mid\boldsymbol{X}_{i},M_{i}\right) & =\Pr\left(C_{i,k},C_{i,d},N_{i,k}\mid\boldsymbol{X}_{i},M_{i}\right)\nonumber \\
 & =\Pr\left(C_{i,k},C_{i,d}\mid N_{i,k},\boldsymbol{X}_{i},M_{i}\right)\Pr\left(N_{i,k}\mid\boldsymbol{X}_{i},M_{i}\right).\label{eq: Ck Cd conditional}
\end{align}
Compute $\Pr\left(N_{i,k}\mid\boldsymbol{X}_{i},M_{i}\right)$ as
\[
\Pr\left(N_{i,k}\mid\boldsymbol{X}_{i},M_{i}\right)=\frac{M_{i}!}{N_{i,k}!\left(M_{i}-N_{i,k}\right)!}\left(\frac{e^{\eta^{\ast}_{i,k}}+e^{\eta^{\ast}_{i,d}}}{\Lambda^{\ast}_{i}}\right)^{N_{ik}}\left(\frac{\Lambda^{\ast}_{i}-e^{\eta^{\ast}_{i,k}}-e^{\eta^{\ast}_{i,d}}}{\Lambda^{\ast}_{i}}\right)^{M_{i}-N_{i,k}}.
\]
Together with Equations (\ref{eq: Ck Cd probability}) and (\ref{eq: Ck Cd conditional}),
we obtain that 
\begin{align*}
\Pr\left(C_{i,k},C_{i,d}\mid N_{i,k},\boldsymbol{X}_{i},M_{i}\right) & =\frac{N_{i,k}!}{C_{i,k}!C_{i,d}!}\left(\frac{e^{\eta^{\ast}_{i,k}}}{e^{\eta^{\ast}_{i,k}}+e^{\eta^{\ast}_{i,d}}}\right)^{C_{i,k}}\left(\frac{e^{\eta^{\ast}_{i,d}}}{e^{\eta^{\ast}_{i,k}}+e^{\eta^{\ast}_{i,d}}}\right)^{C_{i,d}}\\
 & =\Pr\left(C_{ik},C_{id}\mid N_{ik},\boldsymbol{X}_{i}\right).
\end{align*}
By replacing $e^{\eta^{\ast}_{i,d}}$ with 1 because $\boldsymbol{\theta}^{\ast}_{d}=\mathbf{0}$,
we obtain the claimed result. \qed

\paragraph{Proof of Lemma \protect\ref{lem:: consistency under stronger assumption}:}

Under the assumption that $\Pr\left(M_{i}\mid\boldsymbol{X}_{i}\right)=\mathrm{Po}\left(\sum^{d}_{k=1}e^{\eta^{\ast}_{i,k}}\right),$
we can obtain that 
\begin{align*}
\Pr\left(\boldsymbol{C}_{i}\mid\boldsymbol{X}_{i}\right) & =\Pr\left(\boldsymbol{C}_{i}\mid\boldsymbol{X}_{i},M_{i}\right)\mathrm{Po}\left(\sum^{d}_{k=1}e^{\eta^{\ast}_{i,k}}\right)\\
 & =\prod^{d}_{k=1}\mathrm{Po}\left(e^{\eta^{\ast}_{i,k}}\right)=\prod^{d}_{k=1}\frac{e^{\eta^{\ast}_{i,k}}{}^{C_{i,k}}e^{-e^{\eta^{\ast}_{i,k}}}}{C_{i,k}!}=\frac{\prod^{d}_{k=1}e^{\eta^{\ast}_{i,k}}{}^{C_{i,k}}}{C_{i,1}!\cdots C_{i,d}!}e^{-\sum^{d}_{k=1}e^{\eta^{\ast}_{i,k}}}.
\end{align*}
The log-likelihood function is written as 
\begin{eqnarray*}
\log\left[\prod^{n}_{i=1}\frac{\prod^{d}_{k=1}e^{\eta_{i,k}}{}^{C_{i,k}}}{C_{i,1}!\cdots C_{i,d}!}e^{-\sum^{d}_{k=1}e^{\eta_{i,k}}}\right] & \sim & \sum^{n}_{i=1}\left[\sum^{d}_{k=1}C_{i,k}\eta_{i,k}-\sum^{d}_{k=1}e^{\eta_{i,k}}\right]=\sum^{n}_{i=1}\sum^{d}_{k=1}\left(C_{i,k}\eta_{i,k}-e^{\eta_{i,k}}\right)\\
 & = & ql_{C\mid X}\left(\boldsymbol{\theta},\mathbf{0}\right).
\end{eqnarray*}
Therefore, $\widehat{\boldsymbol{\theta}}_{P}$ maximizes the true
log-likelihood function based upon $\Pr\left(\boldsymbol{C}_{i}\mid\boldsymbol{X}_{i}\right)$.
The lemma follows. \qed

\begin{lemma} \label{lem:: convexity of L} Under Assumption \ref{Assm:: consistency 1},
$-L_{C\mid X,M}\left(\boldsymbol{\theta}\right)$ is convex in $\boldsymbol{\theta}$.
\end{lemma}

\paragraph{Proof of Lemma \ref{lem:: convexity of L}:}

By the definitions of $l_{C\mid X,M}\left(\boldsymbol{\theta}\right)$
and $L_{C\mid X,M}\left(\boldsymbol{\theta}\right)$ and Assumption
\ref{Assm:: consistency 1}, we have that 
\begin{eqnarray*}
-\frac{1}{n}l_{C\mid X,M}\left(\boldsymbol{\theta}\right) & = & -\frac{1}{n}\sum^{n}_{i=1}\left[C_{i,1}\eta_{i1}+\cdots+C_{i,d}\eta_{i,d}-M_{i}\log\left(\sum^{d}_{k=1}e^{\eta_{i,k}}\right)\right]\\
 & = & -\frac{1}{n}\sum^{n}_{i=1}\left[C_{i,1}\boldsymbol{X}^{\prime}_{i}\boldsymbol{\theta}_{1}+\cdots+C_{i,d}\boldsymbol{X}^{\prime}_{i}\boldsymbol{\theta}_{d}-M_{i}\log\left(\sum^{d}_{k=1}e^{\boldsymbol{X}^{\prime}_{i}\boldsymbol{\theta}_{k}}\right)\right]\\
 & \stackrel{p}{\rightarrow} & -\left(\mathbb{E}\left[C_{1}\boldsymbol{X}^{\prime}\boldsymbol{\theta}_{1}\right]+\cdots+\mathbb{E}\left[C_{d}\boldsymbol{X}^{\prime}\boldsymbol{\theta}_{d}\right]-\mathbb{E}\left[M\log\left(\sum^{d}_{k=1}e^{\boldsymbol{X}^{\prime}\boldsymbol{\theta}_{k}}\right)\right]\right)\\
 & \equiv & -L_{C\mid X,M}\left(\boldsymbol{\theta}\right)\\
 & = & -\sum^{d}_{k=1}\mathbb{E}\left[C_{k}\boldsymbol{X}^{\prime}\boldsymbol{\theta}_{k}\right]+\mathbb{E}\left[M\log\left(\sum^{d}_{k=1}e^{\boldsymbol{X}^{\prime}\boldsymbol{\theta}_{k}}\right)\right].
\end{eqnarray*}
The first term is convex in $\boldsymbol{\theta}$ because it only
involves linear functions. It has been shown in Section 3.1.5 in \citet{boyd2004convex}
that $\log\left(\sum^{d}_{k=1}e^{\boldsymbol{X}^{\prime}\boldsymbol{\theta}_{k}}\right)$
is convex in $\boldsymbol{X}^{\prime}\boldsymbol{\theta}_{1},\ldots,\boldsymbol{X}^{\prime}\boldsymbol{\theta}_{d}$.
Because $\boldsymbol{X}^{\prime}\boldsymbol{\theta}_{k}$ is a linear
function of $\theta_{k}$ and sums of convex functions are convex,
the second term is convex in $\boldsymbol{\theta}$. \qed

\begin{lemma} \label{lem:: unique theta star for Q infinity} Under
Assumptions \ref{Assmp:: Maintained assumption} and \ref{Assm:: consistency 1},
$\boldsymbol{\theta}^{\ast}$ is the unique fixed point of $\overline{\boldsymbol{\theta}}\left(\boldsymbol{\theta}\right)$.
\end{lemma}

\paragraph{Proof of Lemma \ref{lem:: unique theta star for Q infinity}:}

By definition, $\overline{\boldsymbol{\theta}}\left(\boldsymbol{\vartheta}\right)\equiv\arg\min_{\boldsymbol{\theta}\in\Theta}Q^{\dagger}\left(\boldsymbol{\theta},\boldsymbol{\vartheta}\right)$.
Thus, if we can show that $\frac{\partial Q^{\dagger}\left(\boldsymbol{\theta},\boldsymbol{\vartheta}\right)}{\partial\boldsymbol{\theta}}\mid_{\boldsymbol{\vartheta}=\boldsymbol{\theta}}=\mathbf{0}$
holds only at $\boldsymbol{\theta}=\boldsymbol{\theta}^{\ast}$, then
$\boldsymbol{\theta}^{\ast}$ is the unique fixed point of $\overline{\boldsymbol{\theta}}\left(\cdot\right)$.
Taking the first order derivative of $-L_{C\mid X,M}\left(\cdot\right)$,
we obtain that for any $\dot{\boldsymbol{\theta}}\in\Theta$, 
\begin{align*}
-\frac{d}{d\boldsymbol{\theta}}L_{C\mid X,M}\left(\boldsymbol{\theta}\right)\mid_{\boldsymbol{\theta}=\dot{\boldsymbol{\theta}}}= & \left[\mathbb{E}\left[\frac{Me^{\boldsymbol{X}^{\prime}\dot{\boldsymbol{\theta}}_{1}}\boldsymbol{X}^{\prime}}{\sum^{d}_{k=1}e^{\boldsymbol{X}^{\prime}\dot{\boldsymbol{\theta}}_{k}}}-C_{1}\boldsymbol{X}^{\prime}\right],\ldots,\mathbb{E}\left[\frac{Me^{\boldsymbol{X}^{\prime}\boldsymbol{\dot{\theta}}_{d-1}}\boldsymbol{X}^{\prime}}{\sum^{d}_{k=1}e^{\boldsymbol{X}^{\prime}\dot{\boldsymbol{\theta}}_{k}}}-C_{d-1}\boldsymbol{X}^{\prime}\right]\right]^{\prime}\\
= & \frac{\partial Q^{\dagger}\left(\boldsymbol{\theta},\boldsymbol{\vartheta}\right)}{\partial\boldsymbol{\theta}}\mid_{\left(\boldsymbol{\theta},\boldsymbol{\vartheta}\right)=\left(\dot{\boldsymbol{\theta}},\dot{\boldsymbol{\theta}}\right)}.
\end{align*}
Therefore, proving that $\frac{\partial Q^{\dagger}\left(\boldsymbol{\theta},\boldsymbol{\vartheta}\right)}{\partial\boldsymbol{\theta}}\mid_{\boldsymbol{\vartheta}=\boldsymbol{\theta}}=\mathbf{0}$
holds only at $\boldsymbol{\theta}=\boldsymbol{\theta}^{\ast}$ is
equivalent to showing that $\frac{d}{d\boldsymbol{\theta}}L_{C\mid X,M}\left(\boldsymbol{\theta}\right)=\mathbf{0}$
only at $\boldsymbol{\theta}^{\ast}$. By Lemma \ref{lem:: convexity of L},
$-L_{C\mid X,M}\left(\boldsymbol{\theta}\right)$ is convex in $\boldsymbol{\theta}$.
The set $\Theta$ is convex by Assumption \ref{Assm:: consistency 1}.
For a convex function over a convex set, any local minimum is also
a global minimum. Along with Assumption \ref{Assmp:: Maintained assumption}
that $\boldsymbol{\theta}^{\ast}$ is the unique solution to $\arg\min_{\boldsymbol{\theta}\in\Theta}-L_{C\mid X,M}\left(\boldsymbol{\theta}\right)$,
we obtain that $\frac{\partial}{\partial\boldsymbol{\theta}}L_{C\mid X,M}\left(\boldsymbol{\theta}\right)=\mathbf{0}$
holds only at $\boldsymbol{\theta}=\boldsymbol{\theta}^{\ast}$. \qed

\paragraph{Proof of Lemma \ref{lem:: properties of Qn Qstar}:}

Part (i) follows by applying Theorem 2 in \citet{jennrich1969asymptotic}
on the uniform law of large numbers with conditions satisfied by Assumption
\ref{Assm:: consistency 1}.

Since $Q^{\dagger}\left(\boldsymbol{\theta},\boldsymbol{\vartheta}\right)$
is continuous in $\boldsymbol{\theta}$ and $\Theta$ is compact by
Assumption \ref{Assm:: consistency 1} (i), $Q^{\dagger}\left(\boldsymbol{\theta},\boldsymbol{\vartheta}\right)$
achieves its minimum in $\Theta$ for any $\boldsymbol{\vartheta}$.
For each alternative $k=1,\ldots,d-1$, the corresponding block of
the Hessian matrix of $Q^{\dagger}\left(\boldsymbol{\theta},\boldsymbol{\vartheta}\right)$
for any given $\boldsymbol{\vartheta}$ is 
\[
H_{k}\left(\boldsymbol{\theta}_{k},\boldsymbol{\vartheta}\right)=\mathbb{E}\left[\frac{Me^{\boldsymbol{X}^{\prime}\boldsymbol{\theta}_{k}}}{\sum^{d}_{l=1}e^{\boldsymbol{X}^{\prime}\boldsymbol{\vartheta}_{l}}}\boldsymbol{X}\boldsymbol{X}^{\prime}\right].
\]
Assumption \ref{Assmp:: Maintained assumption} implies for every
nonzero vector $a$, we have $\Pr\left(\boldsymbol{X}^{\prime}a\neq0\right)>0$.
Otherwise, a sufficiently small perturbation of one coefficient vector
in the direction $a$ would leave all choice probabilities unchanged,
contradicting Assumption \ref{Assmp:: Maintained assumption}. As
a result, for any $a\neq\mathbf{0}$, we have 
\[
a^{\prime}H_{k}\left(\boldsymbol{\theta}_{k},\boldsymbol{\vartheta}\right)a=\mathbb{E}\left[\frac{Me^{\boldsymbol{X}^{\prime}\boldsymbol{\theta}_{k}}}{\sum^{d}_{l=1}e^{\boldsymbol{X}^{\prime}\boldsymbol{\vartheta}_{l}}}\left(\boldsymbol{X}^{\prime}a\right)^{2}\right]>0.
\]
Therefore, each term in the expression of $Q^{\dagger}\left(\boldsymbol{\theta},\boldsymbol{\vartheta}\right)$
is strictly convex, which implies that $Q^{\dagger}\left(\boldsymbol{\theta},\boldsymbol{\vartheta}\right)$
is strictly convex in $\boldsymbol{\theta}$ for any $\boldsymbol{\vartheta}$.
Because $\Theta$ is a convex set, we have that $Q^{\dagger}\left(\boldsymbol{\theta},\boldsymbol{\vartheta}\right)$
has a unique minimizer for any $\boldsymbol{\vartheta}$. Part (ii)
holds.

For part (iii), because $Q^{\dagger}\left(\boldsymbol{\theta},\boldsymbol{\vartheta}\right)$
is strictly convex in $\boldsymbol{\theta}$ for any $\boldsymbol{\vartheta}\in\Theta$
and $\Theta$ is convex, Berge’s maximum theorem implies the continuity.

To prove part (iv), we take the first-order partial derivative of
$Q^{\dagger}\left(\boldsymbol{\theta},\boldsymbol{\vartheta}\right)$
with respect to $\boldsymbol{\theta}$ and obtain that 
\begin{eqnarray*}
\frac{\partial Q^{\dagger}\left(\boldsymbol{\theta},\boldsymbol{\vartheta}\right)}{\partial\boldsymbol{\theta}} & = & \left[\mathbb{E}\left[\frac{Me^{\boldsymbol{X}^{\prime}\boldsymbol{\theta}_{1}}\boldsymbol{X}^{\prime}}{\sum^{d}_{k=1}e^{\boldsymbol{X}^{\prime}\boldsymbol{\vartheta}_{k}}}-C_{1}\boldsymbol{X}^{\prime}\right],\ldots,\mathbb{E}\left[\frac{Me^{\boldsymbol{X}^{\prime}\boldsymbol{\theta}_{d-1}}\boldsymbol{X}^{\prime}}{\sum^{d}_{k=1}e^{\boldsymbol{X}^{\prime}\boldsymbol{\vartheta}_{k}}}-C_{d-1}\boldsymbol{X}^{\prime}\right]\right]^{\prime}.
\end{eqnarray*}
Since $\mathbb{E}\left[C_{k}\mid\boldsymbol{X},M\right]=\frac{Me^{\boldsymbol{X}^{\prime}\boldsymbol{\theta}^{\ast}_{k}}}{\sum^{d}_{k=1}e^{\boldsymbol{X}^{\prime}\boldsymbol{\theta}^{\ast}_{k}}}$,
we have that 
\begin{align}
 & \frac{\partial Q^{\dagger}\left(\boldsymbol{\theta},\boldsymbol{\vartheta}\right)}{\partial\boldsymbol{\theta}}\mid_{\left(\boldsymbol{\theta},\boldsymbol{\vartheta}\right)=\left(\boldsymbol{\theta}^{\ast},\boldsymbol{\theta}^{\ast}\right)}\nonumber \\
 & =\left[\mathbb{E}\left[\mathbb{E}\left[\frac{Me^{\boldsymbol{X}^{\prime}\boldsymbol{\theta}^{\ast}_{1}}\boldsymbol{X}^{\prime}}{\sum^{d}_{k=1}e^{\boldsymbol{X}^{\prime}\boldsymbol{\theta}^{\ast}_{k}}}-C_{1}\boldsymbol{X}^{\prime}\mid\boldsymbol{X},M\right]\right],\ldots,\mathbb{E}\left[\mathbb{E}\left[\frac{Me^{\boldsymbol{X}^{\prime}\boldsymbol{\theta}^{\ast}_{d-1}}\boldsymbol{X}^{\prime}}{\sum^{d}_{k=1}e^{\boldsymbol{X}^{\prime}\boldsymbol{\theta}^{\ast}_{k}}}-C_{d-1}\boldsymbol{X}^{\prime}\mid\boldsymbol{X},M\right]\right]\right]^{\prime}\nonumber \\
 & =\left[\mathbb{E}\left[\frac{Me^{\boldsymbol{X}^{\prime}\boldsymbol{\theta}^{\ast}_{1}}\boldsymbol{X}^{\prime}}{\sum^{d}_{k=1}e^{\boldsymbol{X}^{\prime}\boldsymbol{\theta}^{\ast}_{k}}}-\frac{Me^{\boldsymbol{X}^{\prime}\boldsymbol{\theta}^{\ast}_{1}}\boldsymbol{X}^{\prime}}{\sum^{d}_{k=1}e^{\boldsymbol{X}^{\prime}\boldsymbol{\theta}^{\ast}_{k}}}\right],\ldots,\mathbb{E}\left[\frac{Me^{\boldsymbol{X}^{\prime}\boldsymbol{\theta}^{\ast}_{d-1}}\boldsymbol{X}^{\prime}}{\sum^{d}_{k=1}e^{\boldsymbol{X}^{\prime}\boldsymbol{\theta}^{\ast}_{k}}}-\frac{Me^{\boldsymbol{X}^{\prime}\boldsymbol{\theta}^{\ast}_{d-1}}\boldsymbol{X}^{\prime}}{\sum^{d}_{k=1}e^{\boldsymbol{X}^{\prime}\boldsymbol{\theta}^{\ast}_{k}}}\right]\right]^{\prime}\nonumber \\
 & =\mathbf{0}.\label{eq:Q infinity derivative}
\end{align}
Because $Q^{\dagger}\left(\boldsymbol{\theta},\boldsymbol{\vartheta}\right)$
is strictly convex in $\boldsymbol{\theta}$ for any $\boldsymbol{\vartheta}$,
$Q^{\dagger}\left(\boldsymbol{\theta},\boldsymbol{\theta}^{\ast}\right)$
is certainly strictly convex in $\boldsymbol{\theta}$. Combining
with (\ref{eq:Q infinity derivative}), we obtain that $\boldsymbol{\theta}^{\ast}=\arg\min_{\boldsymbol{\theta}\in\Theta}Q^{\dagger}\left(\boldsymbol{\theta},\boldsymbol{\theta}^{\ast}\right)$,
which implies that $\overline{\boldsymbol{\theta}}\left(\boldsymbol{\theta}^{\ast}\right)=\boldsymbol{\theta}^{\ast}$.

Part (v) is proved by Lemma \ref{lem:: unique theta star for Q infinity}.
\qed

\paragraph{Proof of Lemma \ref{lem:: properties of Fn Fstar}:}

By the definition of $Q^{\dagger}_{n}\left(\boldsymbol{\theta},\boldsymbol{\vartheta}\right)$,
we have that $F_{n}\left(\boldsymbol{\theta}\right)\equiv Q_{n}\left(\boldsymbol{\theta},\overline{\boldsymbol{\mu}}_{n}\left(\boldsymbol{\theta}\right)\right)$.
It holds that at any iteration $s\geq1$, 
\begin{align*}
F_{n}\left(\widehat{\boldsymbol{\theta}}^{\left(s\right)}\right) & =Q_{n}\left(\widehat{\boldsymbol{\theta}}^{\left(s\right)},\overline{\boldsymbol{\mu}}_{n}\left(\widehat{\boldsymbol{\theta}}^{\left(s\right)}\right)\right)=\min_{\boldsymbol{\mu}}Q_{n}\left(\widehat{\boldsymbol{\theta}}^{\left(s\right)},\boldsymbol{\mu}\right)\leq Q_{n}\left(\widehat{\boldsymbol{\theta}}^{\left(s\right)},\overline{\boldsymbol{\mu}}_{n}\left(\widehat{\boldsymbol{\theta}}^{\left(s-1\right)}\right)\right)\\
 & \leq Q_{n}\left(\widehat{\boldsymbol{\theta}}^{\left(s-1\right)},\overline{\boldsymbol{\mu}}_{n}\left(\widehat{\boldsymbol{\theta}}^{\left(s-1\right)}\right)\right)=F_{n}\left(\widehat{\boldsymbol{\theta}}^{\left(s-1\right)}\right).
\end{align*}
The first equality follows because $\overline{\boldsymbol{\mu}}_{n}\left(\widehat{\boldsymbol{\theta}}^{\left(s\right)}\right)$
minimizes $Q_{n}\left(\widehat{\boldsymbol{\theta}}^{\left(s\right)},\boldsymbol{\mu}\right)$
over $\boldsymbol{\mu}$; while the second inequality follows because
$\widehat{\boldsymbol{\theta}}^{\left(s\right)}$ minimizes $Q_{n}\left(\boldsymbol{\theta},\overline{\boldsymbol{\mu}}_{n}\left(\widehat{\boldsymbol{\theta}}^{\left(s-1\right)}\right)\right)$
over $\boldsymbol{\theta}\in\Theta$. This proves the first part of
the lemma.

Part (ii) of the lemma follows from Lemma \ref{lem:: properties of Qn Qstar}
(i), which implies that 
\[
\sup_{\boldsymbol{\theta}\in\Theta}\left|F_{n}\left(\boldsymbol{\theta}\right)-F\left(\boldsymbol{\theta}\right)\right|\leq\sup_{\boldsymbol{\theta},\boldsymbol{\vartheta}\in\Theta}\left|Q^{\dagger}_{n}\left(\boldsymbol{\theta},\boldsymbol{\vartheta}\right)-Q^{\dagger}\left(\boldsymbol{\theta},\boldsymbol{\vartheta}\right)\right|=o_{p}\left(1\right).
\]

For Part (iii), by the proof of Lemma \ref{lem:: Equivalence conditional and unconditional},
we have that 
\[
F_{n}\left(\boldsymbol{\theta}\right)=Q^{\ast}_{n}\left(\boldsymbol{\theta}\right)-\frac{1}{n}\sum^{n}_{i=1}\left[M_{i}\log M_{i}-M_{i}\right],
\]
where the second term on the right-hand side is independent of $\boldsymbol{\theta}$.
As a result, we have 
\[
\arg\min_{\boldsymbol{\theta}\in\Theta}F_{n}\left(\boldsymbol{\theta}\right)=\arg\min_{\boldsymbol{\theta}\in\Theta}Q^{\ast}_{n}\left(\boldsymbol{\theta}\right)=\widetilde{\boldsymbol{\theta}}.
\]
Taking the probability limit on both sides of the equality, we obtain
that 
\[
F\left(\boldsymbol{\theta}\right)=-L_{C\mid X,M}\left(\boldsymbol{\theta}\right)-\mathbb{E}\left[M\log M-M\right].
\]
By Assumption \ref{Assmp:: Maintained assumption}, it holds that
$\boldsymbol{\theta}^{\ast}$ is the unique minimizer of $F\left(\boldsymbol{\theta}\right)$.\qed

\begin{lemma} \label{lem:: joint/separate min}Let $G\left(u,v\right)$
be some function that is continuously differentiable and jointly convex
in $\mathcal{U}\times\mathbb{R}^{m}$, where $\mathcal{U}$ is convex.
For some given $\left(u^{\ast},v^{\ast}\right)$, if it holds that
$u^{\ast}\in\arg\min_{u\in\mathcal{U}}G\left(u,v^{\ast}\right)$ and
$v^{\ast}\in\arg\min_{v\in\mathbb{R}^{m}}G\left(u^{\ast},v\right)$,
then it also holds that $\left(u^{\ast},v^{\ast}\right)\in\arg\min_{u\in\mathcal{U},v\in\mathbb{R}^{m}}G\left(u,v\right)$.
\end{lemma}

\paragraph{Proof of Lemma \ref{lem:: joint/separate min}:}

Using the fact that $u^{\ast}\in\arg\min_{u\in\mathcal{U}}G\left(u,v^{\ast}\right)$
and $\mathcal{U}$ is convex, the variational inequality implies that
for every $u\in\mathcal{U}$, we have $\nabla_{u}G\left(u^{\ast},v^{\ast}\right)^{\prime}\left(u-u^{\ast}\right)\geq0$.
At the same time, because $v^{\ast}\in\arg\min_{v\in\mathbb{R}^{m}}G\left(u^{\ast},v\right)$,
we have $\nabla_{v}G\left(u^{\ast},v^{\ast}\right)=\boldsymbol{0}$.
The convexity of $G\left(\cdot,\cdot\right)$ provides that for every
$\left(u,v\right)\in\mathcal{U}\times\mathbb{R}^{m}$, it holds that
\[
G\left(u,v\right)\geq G\left(u^{\ast},v^{\ast}\right)+\nabla_{u}G\left(u^{\ast},v^{\ast}\right)^{\prime}\left(u-u^{\ast}\right)+\nabla_{v}G\left(u^{\ast},v^{\ast}\right)^{\prime}\left(v-v^{\ast}\right)\geq G\left(u^{\ast},v^{\ast}\right).
\]
The lemma then follows. \qed

\begin{lemma} \label{lem:: Dn properties}For any $\boldsymbol{\vartheta}\in\Theta$,
define $\mathcal{D}_{n}\left(\boldsymbol{\vartheta}\right)\equiv F_{n}\left(\boldsymbol{\vartheta}\right)-\min{}_{\boldsymbol{\theta}\in\Theta}Q^{\dagger}_{n}\left(\boldsymbol{\theta},\boldsymbol{\vartheta}\right)$
and $\mathcal{D}\left(\boldsymbol{\vartheta}\right)\equiv F\left(\boldsymbol{\vartheta}\right)-\min{}_{\boldsymbol{\theta}\in\Theta}Q^{\dagger}\left(\boldsymbol{\theta},\boldsymbol{\vartheta}\right)$.
Under Assumptions \ref{Assmp:: Maintained assumption}, \ref{Assm:: consistency 0},
and \ref{Assm:: consistency 1}, the following results hold:

(i) $\sup_{\boldsymbol{\vartheta}\in\Theta}\left|\mathcal{D}_{n}\left(\boldsymbol{\vartheta}\right)-\mathcal{D}\left(\boldsymbol{\vartheta}\right)\right|\leq2\sup_{\boldsymbol{\theta},\boldsymbol{\vartheta}\in\Theta}\left|Q^{\dagger}_{n}\left(\boldsymbol{\theta},\boldsymbol{\vartheta}\right)-Q^{\dagger}\left(\boldsymbol{\theta},\boldsymbol{\vartheta}\right)\right|=o_{p}\left(1\right)$.

(ii) Function $\mathcal{D}\left(\boldsymbol{\vartheta}\right)$ is
continuous and nonnegative on $\Theta$.

(iii) $\mathcal{D}\left(\boldsymbol{\vartheta}\right)=0$ if and only
if $\boldsymbol{\vartheta}=\boldsymbol{\theta}^{\ast}$.

(iv) For any $r>0$, $\inf_{\left\Vert \boldsymbol{\vartheta}-\boldsymbol{\theta}^{\ast}\right\Vert \geq r}\mathcal{D}\left(\boldsymbol{\vartheta}\right)>0$.

(v) $\mathcal{D}_{n}\left(\widehat{\boldsymbol{\theta}}^{\left(s\right)}\right)\leq F_{n}\left(\widehat{\boldsymbol{\theta}}^{\left(s\right)}\right)-F_{n}\left(\widehat{\boldsymbol{\theta}}^{\left(s+1\right)}\right)$
for every $s\geq0$.\end{lemma}

\paragraph{Proof of Lemma \ref{lem:: Dn properties}:}

Define $\Delta_{n}\equiv\sup_{\boldsymbol{\theta},\boldsymbol{\vartheta}\in\Theta}\left|Q^{\dagger}_{n}\left(\boldsymbol{\theta},\boldsymbol{\vartheta}\right)-Q^{\dagger}\left(\boldsymbol{\theta},\boldsymbol{\vartheta}\right)\right|$.
Since $\Theta$ is compact, it holds that 
\[
\sup_{\boldsymbol{\vartheta}\in\Theta}\left|\min{}_{\boldsymbol{\theta}\in\Theta}Q^{\dagger}_{n}\left(\boldsymbol{\theta},\boldsymbol{\vartheta}\right)-\min{}_{\boldsymbol{\theta}\in\Theta}Q^{\dagger}\left(\boldsymbol{\theta},\boldsymbol{\vartheta}\right)\right|\leq\Delta_{n}.
\]
By definition, it also holds that $\sup_{\boldsymbol{\theta}\in\Theta}\left|F_{n}\left(\boldsymbol{\theta}\right)-F\left(\boldsymbol{\theta}\right)\right|\leq\Delta_{n}$.
Combining the two inequalities, we have $\sup_{\boldsymbol{\vartheta}\in\Theta}\left|\mathcal{D}_{n}\left(\boldsymbol{\vartheta}\right)-\mathcal{D}\left(\boldsymbol{\vartheta}\right)\right|\leq2\Delta_{n}$.
By Lemma \ref{lem:: properties of Qn Qstar} (i), we have that $\Delta_{n}\stackrel{p}{\rightarrow}0$.
The first part of the lemma holds.

The continuity of $F\left(\cdot\right)$ and $Q^{\dagger}\left(\cdot\right)$
follows from their definitions. In particular, $Q^{\dagger}\left(\boldsymbol{\theta},\boldsymbol{\vartheta}\right)$
is jointly continuous in $\left(\boldsymbol{\theta},\boldsymbol{\vartheta}\right)$.
Together with the compactness of $\Theta$, Berge’s maximum theorem
implies that $\mathcal{D}\left(\boldsymbol{\vartheta}\right)$ is
continuous. By definition, we have 
\[
\mathcal{D}\left(\boldsymbol{\vartheta}\right)=F\left(\boldsymbol{\vartheta}\right)-\min{}_{\boldsymbol{\theta}\in\Theta}Q^{\dagger}\left(\boldsymbol{\theta},\boldsymbol{\vartheta}\right)=Q^{\dagger}\left(\boldsymbol{\vartheta},\boldsymbol{\vartheta}\right)-\min{}_{\boldsymbol{\theta}\in\Theta}Q^{\dagger}\left(\boldsymbol{\theta},\boldsymbol{\vartheta}\right)\geq0.
\]
This proves part (ii) of the lemma.

Because $Q^{\dagger}\left(\boldsymbol{\theta},\boldsymbol{\vartheta}\right)$
has a unique minimizer for any $\boldsymbol{\vartheta}\in\Theta$
by Lemma \ref{lem:: properties of Qn Qstar} (ii), we have that $\mathcal{D}\left(\boldsymbol{\vartheta}\right)=0$
if and only if $\overline{\boldsymbol{\theta}}\left(\boldsymbol{\vartheta}\right)=\boldsymbol{\vartheta}$,
where $\overline{\boldsymbol{\theta}}\left(\boldsymbol{\vartheta}\right)$
is defined in Lemma \ref{lem:: properties of Qn Qstar}. Because $\boldsymbol{\theta}^{\ast}$
is the unique fixed point of $\overline{\boldsymbol{\theta}}\left(\cdot\right)$,
we have that $\overline{\boldsymbol{\theta}}\left(\boldsymbol{\vartheta}\right)=\boldsymbol{\vartheta}$
if and only if $\boldsymbol{\vartheta}=\boldsymbol{\theta}^{\ast}$.
Hence, $\mathcal{D}\left(\boldsymbol{\vartheta}\right)=0$ if and
only if $\boldsymbol{\vartheta}=\boldsymbol{\theta}^{\ast}$, which
proves the third part of the lemma. Because $\mathcal{D}\left(\cdot\right)$
is continuous and $\Theta$ is compact, it holds that $\inf_{\left\Vert \boldsymbol{\vartheta}-\boldsymbol{\theta}^{\ast}\right\Vert \geq r}\mathcal{D}\left(\boldsymbol{\vartheta}\right)>0$
for any $r>0$. Part (iv) of the lemma holds.

By definition, we have $\mathcal{D}_{n}\left(\boldsymbol{\vartheta}\right)\geq0$.
Moreover, by the definition of $\overline{\boldsymbol{\mu}}_{n}\left(\cdot\right)$,
it holds that 
\begin{align*}
\min{}_{\boldsymbol{\theta}\in\Theta}Q_{n}\left(\boldsymbol{\theta},\overline{\boldsymbol{\mu}}_{n}\left(\widehat{\boldsymbol{\theta}}^{\left(s\right)}\right)\right) & =Q_{n}\left(\widehat{\boldsymbol{\theta}}^{\left(s+1\right)},\overline{\boldsymbol{\mu}}_{n}\left(\widehat{\boldsymbol{\theta}}^{\left(s\right)}\right)\right)\\
 & \geq Q_{n}\left(\widehat{\boldsymbol{\theta}}^{\left(s+1\right)},\overline{\boldsymbol{\mu}}_{n}\left(\widehat{\boldsymbol{\theta}}^{\left(s+1\right)}\right)\right)=F_{n}\left(\widehat{\boldsymbol{\theta}}^{\left(s+1\right)}\right).
\end{align*}
Therefore, we obtain that 
\[
0\leq\mathcal{D}_{n}\left(\widehat{\boldsymbol{\theta}}^{\left(s\right)}\right)=F_{n}\left(\widehat{\boldsymbol{\theta}}^{\left(s\right)}\right)-\min{}_{\boldsymbol{\theta}\in\Theta}Q_{n}\left(\boldsymbol{\theta},\overline{\boldsymbol{\mu}}_{n}\left(\widehat{\boldsymbol{\theta}}^{\left(s\right)}\right)\right)\leq F_{n}\left(\widehat{\boldsymbol{\theta}}^{\left(s\right)}\right)-F_{n}\left(\widehat{\boldsymbol{\theta}}^{\left(s+1\right)}\right).
\]
Hence, we have proved the last part of the lemma. \qed

\paragraph{Proof of Theorem \ref{Thm:: Convergence property of algorithm}:}

By Lemma \ref{lem:: properties of Fn Fstar} (i), we have that $F_{n}\left(\widehat{\boldsymbol{\theta}}^{\left(s+1\right)}\right)\leq F_{n}\left(\widehat{\boldsymbol{\theta}}^{\left(s\right)}\right)$
for every $s\geq0$. Because $F_{n}\left(\cdot\right)$ is continuous
on the compact set $\Theta$, it is bounded below. Therefore, the
sequence $\left\{ F_{n}\left(\widehat{\boldsymbol{\theta}}^{\left(s\right)}\right)\right\} $
must converge as $s\rightarrow\infty$ and 
\begin{equation}
F_{n}\left(\widehat{\boldsymbol{\theta}}^{\left(s\right)}\right)-F_{n}\left(\widehat{\boldsymbol{\theta}}^{\left(s+1\right)}\right)\rightarrow0.\label{eq:F theta s+1 minus F theta s}
\end{equation}
By Lemma \ref{lem:: Dn properties} (v), we have that $\mathcal{D}_{n}\left(\widehat{\boldsymbol{\theta}}^{\left(s\right)}\right)\leq F_{n}\left(\widehat{\boldsymbol{\theta}}^{\left(s\right)}\right)-F_{n}\left(\widehat{\boldsymbol{\theta}}^{\left(s+1\right)}\right)$,
where $\mathcal{D}_{n}\left(\cdot\right)$ is defined in the lemma.
By (\ref{eq:F theta s+1 minus F theta s}), we have that $\mathcal{D}_{n}\left(\widehat{\boldsymbol{\theta}}^{\left(s\right)}\right)\rightarrow0$
as $s\rightarrow\infty$.

By the explicit expression of $\overline{\boldsymbol{\mu}}_{n}\left(\cdot\right)$,
it is continuous. Because $Q_{n}\left(\boldsymbol{\theta},\boldsymbol{\mu}\right)$
is continuous and $\Theta$ is compact, the maximum theorem implies
that $\boldsymbol{\vartheta}\rightarrow\min{}_{\boldsymbol{\theta}\in\Theta}Q_{n}\left(\boldsymbol{\theta},\overline{\boldsymbol{\mu}}_{n}\left(\boldsymbol{\vartheta}\right)\right)$
is also continuous. Because $F_{n}\left(\cdot\right)$ is continuous,
we can conclude that $\mathcal{D}_{n}\left(\cdot\right)$ is continuous.

Because $\Theta$ is compact, the IDC sequence $\left\{ \widehat{\boldsymbol{\theta}}^{\left(s\right)}\right\} $
has a convergent subsequence. Let $\left\{ \widehat{\boldsymbol{\theta}}^{\left(s_{j}\right)}\right\} $
be the convergent subsequence with limit $\boldsymbol{\theta}^{\infty}$.
The continuity of $\mathcal{D}_{n}\left(\cdot\right)$ together with
the fact that $\mathcal{D}_{n}\left(\widehat{\boldsymbol{\theta}}^{\left(s\right)}\right)\rightarrow0$
imply that $\mathcal{D}_{n}\left(\boldsymbol{\theta}^{\infty}\right)=0$.
As a result, $\boldsymbol{\theta}^{\infty}\in\arg\min{}_{\boldsymbol{\theta}\in\Theta}Q_{n}\left(\boldsymbol{\theta},\overline{\boldsymbol{\mu}}_{n}\left(\boldsymbol{\theta}^{\infty}\right)\right)$.
By the definition of $\overline{\boldsymbol{\mu}}_{n}\left(\cdot\right)$,
we have $\overline{\boldsymbol{\mu}}_{n}\left(\boldsymbol{\theta}^{\infty}\right)\in\arg\min{}_{\boldsymbol{\mu}\in\mathbb{R}^{n}}Q_{n}\left(\boldsymbol{\theta}^{\infty},\boldsymbol{\mu}\right)$.
The convexity of $Q_{n}\left(\boldsymbol{\theta},\boldsymbol{\mu}\right)$
in $\boldsymbol{\theta}$ and $\boldsymbol{\mu}$ jointly can be verified
using its expression. In consequence, Lemma \ref{lem:: joint/separate min}
shows that $\left(\boldsymbol{\theta}^{\infty},\overline{\boldsymbol{\mu}}_{n}\left(\boldsymbol{\theta}^{\infty}\right)\right)$
is the global minimizer of $Q_{n}\left(\boldsymbol{\theta},\boldsymbol{\mu}\right)$.
By definition, we have that $\boldsymbol{\theta}^{\infty}$ also minimizes
$F_{n}\left(\cdot\right)$.

By Lemma \ref{lem:: properties of Fn Fstar} (iii), $F_{n}\left(\cdot\right)$
is uniquely minimized at $\widetilde{\boldsymbol{\theta}}$. Therefore,
we have $\boldsymbol{\theta}^{\infty}=\widetilde{\boldsymbol{\theta}}$,
which implies that every convergent subsequence of $\left\{ \widehat{\boldsymbol{\theta}}^{\left(s\right)}\right\} $
has the same limit $\widetilde{\boldsymbol{\theta}}$. Because $\Theta$
is compact by Assumption \ref{Assm:: consistency 1}, the full sequence
also converges. Hence, it holds that $\widehat{\boldsymbol{\theta}}^{\left(s\right)}\rightarrow\widetilde{\boldsymbol{\theta}}$
as $s\rightarrow\infty$.\qed

\paragraph{Proof of Theorem \ref{thm:: consistency-fixed S}:}

Because $\widehat{\boldsymbol{\theta}}^{\left(0\right)}\stackrel{p}{\rightarrow}\boldsymbol{\theta}^{\ast}$
and $F\left(\cdot\right)$ is continuous, we have $\left|F\left(\widehat{\boldsymbol{\theta}}^{\left(0\right)}\right)-F\left(\boldsymbol{\theta}^{\ast}\right)\right|=o_{p}\left(1\right)$.
Combining with Lemma \ref{lem:: properties of Fn Fstar} (ii), it
holds that 
\begin{equation}
\left|F_{n}\left(\widehat{\boldsymbol{\theta}}^{\left(0\right)}\right)-F_{n}\left(\boldsymbol{\theta}^{\ast}\right)\right|\leq2\sup_{\boldsymbol{\theta}\in\Theta}\left|F_{n}\left(\boldsymbol{\theta}\right)-F\left(\boldsymbol{\theta}\right)\right|+\left|F\left(\widehat{\boldsymbol{\theta}}^{\left(0\right)}\right)-F\left(\boldsymbol{\theta}^{\ast}\right)\right|=o_{p}\left(1\right),\label{eq:Fn theta zero}
\end{equation}
where the first inequality follows from the triangle inequality.

For any $\varepsilon>0$, let $\delta_{\varepsilon}\equiv\inf_{\boldsymbol{\theta}\in\Theta:\left\Vert \boldsymbol{\theta}-\boldsymbol{\theta}^{\ast}\right\Vert \geq\varepsilon}F\left(\boldsymbol{\theta}\right)-F\left(\boldsymbol{\theta}^{\ast}\right)$.
By Lemma \ref{lem:: properties of Fn Fstar} (iii), the continuity
of $F\left(\cdot\right)$, and the compactness of $\Theta$, we have
that $\delta_{\varepsilon}>0$. Consider the event 
\[
\mathcal{E}_{n,\varepsilon}\equiv\left\{ \sup_{\boldsymbol{\theta}\in\Theta}\left|F_{n}\left(\boldsymbol{\theta}\right)-F\left(\boldsymbol{\theta}\right)\right|<\frac{\delta_{\varepsilon}}{6}\right\} \cap\left\{ \left|F_{n}\left(\widehat{\boldsymbol{\theta}}^{\left(0\right)}\right)-F_{n}\left(\boldsymbol{\theta}^{\ast}\right)\right|<\frac{\delta_{\varepsilon}}{3}\right\} .
\]
By Lemma \ref{lem:: properties of Fn Fstar} (ii) and Equation (\ref{eq:Fn theta zero}),
we have that $\Pr\left(\mathcal{E}_{n,\varepsilon}\right)\rightarrow1$
as $n\rightarrow\infty$ for any given $\varepsilon>0$. On this event,
we have that for every $\boldsymbol{\theta}$ such that $\left\Vert \boldsymbol{\theta}-\boldsymbol{\theta}^{\ast}\right\Vert \geq\varepsilon$,
\[
F_{n}\left(\boldsymbol{\theta}\right)-F_{n}\left(\boldsymbol{\theta}^{\ast}\right)\geq F\left(\boldsymbol{\theta}\right)-F\left(\boldsymbol{\theta}^{\ast}\right)-2\sup_{\boldsymbol{\theta}\in\Theta}\left|F_{n}\left(\boldsymbol{\theta}\right)-F\left(\boldsymbol{\theta}\right)\right|>\delta_{\varepsilon}-\frac{\delta_{\varepsilon}}{3}=\frac{2\delta_{\varepsilon}}{3}.
\]
On the other hand, Lemma \ref{lem:: properties of Fn Fstar} (i) implies
that for every $S\geq0$, 
\[
F_{n}\left(\widehat{\boldsymbol{\theta}}^{\left(S\right)}\right)-F_{n}\left(\boldsymbol{\theta}^{\ast}\right)\leq F_{n}\left(\widehat{\boldsymbol{\theta}}^{\left(0\right)}\right)-F_{n}\left(\boldsymbol{\theta}^{\ast}\right)<\frac{\delta_{\varepsilon}}{3}.
\]
Therefore, on the event $\mathcal{E}_{n,\varepsilon}$, no iterate
can satisfy $\left\Vert \widehat{\boldsymbol{\theta}}^{\left(S\right)}-\boldsymbol{\theta}^{\ast}\right\Vert \geq\varepsilon$,
which implies that 
\[
\Pr\left(\sup_{S\geq0}\left\Vert \widehat{\boldsymbol{\theta}}^{\left(S\right)}-\boldsymbol{\theta}^{\ast}\right\Vert \geq\varepsilon\right)\leq1-\Pr\left(\mathcal{E}_{n,\varepsilon}\right)\rightarrow0.
\]
Hence, we obtain that $\sup_{S\geq0}\left\Vert \widehat{\boldsymbol{\theta}}^{\left(S\right)}-\boldsymbol{\theta}^{\ast}\right\Vert =o_{p}\left(1\right)$.
This completes the proof. \qed

\paragraph{Proof of Theorem \ref{thm:: consistency}:}

Define $R_{n}\equiv\sup_{\boldsymbol{\theta}\in\Theta}F_{n}\left(\boldsymbol{\theta}\right)-\inf_{\boldsymbol{\theta}\in\Theta}F_{n}\left(\boldsymbol{\theta}\right)$
and $R\equiv\sup_{\boldsymbol{\theta}\in\Theta}F\left(\boldsymbol{\theta}\right)-\inf_{\boldsymbol{\theta}\in\Theta}F\left(\boldsymbol{\theta}\right)$.
Because $F\left(\cdot\right)$ is continuous on the compact set $\Theta$,
we have $R<\infty$. Together with Lemma \ref{lem:: properties of Fn Fstar}
(ii), we obtain that $R_{n}=O_{p}\left(1\right)$.

By Lemma \ref{lem:: Dn properties} (v), we have $\mathcal{D}_{n}\left(\widehat{\boldsymbol{\theta}}^{\left(s\right)}\right)\leq F_{n}\left(\widehat{\boldsymbol{\theta}}^{\left(s\right)}\right)-F_{n}\left(\widehat{\boldsymbol{\theta}}^{\left(s+1\right)}\right)$
for every $s\geq0$. Summing up the two sides of the inequalities
from $s=0$ to $s=S_{n}-1$, we have 
\[
\sum^{S_{n}-1}_{s=0}\mathcal{D}_{n}\left(\widehat{\boldsymbol{\theta}}^{\left(s\right)}\right)\leq F_{n}\left(\widehat{\boldsymbol{\theta}}^{\left(0\right)}\right)-F_{n}\left(\widehat{\boldsymbol{\theta}}^{\left(S_{n}\right)}\right)\leq R_{n},
\]
where the last inequality follows from the definition of $R_{n}$.
Let $\tau_{n}$ be the smallest index attaining $\min_{0\leq s<S_{n}}\mathcal{D}_{n}\left(\widehat{\boldsymbol{\theta}}^{\left(s\right)}\right)$.
Then, it holds that 
\[
0\leq\mathcal{D}_{n}\left(\widehat{\boldsymbol{\theta}}^{\left(\tau_{n}\right)}\right)\leq\frac{R_{n}}{S_{n}}=o_{p}\left(1\right),
\]
where the last equality follows from $S_{n}\rightarrow\infty$ and
$R_{n}=O_{p}\left(1\right)$. By Lemma \ref{lem:: Dn properties}
(i), we then have 
\[
\mathcal{D}\left(\widehat{\boldsymbol{\theta}}^{\left(\tau_{n}\right)}\right)\leq\sup_{\boldsymbol{\vartheta}\in\Theta}\left|\mathcal{D}_{n}\left(\boldsymbol{\vartheta}\right)-\mathcal{D}\left(\boldsymbol{\vartheta}\right)\right|+\mathcal{D}_{n}\left(\widehat{\boldsymbol{\theta}}^{\left(\tau_{n}\right)}\right)=o_{p}\left(1\right).
\]
Since $\mathcal{D}\left(\cdot\right)$ is continuous with a unique
zero at $\boldsymbol{\theta}^{\ast}$ by Lemma \ref{lem:: Dn properties}
(ii) and (iii), the standard identification argument implies that
$\widehat{\boldsymbol{\theta}}^{\left(\tau_{n}\right)}\stackrel{p}{\rightarrow}\boldsymbol{\theta}^{\ast}$.

It remains to show that the result holds for $\widehat{\boldsymbol{\theta}}^{\left(S_{n}\right)}$,
where $S_{n}\geq\tau_{n}$. By Lemma \ref{lem:: properties of Fn Fstar}
(i), we have $F_{n}\left(\widehat{\boldsymbol{\theta}}^{\left(S_{n}\right)}\right)\leq F_{n}\left(\widehat{\boldsymbol{\theta}}^{\left(\tau_{n}\right)}\right)$.
Therefore, 
\begin{align*}
F\left(\widehat{\boldsymbol{\theta}}^{\left(S_{n}\right)}\right)-F\left(\widehat{\boldsymbol{\theta}}^{\left(\tau_{n}\right)}\right) & \leq2\sup_{\boldsymbol{\theta}\in\Theta}\left|F_{n}\left(\boldsymbol{\theta}\right)-F\left(\boldsymbol{\theta}\right)\right|+F_{n}\left(\widehat{\boldsymbol{\theta}}^{\left(S_{n}\right)}\right)-F_{n}\left(\widehat{\boldsymbol{\theta}}^{\left(\tau_{n}\right)}\right)\\
 & \leq2\sup_{\boldsymbol{\theta}\in\Theta}\left|F_{n}\left(\boldsymbol{\theta}\right)-F\left(\boldsymbol{\theta}\right)\right|.
\end{align*}
Because $\boldsymbol{\theta}^{\ast}$ is the unique minimizer of $F\left(\cdot\right)$
by Lemma \ref{lem:: properties of Fn Fstar} (iii), it holds that
\begin{align*}
0\leq & F\left(\widehat{\boldsymbol{\theta}}^{\left(S_{n}\right)}\right)-F\left(\boldsymbol{\theta}^{\ast}\right)=F\left(\widehat{\boldsymbol{\theta}}^{\left(S_{n}\right)}\right)-F\left(\widehat{\boldsymbol{\theta}}^{\left(\tau_{n}\right)}\right)+F\left(\widehat{\boldsymbol{\theta}}^{\left(\tau_{n}\right)}\right)-F\left(\boldsymbol{\theta}^{\ast}\right)\\
\leq & 2\sup_{\boldsymbol{\theta}\in\Theta}\left|F_{n}\left(\boldsymbol{\theta}\right)-F\left(\boldsymbol{\theta}\right)\right|+F\left(\widehat{\boldsymbol{\theta}}^{\left(\tau_{n}\right)}\right)-F\left(\boldsymbol{\theta}^{\ast}\right)=o_{p}\left(1\right),
\end{align*}
where the last equality follows from Lemma \ref{lem:: properties of Fn Fstar}
(ii), the continuity of $F\left(\cdot\right)$, and $\widehat{\boldsymbol{\theta}}^{\left(\tau_{n}\right)}\stackrel{p}{\rightarrow}\boldsymbol{\theta}^{\ast}$.
Because $F\left(\cdot\right)$ is continuous on the compact set $\Theta$
and is uniquely minimized at $\boldsymbol{\theta}^{\ast}$, the convergence
of $F\left(\widehat{\boldsymbol{\theta}}^{\left(S_{n}\right)}\right)$
towards $F\left(\boldsymbol{\theta}^{\ast}\right)$ implies that $\widehat{\boldsymbol{\theta}}^{\left(S_{n}\right)}\stackrel{p}{\rightarrow}\boldsymbol{\theta}^{\ast}$.\qed

\begin{lemma} \label{lem:: theta bar n convergence} Under Assumptions
\ref{Assmp:: Maintained assumption}, \ref{Assm:: consistency 0},
\ref{Assm:: consistency 1}, and \ref{ass::Information Dominance},
there exist constants $\epsilon>0$ and $\rho\in\left(0,1\right)$
such that, with probability approaching one, 
\[
\left\Vert \overline{\boldsymbol{\theta}}_{n}\left(\check{\boldsymbol{\vartheta}}\right)-\overline{\boldsymbol{\theta}}_{n}\left(\boldsymbol{\vartheta}\right)\right\Vert \leq\rho\left\Vert \check{\boldsymbol{\vartheta}}-\boldsymbol{\vartheta}\right\Vert 
\]
for every $\boldsymbol{\vartheta},\check{\boldsymbol{\vartheta}}\in\overline{\mathcal{B}}\left(\boldsymbol{\theta}^{\ast},\epsilon\right)$
and $\overline{\boldsymbol{\theta}}_{n}\left(\overline{\mathcal{B}}\left(\boldsymbol{\theta}^{\ast},\epsilon\right)\right)\subset\overline{\mathcal{B}}\left(\boldsymbol{\theta}^{\ast},\epsilon\right)$,
where 
\[\overline{\mathcal{B}}\left(\boldsymbol{\theta}^{\ast},\epsilon\right)\equiv\left\{ \boldsymbol{\theta}\in\Theta:\left\Vert \boldsymbol{\theta}-\boldsymbol{\theta}^{\ast}\right\Vert \leq\epsilon\right\}.\]\end{lemma}

\paragraph{Proof of Lemma \ref{lem:: theta bar n convergence}:}

For every $\eta>0$, define 
\[
\Delta_{\eta}\equiv\inf_{\boldsymbol{\theta},\boldsymbol{\vartheta}\in\Theta:\left\Vert \boldsymbol{\theta}-\overline{\boldsymbol{\theta}}\left(\boldsymbol{\vartheta}\right)\right\Vert \geq\eta}\left\{ Q^{\dagger}\left(\boldsymbol{\theta},\boldsymbol{\vartheta}\right)-Q^{\dagger}\left(\overline{\boldsymbol{\theta}}\left(\boldsymbol{\vartheta}\right),\boldsymbol{\vartheta}\right)\right\} .
\]
By the compactness of $\Theta$, the joint continuity of $Q^{\dagger}\left(\cdot\right)$,
the continuity of $\overline{\boldsymbol{\theta}}\left(\cdot\right)$,
and Lemma \ref{lem:: properties of Qn Qstar} (ii), we have $\Delta_{\eta}>0$.
Further let 
\[
\delta_{n}\equiv\sup_{\boldsymbol{\theta},\boldsymbol{\vartheta}\in\Theta}\left|Q^{\dagger}_{n}\left(\boldsymbol{\theta},\boldsymbol{\vartheta}\right)-Q^{\dagger}\left(\boldsymbol{\theta},\boldsymbol{\vartheta}\right)\right|.
\]
Lemma \ref{lem:: properties of Qn Qstar} (i) implies that $\delta_{n}=o_{p}\left(1\right)$.
On the event $\delta_{n}<\frac{\Delta_{\eta}}{3}$, suppose that for
some $\boldsymbol{\vartheta}\in\Theta$, $\left\Vert \overline{\boldsymbol{\theta}}_{n}\left(\boldsymbol{\vartheta}\right)-\overline{\boldsymbol{\theta}}\left(\boldsymbol{\vartheta}\right)\right\Vert \geq\eta$.
Then, the definition of $\delta_{n}$ implies that 
\begin{align*}
Q^{\dagger}_{n}\left(\overline{\boldsymbol{\theta}}_{n}\left(\boldsymbol{\vartheta}\right),\boldsymbol{\vartheta}\right) & \geq Q^{\dagger}\left(\overline{\boldsymbol{\theta}}_{n}\left(\boldsymbol{\vartheta}\right),\boldsymbol{\vartheta}\right)-\delta_{n}\geq Q^{\dagger}\left(\overline{\boldsymbol{\theta}}\left(\boldsymbol{\vartheta}\right),\boldsymbol{\vartheta}\right)+\Delta_{\eta}-\delta_{n}\\
 & >Q^{\dagger}\left(\overline{\boldsymbol{\theta}}\left(\boldsymbol{\vartheta}\right),\boldsymbol{\vartheta}\right)+\delta_{n}\geq Q^{\dagger}_{n}\left(\overline{\boldsymbol{\theta}}\left(\boldsymbol{\vartheta}\right),\boldsymbol{\vartheta}\right),
\end{align*}
which contradicts the definition of $\overline{\boldsymbol{\theta}}_{n}\left(\boldsymbol{\vartheta}\right)$.
Therefore, the minimizer of $Q^{\dagger}_{n}\left(\boldsymbol{\theta},\boldsymbol{\vartheta}\right)$
given $\boldsymbol{\vartheta}$ must be within the distance $\eta$
from $\overline{\boldsymbol{\theta}}\left(\boldsymbol{\vartheta}\right)$
uniformly in $\boldsymbol{\vartheta}$. Since $\eta$ is arbitrary,
we have 
\begin{equation}
\sup_{\boldsymbol{\vartheta}\in\Theta}\left\Vert \overline{\boldsymbol{\theta}}_{n}\left(\boldsymbol{\vartheta}\right)-\overline{\boldsymbol{\theta}}\left(\boldsymbol{\vartheta}\right)\right\Vert =o_{p}\left(1\right).\label{eq:theta n bar convergence}
\end{equation}

We next establish the local contraction property. By the first-order
conditions and the implicit function theorem, we have that for any
$\boldsymbol{\vartheta}\in\Theta$, 
\begin{align*}
\frac{\partial}{\partial\boldsymbol{\vartheta}}\overline{\boldsymbol{\theta}}_{n}\left(\boldsymbol{\vartheta}\right) & =-\left[\frac{\partial^{2}Q^{\dagger}_{n}\left(\boldsymbol{\theta},\boldsymbol{\vartheta}\right)}{\partial\boldsymbol{\theta}\partial\boldsymbol{\theta}^{\prime}}\mid_{\left(\boldsymbol{\theta},\boldsymbol{\vartheta}\right)=\left(\overline{\boldsymbol{\theta}}_{n}\left(\boldsymbol{\vartheta}\right),\boldsymbol{\vartheta}\right)}\right]^{-1}\frac{\partial^{2}Q^{\dagger}_{n}\left(\boldsymbol{\theta},\boldsymbol{\vartheta}\right)}{\partial\boldsymbol{\theta}\partial\boldsymbol{\vartheta}^{\prime}}\mid_{\left(\boldsymbol{\theta},\boldsymbol{\vartheta}\right)=\left(\overline{\boldsymbol{\theta}}_{n}\left(\boldsymbol{\vartheta}\right),\boldsymbol{\vartheta}\right)},\\
\frac{\partial}{\partial\boldsymbol{\vartheta}}\overline{\boldsymbol{\theta}}\left(\boldsymbol{\vartheta}\right) & =-\left[\frac{\partial^{2}Q^{\dagger}\left(\boldsymbol{\theta},\boldsymbol{\vartheta}\right)}{\partial\boldsymbol{\theta}\partial\boldsymbol{\theta}^{\prime}}\mid_{\left(\boldsymbol{\theta},\boldsymbol{\vartheta}\right)=\left(\overline{\boldsymbol{\theta}}\left(\boldsymbol{\vartheta}\right),\boldsymbol{\vartheta}\right)}\right]^{-1}\frac{\partial^{2}Q^{\dagger}\left(\boldsymbol{\theta},\boldsymbol{\vartheta}\right)}{\partial\boldsymbol{\theta}\partial\boldsymbol{\vartheta}^{\prime}}\mid_{\left(\boldsymbol{\theta},\boldsymbol{\vartheta}\right)=\left(\overline{\boldsymbol{\theta}}\left(\boldsymbol{\vartheta}\right),\boldsymbol{\vartheta}\right)}.
\end{align*}
Lemma \ref{lem:: properties of Qn Qstar} (ii) and its proof imply
that $\frac{\partial^{2}Q^{\dagger}\left(\boldsymbol{\theta},\boldsymbol{\vartheta}\right)}{\partial\boldsymbol{\theta}\partial\boldsymbol{\theta}^{\prime}}\mid_{\left(\boldsymbol{\theta},\boldsymbol{\vartheta}\right)=\left(\overline{\boldsymbol{\theta}}\left(\boldsymbol{\vartheta}\right),\boldsymbol{\vartheta}\right)}$
is positive definite for any $\boldsymbol{\vartheta}\in\Theta$. Following
the argument for proving Lemma \ref{lem:: properties of Qn Qstar}
(i), we can show that 
\begin{align*}
\sup_{\boldsymbol{\theta},\boldsymbol{\vartheta}\in\Theta}\left\Vert \frac{\partial^{2}Q^{\dagger}_{n}\left(\boldsymbol{\theta},\boldsymbol{\vartheta}\right)}{\partial\boldsymbol{\theta}\partial\boldsymbol{\theta}^{\prime}}-\frac{\partial^{2}Q^{\dagger}\left(\boldsymbol{\theta},\boldsymbol{\vartheta}\right)}{\partial\boldsymbol{\theta}\partial\boldsymbol{\theta}^{\prime}}\right\Vert  & =o_{p}\left(1\right)\mbox{ and }\\
\sup_{\boldsymbol{\theta},\boldsymbol{\vartheta}\in\Theta}\left\Vert \frac{\partial^{2}Q^{\dagger}_{n}\left(\boldsymbol{\theta},\boldsymbol{\vartheta}\right)}{\partial\boldsymbol{\theta}\partial\boldsymbol{\vartheta}^{\prime}}-\frac{\partial^{2}Q^{\dagger}\left(\boldsymbol{\theta},\boldsymbol{\vartheta}\right)}{\partial\boldsymbol{\theta}\partial\boldsymbol{\vartheta}^{\prime}}\right\Vert  & =o_{p}\left(1\right).
\end{align*}
Together with Equation (\ref{eq:theta n bar convergence}), we have
that 
\begin{equation}
\sup_{\boldsymbol{\vartheta}\in\Theta}\left\Vert \frac{\partial}{\partial\boldsymbol{\vartheta}}\overline{\boldsymbol{\theta}}_{n}\left(\boldsymbol{\vartheta}\right)-\frac{\partial}{\partial\boldsymbol{\vartheta}}\overline{\boldsymbol{\theta}}\left(\boldsymbol{\vartheta}\right)\right\Vert =o_{p}\left(1\right).\label{eq: theta bar n derivative}
\end{equation}

By Assumption \ref{ass::Information Dominance} and Lemma \ref{lem:: properties of Qn Qstar}
(v), we have that $\left\Vert \frac{\partial}{\partial\boldsymbol{\vartheta}}\overline{\boldsymbol{\theta}}\left(\boldsymbol{\theta}^{\ast}\right)\right\Vert <1$.
By the continuity of $\frac{\partial}{\partial\boldsymbol{\vartheta}}\overline{\boldsymbol{\theta}}\left(\cdot\right)$,
there exist some $\epsilon>0$ and $\rho_{0}\in\left(0,1\right)$
such that $\sup_{\boldsymbol{\vartheta}\in\overline{\mathcal{B}}\left(\boldsymbol{\theta}^{\ast},\epsilon\right)}\left\Vert \frac{\partial}{\partial\boldsymbol{\vartheta}}\overline{\boldsymbol{\theta}}\left(\boldsymbol{\vartheta}\right)\right\Vert <\rho_{0}$,
where $\overline{\mathcal{B}}\left(\boldsymbol{\theta}^{\ast},\epsilon\right)$
is defined in the lemma and is convex and compact. By Equation (\ref{eq: theta bar n derivative}),
we can find another $\rho\in\left(0,1\right)$ such that $\rho_{0}<\rho$
and 
\begin{equation}
\sup_{\boldsymbol{\vartheta}\in\overline{\mathcal{B}}\left(\boldsymbol{\theta}^{\ast},\epsilon\right)}\left\Vert \frac{\partial}{\partial\boldsymbol{\vartheta}}\overline{\boldsymbol{\theta}}_{n}\left(\boldsymbol{\vartheta}\right)\right\Vert \leq\rho\label{eq:theta bar n derivative bounded}
\end{equation}
with probability approaching one. Because $\overline{\mathcal{B}}\left(\boldsymbol{\theta}^{\ast},\epsilon\right)$
is convex, we can apply the multivariate Taylor expansion (\citet{dieudonne2011foundations},
p. 190) and write 
\[
\overline{\boldsymbol{\theta}}_{n}\left(\check{\boldsymbol{\vartheta}}\right)-\overline{\boldsymbol{\theta}}_{n}\left(\boldsymbol{\vartheta}\right)=\left[\int^{1}_{0}\frac{\partial}{\partial\boldsymbol{\vartheta}}\overline{\boldsymbol{\theta}}_{n}\left(\boldsymbol{\vartheta}+\xi\left(\check{\boldsymbol{\vartheta}}-\boldsymbol{\vartheta}\right)\right)d\xi\right]\left(\check{\boldsymbol{\vartheta}}-\boldsymbol{\vartheta}\right)
\]
for any $\boldsymbol{\vartheta},\check{\boldsymbol{\vartheta}}\in\overline{\mathcal{B}}\left(\boldsymbol{\theta}^{\ast},\epsilon\right)$.
Combining with (\ref{eq:theta bar n derivative bounded}), we obtain
that for any $\boldsymbol{\vartheta},\check{\boldsymbol{\vartheta}}\in\overline{\mathcal{B}}\left(\boldsymbol{\theta}^{\ast},\epsilon\right)$,
\[
\left\Vert \overline{\boldsymbol{\theta}}_{n}\left(\check{\boldsymbol{\vartheta}}\right)-\overline{\boldsymbol{\theta}}_{n}\left(\boldsymbol{\vartheta}\right)\right\Vert \leq\rho\left\Vert \check{\boldsymbol{\vartheta}}-\boldsymbol{\vartheta}\right\Vert .
\]
This proves the first claim in the lemma.

Applying the same argument, we can obtain that for any $\boldsymbol{\vartheta},\check{\boldsymbol{\vartheta}}\in\overline{\mathcal{B}}\left(\boldsymbol{\theta}^{\ast},\epsilon\right)$,
\[
\left\Vert \overline{\boldsymbol{\theta}}\left(\check{\boldsymbol{\vartheta}}\right)-\overline{\boldsymbol{\theta}}\left(\boldsymbol{\vartheta}\right)\right\Vert \leq\rho_{0}\left\Vert \check{\boldsymbol{\vartheta}}-\boldsymbol{\vartheta}\right\Vert 
\]
because $\sup_{\boldsymbol{\vartheta}\in\overline{\mathcal{B}}\left(\boldsymbol{\theta}^{\ast},\epsilon\right)}\left\Vert \frac{\partial}{\partial\boldsymbol{\vartheta}}\overline{\boldsymbol{\theta}}\left(\boldsymbol{\vartheta}\right)\right\Vert <\rho_{0}$.
Since $\overline{\boldsymbol{\theta}}\left(\boldsymbol{\theta}^{\ast}\right)=\boldsymbol{\theta}^{\ast}$,
we have that for any $\boldsymbol{\vartheta}\in\overline{\mathcal{B}}\left(\boldsymbol{\theta}^{\ast},\epsilon\right)$,
\[
\left\Vert \overline{\boldsymbol{\theta}}\left(\boldsymbol{\vartheta}\right)-\boldsymbol{\theta}^{\ast}\right\Vert \leq\rho_{0}\left\Vert \boldsymbol{\vartheta}-\boldsymbol{\theta}^{\ast}\right\Vert \leq\rho_{0}\epsilon.
\]
Furthermore, Equation (\ref{eq:theta n bar convergence}) implies
that, with probability approaching one, 
\[
\sup_{\boldsymbol{\vartheta}\in\overline{\mathcal{B}}\left(\boldsymbol{\theta}^{\ast},\epsilon\right)}\left\Vert \overline{\boldsymbol{\theta}}_{n}\left(\boldsymbol{\vartheta}\right)-\overline{\boldsymbol{\theta}}\left(\boldsymbol{\vartheta}\right)\right\Vert <\left(1-\rho_{0}\right)\epsilon.
\]
As a result, we have $\left\Vert \overline{\boldsymbol{\theta}}_{n}\left(\boldsymbol{\vartheta}\right)-\boldsymbol{\theta}^{\ast}\right\Vert <\epsilon$
for any $\boldsymbol{\vartheta}\in\overline{\mathcal{B}}\left(\boldsymbol{\theta}^{\ast},\epsilon\right)$
with probability approaching one. The second claim then follows.\qed

\begin{lemma} \label{lem:: theta hat s+1 minus theta hat s} Under
the conditions in Theorem \ref{Thm:: equivalence IDC and theta tilda},
it holds that 
\[
\sqrt{n}\left(\widehat{\boldsymbol{\theta}}^{\left(S_{n}\right)}-\widehat{\boldsymbol{\theta}}^{\left(S_{n}-1\right)}\right)\stackrel{p}{\rightarrow}0.
\]
\end{lemma}

\paragraph{Proof of Lemma \ref{lem:: theta hat s+1 minus theta hat s}:}

Suppose that the initial estimator is consistent. By Theorem \ref{thm:: consistency-fixed S},
we have that $\sup_{S\geq0}\left\Vert \widehat{\boldsymbol{\theta}}^{\left(S\right)}-\boldsymbol{\theta}^{\ast}\right\Vert =o_{p}\left(1\right)$.
Therefore, with probability approaching one, all IDC iterates belong
to $\overline{\mathcal{B}}\left(\boldsymbol{\theta}^{\ast},\epsilon\right)$
defined in Lemma \ref{lem:: theta bar n convergence}. Applying Lemma
\ref{lem:: theta bar n convergence}, it holds that 
\begin{align*}
\left\Vert \widehat{\boldsymbol{\theta}}^{\left(S_{n}\right)}-\widehat{\boldsymbol{\theta}}^{\left(S_{n}-1\right)}\right\Vert  & =\left\Vert \overline{\boldsymbol{\theta}}_{n}\left(\widehat{\boldsymbol{\theta}}^{\left(S_{n}-1\right)}\right)-\overline{\boldsymbol{\theta}}_{n}\left(\widehat{\boldsymbol{\theta}}^{\left(S_{n}-2\right)}\right)\right\Vert \leq\rho\left\Vert \widehat{\boldsymbol{\theta}}^{\left(S_{n}-1\right)}-\widehat{\boldsymbol{\theta}}^{\left(S_{n}-2\right)}\right\Vert \\
 & \leq\rho^{S_{n}-1}\left\Vert \widehat{\boldsymbol{\theta}}^{\left(1\right)}-\widehat{\boldsymbol{\theta}}^{\left(0\right)}\right\Vert .
\end{align*}
Since both $\widehat{\boldsymbol{\theta}}^{\left(1\right)}$ and $\widehat{\boldsymbol{\theta}}^{\left(0\right)}$
are consistent estimator, we have $\left\Vert \widehat{\boldsymbol{\theta}}^{\left(1\right)}-\widehat{\boldsymbol{\theta}}^{\left(0\right)}\right\Vert =o_{p}\left(1\right)$.
Because $\rho<1$ and $\log\left(n\right)=o\left(S_{n}\right)$, it
holds that $\sqrt{n}\rho^{S_{n}}\rightarrow0$ as $n\rightarrow\infty$.
Hence, if $\widehat{\boldsymbol{\theta}}^{\left(0\right)}$ is a consistent
estimator and $\log\left(n\right)=o\left(S_{n}\right)$, then $\sqrt{n}\left(\widehat{\boldsymbol{\theta}}^{\left(S_{n}\right)}-\widehat{\boldsymbol{\theta}}^{\left(S_{n}-1\right)}\right)\stackrel{p}{\rightarrow}0$.

For any initial estimator, let $m_{n}=\left\lfloor S_{n}/2\right\rfloor $.
Because $S_{n}>n^{\delta}$ for $\delta>0$, we have $m_{n}\rightarrow\infty$.
Theorem \ref{thm:: consistency} provides that $\widehat{\boldsymbol{\theta}}^{\left(m_{n}\right)}-\boldsymbol{\theta}^{\ast}=o_{p}\left(1\right)$.
Therefore, with probability approaching one, $\widehat{\boldsymbol{\theta}}^{\left(m_{n}\right)}$
belongs to $\overline{\mathcal{B}}\left(\boldsymbol{\theta}^{\ast},\epsilon\right)$.
By Lemma \ref{lem:: theta bar n convergence}, all subsequent iterates
after $\widehat{\boldsymbol{\theta}}^{\left(m_{n}\right)}$ will remain
in $\overline{\mathcal{B}}\left(\boldsymbol{\theta}^{\ast},\epsilon\right)$.
Therefore, we have 
\begin{align*}
\left\Vert \widehat{\boldsymbol{\theta}}^{\left(S_{n}\right)}-\widehat{\boldsymbol{\theta}}^{\left(S_{n}-1\right)}\right\Vert  & =\left\Vert \overline{\boldsymbol{\theta}}_{n}\left(\widehat{\boldsymbol{\theta}}^{\left(S_{n}-1\right)}\right)-\overline{\boldsymbol{\theta}}_{n}\left(\widehat{\boldsymbol{\theta}}^{\left(S_{n}-2\right)}\right)\right\Vert \leq\rho\left\Vert \widehat{\boldsymbol{\theta}}^{\left(S_{n}-1\right)}-\widehat{\boldsymbol{\theta}}^{\left(S_{n}-2\right)}\right\Vert \\
 & \leq\rho^{S_{n}-m_{n}-1}\left\Vert \widehat{\boldsymbol{\theta}}^{\left(m_{n}+1\right)}-\widehat{\boldsymbol{\theta}}^{\left(m_{n}\right)}\right\Vert .
\end{align*}
Because both $\widehat{\boldsymbol{\theta}}^{\left(m_{n}+1\right)}$
and $\widehat{\boldsymbol{\theta}}^{\left(m_{n}\right)}$ are consistent,
we have $\left\Vert \widehat{\boldsymbol{\theta}}^{\left(m_{n}+1\right)}-\widehat{\boldsymbol{\theta}}^{\left(m_{n}\right)}\right\Vert =o_{p}\left(1\right)$.
Together with the fact that $S_{n}>n^{\delta}$, we have 
\[
\sqrt{n}\left\Vert \widehat{\boldsymbol{\theta}}^{\left(S_{n}\right)}-\widehat{\boldsymbol{\theta}}^{\left(S_{n}-1\right)}\right\Vert \leq\sqrt{n}\rho^{S_{n}/2-1}\left\Vert \widehat{\boldsymbol{\theta}}^{\left(m_{n}+1\right)}-\widehat{\boldsymbol{\theta}}^{\left(m_{n}\right)}\right\Vert =o_{p}\left(1\right).
\]
Hence, for arbitrary initial estimator, $\sqrt{n}\left(\widehat{\boldsymbol{\theta}}^{\left(S_{n}\right)}-\widehat{\boldsymbol{\theta}}^{\left(S_{n}-1\right)}\right)\stackrel{p}{\rightarrow}0$
if $S_{n}>n^{\delta}$ for some $\delta>0$.\qed

\paragraph{Proof of Theorem \ref{Thm:: equivalence IDC and theta tilda}:}

We aim to show that $\widehat{\boldsymbol{\theta}}^{\left(S_{n}\right)}$
has the same influence function as $\widetilde{\boldsymbol{\theta}}$.
By the definition of the IDC estimator in Section \ref{sec_sub:Iterative-Distributed-Computing},
we have that 
\[
\frac{\partial}{\partial\boldsymbol{\theta}}Q_{n}\left(\boldsymbol{\theta},\overline{\boldsymbol{\mu}}_{n}\left(\widehat{\boldsymbol{\theta}}^{\left(S_{n}-1\right)}\right)\right)\mid_{\boldsymbol{\theta}=\widehat{\boldsymbol{\theta}}^{\left(S_{n}\right)}}=\mathbf{0}.
\]
Applying the multivariate Taylor expansion (\citet{dieudonne2011foundations},
p. 190) to the function on the left-hand-side of the above equation
round $\widehat{\boldsymbol{\theta}}^{\left(S_{n}-1\right)}$, we
can obtain that 
\begin{align}
\mathbf{0}= & \frac{\partial}{\partial\boldsymbol{\theta}}Q_{n}\left(\boldsymbol{\theta},\overline{\boldsymbol{\mu}}_{n}\left(\widehat{\boldsymbol{\theta}}^{\left(S_{n}-1\right)}\right)\right)\mid_{\boldsymbol{\theta}=\widehat{\boldsymbol{\theta}}^{\left(S_{n}-1\right)}}\nonumber \\
 & +\left[\int^{1}_{0}\frac{\partial^{2}}{\partial\boldsymbol{\theta}\partial\boldsymbol{\theta}^{\prime}}Q_{n}\left(\boldsymbol{\theta},\overline{\boldsymbol{\mu}}_{n}\left(\widehat{\boldsymbol{\theta}}^{(S_{n}-1)}\right)\right)\mid_{\boldsymbol{\theta}=\widehat{\boldsymbol{\theta}}^{\left(S_{n}-1\right)}+\xi\left(\widehat{\boldsymbol{\theta}}^{\left(S_{n}\right)}-\widehat{\boldsymbol{\theta}}^{\left(S_{n}-1\right)}\right)}d\xi\right]\left(\widehat{\boldsymbol{\theta}}^{\left(S_{n}\right)}-\widehat{\boldsymbol{\theta}}^{\left(S_{n}-1\right)}\right).\label{eq:Taylor expansion}
\end{align}
The definition of $Q_{n}\left(\boldsymbol{\theta},\boldsymbol{\mu}\right)$
implies that 
\begin{align*}
 & \frac{\partial}{\partial\boldsymbol{\theta}}Q_{n}\left(\boldsymbol{\theta},\overline{\boldsymbol{\mu}}_{n}\left(\widehat{\boldsymbol{\theta}}^{\left(S_{n}-1\right)}\right)\right)\mid_{\boldsymbol{\theta}=\widehat{\boldsymbol{\theta}}^{\left(S_{n}-1\right)}}\\
 & =-\frac{1}{n}\frac{d}{d\boldsymbol{\theta}}l_{C\mid X,M}\left(\widehat{\boldsymbol{\theta}}^{\left(S_{n}-1\right)}\right)-\frac{1}{n}\frac{\partial}{\partial\boldsymbol{\theta}}f\left(\boldsymbol{\theta},\overline{\boldsymbol{\mu}}_{n}\left(\widehat{\boldsymbol{\theta}}^{\left(S_{n}-1\right)}\right)\right)\mid_{\boldsymbol{\theta}=\widehat{\boldsymbol{\theta}}^{\left(S_{n}-1\right)}}.
\end{align*}
Applying the expression of $f\left(\boldsymbol{\theta},\boldsymbol{\mu}\right)$,
it can be shown that 
\[
\frac{\partial}{\partial\boldsymbol{\theta}}f\left(\boldsymbol{\theta},\overline{\boldsymbol{\mu}}_{n}\left(\widehat{\boldsymbol{\theta}}^{\left(S_{n}-1\right)}\right)\right)\mid_{\boldsymbol{\theta}=\widehat{\boldsymbol{\theta}}^{\left(S_{n}-1\right)}}=\mathbf{0}.
\]
Therefore, Equation (\ref{eq:Taylor expansion}) can be rewritten
as 
\begin{align*}
\mathbf{0} & =-\frac{1}{n}\frac{d}{d\boldsymbol{\theta}}l_{C\mid X,M}\left(\widehat{\boldsymbol{\theta}}^{\left(S_{n}-1\right)}\right)\\
 & +\left[\int^{1}_{0}\frac{\partial^{2}}{\partial\boldsymbol{\theta}\partial\boldsymbol{\theta}^{\prime}}Q_{n}\left(\boldsymbol{\theta},\overline{\boldsymbol{\mu}}_{n}\left(\widehat{\boldsymbol{\theta}}^{(S_{n}-1)}\right)\right)\mid_{\boldsymbol{\theta}=\widehat{\boldsymbol{\theta}}^{\left(S_{n}-1\right)}+\xi\left(\widehat{\boldsymbol{\theta}}^{\left(S_{n}\right)}-\widehat{\boldsymbol{\theta}}^{\left(S_{n}-1\right)}\right)}d\xi\right]\left(\widehat{\boldsymbol{\theta}}^{\left(S_{n}\right)}-\widehat{\boldsymbol{\theta}}^{\left(S_{n}-1\right)}\right).
\end{align*}
The integrated Hessian in the above expression is $O_{p}\left(1\right)$
by the proof of Lemma \ref{lem:: theta bar n convergence}. By Lemma
\ref{lem:: theta hat s+1 minus theta hat s}, we have $\widehat{\boldsymbol{\theta}}^{\left(S_{n}\right)}-\widehat{\boldsymbol{\theta}}^{\left(S_{n}-1\right)}=o_{p}\left(n^{-1/2}\right)$.
This implies that 
\[
\frac{1}{n}\frac{d}{d\boldsymbol{\theta}}l_{C\mid X,M}\left(\widehat{\boldsymbol{\theta}}^{\left(S_{n}-1\right)}\right)=o_{p}\left(n^{-1/2}\right).
\]
Applying the multivariate Taylor expansion to the left-hand side of
the above equality, we obtain that 
\begin{align}
o_{p}\left(n^{-1/2}\right)= & \frac{1}{n}\frac{d}{d\boldsymbol{\theta}}l_{C\mid X,M}\left(\boldsymbol{\theta}^{\ast}\right)\nonumber \\
 & +\left[\int^{1}_{0}\frac{1}{n}\frac{\partial^{2}}{\partial\boldsymbol{\theta}\partial\boldsymbol{\theta}^{\prime}}l_{C\mid X,M}\left(\boldsymbol{\theta}^{\ast}+\xi\left(\widehat{\boldsymbol{\theta}}^{\left(S_{n}-1\right)}-\boldsymbol{\theta}^{\ast}\right)\right)d\xi\right]\left(\widehat{\boldsymbol{\theta}}^{\left(S_{n}-1\right)}-\boldsymbol{\theta}^{\ast}\right).\label{eq:Influence function theta S-1}
\end{align}

By Theorems \ref{thm:: consistency-fixed S} and \ref{thm:: consistency},
we have that $\widehat{\boldsymbol{\theta}}^{\left(S_{n}-1\right)}\stackrel{p}{\rightarrow}\boldsymbol{\theta}^{\ast}$
as $n\rightarrow\infty$. Together with the uniform convergence and
continuity of $\frac{1}{n}\frac{\partial^{2}}{\partial\boldsymbol{\theta}\partial\boldsymbol{\theta}^{\prime}}l_{C\mid X,M}\left(\cdot\right)$,
it holds that 
\[
-\int^{1}_{0}\frac{1}{n}\frac{\partial^{2}}{\partial\boldsymbol{\theta}\partial\boldsymbol{\theta}^{\prime}}l_{C\mid X,M}\left(\boldsymbol{\theta}^{\ast}+\xi\left(\widehat{\boldsymbol{\theta}}^{\left(S_{n}-1\right)}-\boldsymbol{\theta}^{\ast}\right)\right)d\xi=-\frac{1}{n}\frac{\partial^{2}}{\partial\boldsymbol{\theta}\partial\boldsymbol{\theta}^{\prime}}l_{C\mid X,M}\left(\boldsymbol{\theta}^{\ast}\right)+o_{p}\left(1\right)\stackrel{p}{\rightarrow}\mathcal{I}\left(\boldsymbol{\theta}^{\ast}\right),
\]
where $\mathcal{I}\left(\boldsymbol{\theta}^{\ast}\right)$ is the
Fisher information matrix defined in Section \ref{subsec:Asymptotic-Distributions-Inference}.
Since the Fisher information matrix is positive definite, the negative
integrated Hessian is non-singular with probability approaching one.
Because matrix inversion is continuous (at non-singular matrices),
we have that 
\[
\left[-\int^{1}_{0}\frac{1}{n}\frac{\partial^{2}}{\partial\boldsymbol{\theta}\partial\boldsymbol{\theta}^{\prime}}l_{C\mid X,M}\left(\boldsymbol{\theta}^{\ast}+\xi\left(\widehat{\boldsymbol{\theta}}^{\left(S_{n}-1\right)}-\boldsymbol{\theta}^{\ast}\right)\right)d\xi\right]^{-1}\stackrel{p}{\rightarrow}\mathcal{I}^{-1}\left(\boldsymbol{\theta}^{\ast}\right).
\]
Using this result to (\ref{eq:Influence function theta S-1}), we
obtain that 
\[
\widehat{\boldsymbol{\theta}}^{\left(S_{n}-1\right)}-\boldsymbol{\theta}^{\ast}=\mathcal{I}^{-1}\left(\boldsymbol{\theta}^{\ast}\right)\frac{1}{n}\frac{d}{d\boldsymbol{\theta}}l_{C\mid X,M}\left(\boldsymbol{\theta}^{\ast}\right)+o_{p}\left(n^{-1/2}\right).
\]
Since $\widehat{\boldsymbol{\theta}}^{\left(S_{n}\right)}-\boldsymbol{\theta}^{\ast}=\widehat{\boldsymbol{\theta}}^{\left(S_{n}-1\right)}-\boldsymbol{\theta}^{\ast}+o_{p}\left(n^{-1/2}\right)$
by Lemma \ref{lem:: theta hat s+1 minus theta hat s}, it holds that
\begin{align*}
\widehat{\boldsymbol{\theta}}^{\left(S_{n}\right)}-\boldsymbol{\theta}^{\ast} & =\mathcal{I}^{-1}\left(\boldsymbol{\theta}^{\ast}\right)\frac{1}{n}\frac{d}{d\boldsymbol{\theta}}l_{C\mid X,M}\left(\boldsymbol{\theta}^{\ast}\right)+o_{p}\left(n^{-1/2}\right).
\end{align*}
It can be seen that $\widehat{\boldsymbol{\theta}}^{\left(S_{n}\right)}$
and the maximum likelihood estimator $\widetilde{\boldsymbol{\theta}}$
have the same influence function. Hence, under the assumptions in
either part (i) or (ii) of the theorem, we have $\widehat{\boldsymbol{\theta}}^{\left(S_{n}\right)}-\widetilde{\boldsymbol{\theta}}=o_{p}\left(n^{-1/2}\right)$.
\qed

\paragraph{Proof of Corollary \ref{Cor:: Asymptotics}:}

The result directly follows from Theorem \ref{Thm:: equivalence IDC and theta tilda}
and the standard result on the asymptotic distribution of the maximum
likelihood estimator. \qed

\paragraph{Proof of Theorem \ref{thm:: Bootstrap}:}

The proof of the theorem follows from the discussion in Section \ref{subsec:Asymptotic-Distributions-Inference}.\qed

\paragraph{Proof of Corollary \ref{corr:: Constrained IDC Consistency}:}

For either of the two constraints, the consistency of the unconstrained
estimator $\check{\boldsymbol{\theta}}$ follows from Lemma \ref{lem:: Ck Cd distribution}
because the MNL is still correctly specified. Therefore, it suffices
to prove that $\widehat{\boldsymbol{\theta}}^{\left(1\right)}\stackrel{p}{\rightarrow}\boldsymbol{\theta}^{\ast}$.

For the first set of constraints $\theta^{\ast}_{k,1}=\theta^{\ast}_{k,2}$
for $k=1,\ldots,d$ , we have that 
\[
\widehat{\boldsymbol{\theta}}^{\left(1\right)}_{k}=\arg\min_{\boldsymbol{\theta}_{k}\in\varTheta_{k},\theta_{k,1}=\theta_{k,2}}Q_{k,n}\left(\boldsymbol{\theta}_{k},\overline{\boldsymbol{\mu}}_{n}\left(\check{\boldsymbol{\theta}}\right)\right).
\]
Following the same argument as in the proof of Theorem \ref{thm:: consistency-fixed S},
we have $\widehat{\boldsymbol{\theta}}^{\left(1\right)}_{k}\stackrel{p}{\rightarrow}\boldsymbol{\theta}^{\ast}_{k}$
because the constraint $\theta_{k,1}=\theta_{k,2}$ is correct. Thus,
it holds that $\widehat{\boldsymbol{\theta}}^{\left(1\right)}\stackrel{p}{\rightarrow}\boldsymbol{\theta}^{\ast}$.

For the second set of constraints $\theta^{\ast}_{1,1}=\cdots=\theta^{\ast}_{q,1}$,
where $q\leq d$, each component of $\widehat{\boldsymbol{\theta}}^{\left(1\right)}$
is computed as 
\begin{align*}
\widehat{\boldsymbol{\theta}}^{\left(1\right)}_{k,-1} & =\arg\min_{\boldsymbol{\theta}_{k,-1}}Q_{k,n}\left(\check{\theta}_{k,1},\boldsymbol{\theta}_{k,-1},\overline{\boldsymbol{\mu}}_{n}\left(\check{\boldsymbol{\theta}}\right)\right)\textrm{ for }1\leq k\leq q,\\
\widehat{\boldsymbol{\theta}}^{\left(1\right)}_{k} & =\arg\min_{\boldsymbol{\theta}_{k}}Q_{k,n}\left(\boldsymbol{\theta}_{k},\overline{\boldsymbol{\mu}}_{n}\left(\check{\boldsymbol{\theta}}\right)\right)\textrm{ for }q+1\leq k\leq d,\textrm{ and}\\
\widehat{\theta}^{\left(1\right)}_{1,1} & =\arg\min_{\theta_{1,1}}\sum^{q}_{k=1}Q_{k,n}\left(\theta_{1,1},\widehat{\boldsymbol{\theta}}^{\left(1\right)}_{k,-1},\overline{\boldsymbol{\mu}}_{n}\left(\check{\boldsymbol{\theta}}\right)\right).
\end{align*}
The consistency of $\widehat{\boldsymbol{\theta}}^{\left(1\right)}_{k}$
for $q+1\leq k\leq d$ follows from the proof of Theorem \ref{thm:: consistency-fixed S}.
As for $\widehat{\boldsymbol{\theta}}^{\left(1\right)}_{k,-1}$ for
$1\leq k\leq q$, the first order condition provides that 
\[
\frac{\partial}{\partial\boldsymbol{\theta}_{k,-1}}Q_{k,n}\left(\check{\theta}_{k,1},\boldsymbol{\theta}_{k,-1},\overline{\boldsymbol{\mu}}_{n}\left(\check{\boldsymbol{\theta}}\right)\right)\mid_{\boldsymbol{\theta}_{k,-1}=\widehat{\boldsymbol{\theta}}^{\left(1\right)}_{k,-1}}=\mathbf{0}.
\]
By the mean value theorem, we have that 
\begin{align*}
\mathbf{0} & =\frac{1}{n}\frac{\partial}{\partial\boldsymbol{\theta}_{k,-1}}Q_{k,n}\left(\check{\theta}_{k,1},\boldsymbol{\theta}_{k,-1},\overline{\boldsymbol{\mu}}_{n}\left(\check{\boldsymbol{\theta}}\right)\right)\mid_{\boldsymbol{\theta}_{k,-1}=\boldsymbol{\theta}^{\ast}_{k,-1}}\\
 & +\frac{1}{n}\frac{\partial^{2}}{\partial\boldsymbol{\theta}_{k,-1}\partial\boldsymbol{\theta}^{\prime}_{k,-1}}Q_{k,n}\left(\check{\theta}_{k,1},\boldsymbol{\theta}_{k,-1},\overline{\boldsymbol{\mu}}_{n}\left(\check{\boldsymbol{\theta}}\right)\right)\mid_{\boldsymbol{\theta}_{k,-1}=\boldsymbol{\theta}^{\star}_{k,-1}}\left(\widehat{\boldsymbol{\theta}}^{\left(1\right)}_{k,-1}-\boldsymbol{\theta}^{\ast}_{k,-1}\right)\\
 & \equiv A_{n}+B_{n}\left(\widehat{\boldsymbol{\theta}}^{\left(1\right)}_{k,-1}-\boldsymbol{\theta}^{\ast}_{k,-1}\right),
\end{align*}
where $\boldsymbol{\theta}^{\star}_{k,-1}$ lies between $\widehat{\boldsymbol{\theta}}^{\left(1\right)}_{k,-1}$
and $\boldsymbol{\theta}^{\ast}_{k,-1}$. Since $\check{\boldsymbol{\theta}}\stackrel{p}{\rightarrow}\boldsymbol{\theta}^{\ast}$
and in particular, $\check{\theta}_{k,1}\stackrel{p}{\rightarrow}\theta^{\ast}_{k,1}$,
we have that 
\begin{align*}
A_{n} & \stackrel{p}{\rightarrow}\mathbb{E}\left[\frac{Me^{\boldsymbol{X}^{\prime}\boldsymbol{\theta}^{\ast}_{1}}}{\sum^{d}_{k=1}e^{\boldsymbol{X}^{\prime}\boldsymbol{\theta}^{\ast}_{k}}}\boldsymbol{X}_{-1}-C_{k}\boldsymbol{X}_{-1}\right]\\
 & =\mathbb{E}\left[\mathbb{E}\left[\frac{Me^{\boldsymbol{X}^{\prime}\boldsymbol{\theta}^{\ast}_{1}}}{\sum^{d}_{k=1}e^{\boldsymbol{X}^{\prime}\boldsymbol{\theta}^{\ast}_{k}}}\boldsymbol{X}_{-1}-C_{k}\boldsymbol{X}_{-1}\mid\boldsymbol{X}_{-1},M\right]\right]=\mathbf{0},
\end{align*}
where $\boldsymbol{X}_{-1}$ denotes the subvector of $\boldsymbol{X}$
excluding its first element. By a similar argument, it can be shown
that $B_{n}$ converges in probability to a non-singular matrix. Therefore,
it must hold that $\widehat{\boldsymbol{\theta}}^{\left(1\right)}_{k,-1}\stackrel{p}{\rightarrow}\boldsymbol{\theta}^{\ast}_{k,-1}$.
Lastly, the consistency of $\widehat{\theta}^{\left(1\right)}_{1,1}$
follows from the same argument above together with the result that
both $\widehat{\boldsymbol{\theta}}^{\left(1\right)}_{k,-1}$ for
$1\leq k\leq q$ and $\check{\boldsymbol{\theta}}$ are consistent.
Hence, each element of $\widehat{\boldsymbol{\theta}}^{\left(1\right)}$
is consistent and $\widehat{\boldsymbol{\theta}}^{\left(1\right)}\stackrel{p}{\rightarrow}\boldsymbol{\theta}^{\ast}$
as $n\rightarrow\infty$. \qed

\end{appendices} \newpage{}  \bibliographystyle{ecta}
\bibliography{DMR}

@article{tibshirani1996regression,
	author = {Tibshirani, Robert},
	journal = {Journal of the Royal Statistical Society Series B: Statistical Methodology},
	number = {1},
	pages = {267--288},
	publisher = {Oxford University Press},
	title = {Regression shrinkage and selection via the lasso},
	volume = {58},
	year = {1996}}

@book{boyd2004convex,
	author = {Boyd, Stephen and Vandenberghe, Lieven},
	publisher = {Cambridge university press},
	title = {Convex optimization},
	year = {2004}}

@book{horowitz2012semiparametric,
	author = {Horowitz, Joel L},
	publisher = {Springer Science \& Business Media},
	title = {Semiparametric methods in econometrics},
	volume = {131},
	year = {2012}}

@book{lehmann1998theory,
	author = {Lehmann, Erich Leo and Casella, George},
	publisher = {Springer},
	title = {Theory of point estimation},
	year = {1998}}

@article{mcfadden1978modeling,
	author = {McFadden, Daniel},
	journal = {Transportation Research Record},
	number = {673},
	title = {Modeling the choice of residential location},
	year = {1978}}

@article{pan2022note,
	author = {Pan, Rui and Ren, Tunan and Guo, Baishan and Li, Feng and Li, Guodong and Wang, Hansheng},
	journal = {Journal of Business \& Economic Statistics},
	number = {4},
	pages = {1691--1700},
	publisher = {Taylor \& Francis},
	title = {A note on distributed quantile regression by pilot sampling and one-step updating},
	volume = {40},
	year = {2022}}

@article{battey2018distributed,
	author = {Battey, Heather and Fan, Jianqing and Liu, Han and Lu, Junwei and Zhu, Ziwei},
	journal = {Annals of statistics},
	number = {3},
	pages = {1352},
	title = {Distributed testing and estimation under sparse high dimensional models},
	volume = {46},
	year = {2018}}

@article{zhu2021least,
	author = {Zhu, Xuening and Li, Feng and Wang, Hansheng},
	journal = {Journal of Computational and Graphical Statistics},
	number = {4},
	pages = {1004--1018},
	publisher = {Taylor \& Francis},
	title = {Least-square approximation for a distributed system},
	volume = {30},
	year = {2021}}

@article{chang2017distributed,
	author = {Chang, Xiangyu and Lin, Shao-Bo and Zhou, Ding-Xuan},
	journal = {Journal of Machine Learning Research},
	number = {46},
	pages = {1--22},
	title = {Distributed semi-supervised learning with kernel ridge regression},
	volume = {18},
	year = {2017}}

@article{jordan2019communication,
	author = {Jordan, Michael I and Lee, Jason D and Yang, Yun},
	journal = {Journal of the American Statistical Association},
	publisher = {Taylor \& Francis},
	title = {Communication-efficient distributed statistical inference},
	year = {2019}}

@article{zhang2013communication,
	author = {Zhang, Yuchen and Duchi, John C and Wainwright, Martin J},
	journal = {The Journal of Machine Learning Research},
	number = {1},
	pages = {3321--3363},
	publisher = {JMLR. org},
	title = {Communication-efficient algorithms for statistical optimization},
	volume = {14},
	year = {2013}}

@article{wu2025class,
	author = {Wu, Shuyuan and Zhou, Jing and Xu, Ke and Wang, Hansheng},
	journal = {Journal of Computational and Graphical Statistics},
	number = {1},
	pages = {175--186},
	publisher = {Taylor \& Francis},
	title = {Class-Distributed Learning for Multinomial Logistic Regression with High Dimensional Features and a Large Number of Classes},
	volume = {34},
	year = {2025}}

@article{strombom2002switching,
	author = {Strombom, Bruce A and Buchmueller, Thomas C and Feldstein, Paul J},
	journal = {Journal of Health economics},
	number = {1},
	pages = {89--116},
	publisher = {Elsevier},
	title = {Switching costs, price sensitivity and health plan choice},
	volume = {21},
	year = {2002}}

@article{train1987demand,
	author = {Train, Kenneth E and McFadden, Daniel L and Ben-Akiva, Moshe},
	journal = {The RAND Journal of Economics},
	pages = {109--123},
	publisher = {JSTOR},
	title = {The demand for local telephone service: A fully discrete model of residential calling patterns and service choices},
	year = {1987}}

@article{hausman1984econometric,
	author = {Hausman, Jerry and Hall, Bronwyn H and Griliches, Zvi},
	journal = {Econometrica: Journal of the Econometric Society},
	pages = {909--938},
	publisher = {JSTOR},
	title = {Econometric Models for Count Data with an Application to the Patents-R \& D Relationship},
	year = {1984}}

@article{andersen1970asymptotic,
	author = {Andersen, Erling Bernhard},
	journal = {Journal of the Royal Statistical Society Series B: Statistical Methodology},
	number = {2},
	pages = {283--301},
	publisher = {Oxford University Press},
	title = {Asymptotic properties of conditional maximum-likelihood estimators},
	volume = {32},
	year = {1970}}

@article{kasahara2012sequential,
	author = {Kasahara, Hiroyuki and Shimotsu, Katsumi},
	journal = {Econometrica},
	number = {5},
	pages = {2303--2319},
	publisher = {Wiley Online Library},
	title = {Sequential estimation of structural models with a fixed point constraint},
	volume = {80},
	year = {2012}}

@article{aguirregabiria2007sequential,
	author = {Aguirregabiria, Victor and Mira, Pedro},
	journal = {Econometrica},
	number = {1},
	pages = {1--53},
	publisher = {Wiley Online Library},
	title = {Sequential estimation of dynamic discrete games},
	volume = {75},
	year = {2007}}

@article{aguirregabiria2002swapping,
	author = {Aguirregabiria, Victor and Mira, Pedro},
	journal = {Econometrica},
	number = {4},
	pages = {1519--1543},
	publisher = {Wiley Online Library},
	title = {Swapping the nested fixed point algorithm: A class of estimators for discrete Markov decision models},
	volume = {70},
	year = {2002}}

@article{baker1994multinomial,
	author = {Baker, Stuart G},
	journal = {Journal of the Royal Statistical Society: Series D (The Statistician)},
	number = {4},
	pages = {495--504},
	publisher = {Wiley Online Library},
	title = {The multinomial-Poisson transformation},
	volume = {43},
	year = {1994}}

@article{friedman2010regularization,
	author = {Friedman, Jerome and Hastie, Trevor and Tibshirani, Rob},
	journal = {Journal of Statistical Software},
	number = {1},
	pages = {1},
	publisher = {NIH Public Access},
	title = {Regularization paths for generalized linear models via coordinate descent},
	volume = {33},
	year = {2010}}

@book{train2009discrete,
	author = {Train, Kenneth E},
	publisher = {Cambridge university press},
	title = {Discrete choice methods with simulation},
	year = {2009}}

@inproceedings{mcfadden1973conditional,
	author = {McFadden, Daniel},
	booktitle = {Frontiers in Econometrics},
	organization = {edited by P. Zarembka},
	publisher = {New York: Wiley},
	title = {Conditional logit analysis of qualitative choice behavior},
	year = {1973}}

@article{simon2013blockwise,
	author = {Simon, Noah and Friedman, Jerome and Hastie, Trevor},
	journal = {arXiv preprint arXiv:1311.6529},
	title = {A blockwise descent algorithm for group-penalized multiresponse and multinomial regression},
	year = {2013}}

@article{bohning1992multinomial,
	author = {B{\"o}hning, Dankmar},
	journal = {Annals of the institute of Statistical Mathematics},
	number = {1},
	pages = {197--200},
	publisher = {Springer},
	title = {Multinomial logistic regression algorithm},
	volume = {44},
	year = {1992}}

@article{bohning1988monotonicity,
	author = {B{\"o}hning, Dankmar and Lindsay, Bruce G},
	journal = {Annals of the Institute of Statistical Mathematics},
	number = {4},
	pages = {641--663},
	publisher = {Springer},
	title = {Monotonicity of quadratic-approximation algorithms},
	volume = {40},
	year = {1988}}

@article{nibbering2022multiclass,
	author = {Nibbering, Didier and Hastie, Trevor J},
	journal = {Computational Statistics \& Data Analysis},
	pages = {107414},
	publisher = {Elsevier},
	title = {Multiclass-penalized logistic regression},
	volume = {169},
	year = {2022}}

@article{dominitz2005some,
	author = {Dominitz, Jeff and Sherman, Robert P},
	journal = {Econometric Theory},
	number = {4},
	pages = {838--863},
	publisher = {Cambridge University Press},
	title = {Some convergence theory for iterative estimation procedures with an application to semiparametric estimation},
	volume = {21},
	year = {2005}}

@article{recht2011hogwild,
	author = {Recht, Benjamin and Re, Christopher and Wright, Stephen and Niu, Feng},
	journal = {Advances in Neural Information Processing Systems},
	title = {Hogwild!: A lock-free approach to parallelizing stochastic gradient descent},
	volume = {24},
	year = {2011}}

@article{raman2016ds,
	author = {Raman, Parameswaran and Srinivasan, Sriram and Matsushima, Shin and Zhang, Xinhua and Yun, Hyokun and Vishwanathan, SVN},
	journal = {arXiv preprint arXiv:1604.04706},
	title = {DS-MLR: exploiting double separability for scaling up distributed multinomial logistic regression},
	year = {2016}}

@article{fagan2018unbiased,
	author = {Fagan, Francois and Iyengar, Garud},
	journal = {arXiv preprint arXiv:1803.08577},
	title = {Unbiased scalable softmax optimization},
	year = {2018}}

@inproceedings{gopal2013distributed,
	author = {Gopal, Siddharth and Yang, Yiming},
	booktitle = {International Conference on Machine Learning},
	organization = {PMLR},
	pages = {289--297},
	title = {Distributed training of large-scale logistic models},
	year = {2013}}

@article{boyd2011distributed,
	author = {Boyd, Stephen and Parikh, Neal and Chu, Eric and Peleato, Borja and Eckstein, Jonathan},
	journal = {Foundations and Trends{\textregistered} in Machine learning},
	number = {1},
	pages = {1--122},
	publisher = {Now Publishers, Inc.},
	title = {Distributed optimization and statistical learning via the alternating direction method of multipliers},
	volume = {3},
	year = {2011}}

@article{russakovsky2015imagenet,
	author = {Russakovsky, Olga and Deng, Jia and Su, Hao and Krause, Jonathan and Satheesh, Sanjeev and Ma, Sean and Huang, Zhiheng and Karpathy, Andrej and Khosla, Aditya and Bernstein, Michael},
	journal = {International Journal of Computer Vision},
	pages = {211--252},
	publisher = {Springer},
	title = {Imagenet large scale visual recognition challenge},
	volume = {115},
	year = {2015}}

@inproceedings{davidson2010youtube,
	author = {Davidson, James and Liebald, Benjamin and Liu, Junning and Nandy, Palash and Van Vleet, Taylor and Gargi, Ullas and Gupta, Sujoy and He, Yu and Lambert, Mike and Livingston, Blake},
	booktitle = {Proceedings of the Fourth ACM Conference on Recommender Systems},
	pages = {293--296},
	title = {The YouTube video recommendation system},
	year = {2010}}

@book{dieudonne2011foundations,
	author = {Dieudonn{\'e}, Jean},
	publisher = {Read Books Ltd},
	title = {Foundations of Modern Analysis},
	year = {2011}}

@article{jennrich1969asymptotic,
	author = {Jennrich, Robert I},
	journal = {The Annals of Mathematical Statistics},
	number = {2},
	pages = {633--643},
	publisher = {JSTOR},
	title = {Asymptotic properties of non-linear least squares estimators},
	volume = {40},
	year = {1969}}

@article{gentzkow2019measuring,
	author = {Gentzkow, Matthew and Shapiro, Jesse M and Taddy, Matt},
	journal = {Econometrica},
	number = {4},
	pages = {1307--1340},
	publisher = {Wiley Online Library},
	title = {Measuring group differences in high-dimensional choices: method and application to congressional speech},
	volume = {87},
	year = {2019}}

@article{fan2015maximization,
	author = {Fan, Yanqin and Pastorello, Sergio and Renault, Eric},
	journal = {The Econometrics Journal},
	number = {2},
	pages = {147--171},
	publisher = {Oxford University Press Oxford, UK},
	title = {Maximization by parts in extremum estimation},
	volume = {18},
	year = {2015}}

@article{pastorello2003iterative,
	author = {Pastorello, Sergio and Patilea, Valentin and Renault, Eric},
	journal = {Journal of Business \& Economic Statistics},
	number = {4},
	pages = {449--509},
	publisher = {Taylor \& Francis},
	title = {Iterative and recursive estimation in structural nonadaptive models},
	volume = {21},
	year = {2003}}

@article{taddy2015distributed,
	author = {Taddy, Matt},
	journal = {The Annals of Applied Statistics},
	number = {3},
	pages = {1394--1414},
	publisher = {Institute of Mathematical Statistics},
	title = {Distributed multinomial regression},
	volume = {9},
	year = {2015}}

@article{baker2006investor,
	author = {Baker, Malcolm and Wurgler, Jeffrey},
	journal = {Journal of Finance},
	number = {4},
	pages = {1645--1680},
	publisher = {Wiley Online Library},
	title = {Investor sentiment and the cross-section of stock returns},
	volume = {61},
	year = {2006}}

@article{chen2021sentiment,
	author = {Chen, Yong and Han, Bing and Pan, Jing},
	journal = {Journal of Finance},
	number = {4},
	pages = {2001--2033},
	publisher = {Wiley Online Library},
	title = {Sentiment trading and hedge fund returns},
	volume = {76},
	year = {2021}}

@article{taddy2013multinomial,
	author = {Taddy, Matt},
	journal = {Journal of the American Statistical Association},
	number = {503},
	pages = {755--770},
	publisher = {Taylor \& Francis},
	title = {Multinomial inverse regression for text analysis},
	volume = {108},
	year = {2013}}

@article{bucholztaxi2019,
	author = {Buchholz, Nicholas},
	doi = {10.1093/restud/rdab050},
	eprint = {https://academic.oup.com/restud/article-pdf/89/2/556/42748179/rdab050.pdf},
	issn = {0034-6527},
	journal = {The Review of Economic Studies},
	month = {09},
	number = {2},
	pages = {556-591},
	title = {{Spatial equilibrium, search frictions, and dynamic efficiency in the taxi industry}},
	url = {https://doi.org/10.1093/restud/rdab050},
	volume = {89},
	year = {2021}}

@article{pellegrinifotheringham2002,
	author = {Pasquale A. Pellegrini and A. Stewart Fotheringham},
	doi = {10.1191/0309132502ph382ra},
	eprint = {https://doi.org/10.1191/0309132502ph382ra},
	journal = {Progress in Human Geography},
	number = {4},
	pages = {487-510},
	title = {Modelling spatial choice: a review and synthesis in a migration context},
	url = {https://doi.org/10.1191/0309132502ph382ra},
	volume = {26},
	year = {2002}}

@article{bettman1979memory,
	author = {Bettman, James R},
	journal = {Journal of Marketing},
	number = {2},
	pages = {37--53},
	publisher = {SAGE Publications Sage CA: Los Angeles, CA},
	title = {Memory factors in consumer choice: A review},
	volume = {43},
	year = {1979}}

@techreport{NBERwTextSelection,
	author = {Kelly, Bryan T and Manela, Asaf and Moreira, Alan},
	doi = {10.3386/w26517},
	institution = {National Bureau of Economic Research},
	month = {November},
	number = {26517},
	series = {Working Paper Series},
	title = {Text selection},
	type = {Working Paper},
	url = {http://www.nber.org/papers/w26517},
	year = {2019}}

\end{document}